%
%
%
\def\unredoffs{} \def\redoffs{\voffset=-.31truein\hoffset=-.48truein}
\def\speclscape{}
%
%
%
%
%
\newbox\leftpage \newdimen\fullhsize \newdimen\hstitle \newdimen\hsbody
\tolerance=1000\hfuzz=2pt
\catcode`\@=11 
\ifx\hyperdef\UNd@FiNeD\def\hyperdef#1#2#3#4{#4}\def\hyperref#1#2#3#4{#4}\fi
\def\bigans{b }
\def\answ{b }
%
\ifx\answ\bigans\message{(This will come out unreduced.}
\magnification=1200\unredoffs\baselineskip=16pt plus 2pt minus 1pt
\hsbody=\hsize \hstitle=\hsize 
\else\message{(This will be reduced.} \let\l@r=L
\magnification=1000\baselineskip=16pt plus 2pt minus 1pt \vsize=7truein
\redoffs \hstitle=8truein\hsbody=4.75truein\fullhsize=10truein\hsize=\hsbody
\output={\ifnum\pageno=0 
  \shipout\vbox{\speclscape{\hsize\fullhsize\makeheadline}
    \hbox to \fullhsize{\hfill\pagebody\hfill}}\advancepageno
  \else
  \almostshipout{\leftline{\vbox{\pagebody\makefootline}}}\advancepageno
  \fi}
\def\almostshipout#1{\if L\l@r \count1=1 \message{[\the\count0.\the\count1]}
      \global\setbox\leftpage=#1 \global\let\l@r=R
 \else \count1=2
  \shipout\vbox{\speclscape{\hsize\fullhsize\makeheadline}
      \hbox to\fullhsize{\box\leftpage\hfil#1}}  \global\let\l@r=L\fi}
\fi
%
\newcount\yearltd\yearltd=\year\advance\yearltd by -2000

\def\Title#1#2{\nopagenumbers\abstractfont\hsize=\hstitle\rightline{#1}%
\vskip 1in\centerline{\titlefont #2}\abstractfont\vskip .5in\pageno=0}
\def\Date#1{\vfill\leftline{#1}\tenpoint\supereject\global\hsize=\hsbody%
\footline={\hss\tenrm\hyperdef\hypernoname{page}\folio\folio\hss}}%
%

\def\draftmode{\message{ DRAFTMODE }\def\draftdate{{\rm preliminary draft:
\number\month/\number\day/\number\yearltd\ \ \hourmin}}%
\headline={\hfil\draftdate}\writelabels\baselineskip=20pt plus 2pt minus 2pt
 {\count255=\time\divide\count255 by 60 \xdef\hourmin{\number\count255}
  \multiply\count255 by-60\advance\count255 by\time
  \xdef\hourmin{\hourmin:\ifnum\count255<10 0\fi\the\count255}}}
\def\nolabels{\def\wrlabeL##1{}\def\eqlabeL##1{}\def\reflabeL##1{}}
\def\writelabels{\def\wrlabeL##1{\leavevmode\vadjust{\rlap{\smash%
{\line{{\escapechar=` \hfill\rlap{\sevenrm\hskip.03in\string##1}}}}}}}%
\def\eqlabeL##1{{\escapechar-1\rlap{\sevenrm\hskip.05in\string##1}}}%
\def\reflabeL##1{\noexpand\llap{\noexpand\sevenrm\string\string\string##1}}}
\nolabels
%
\global\newcount\secno \global\secno=0
\global\newcount\meqno \global\meqno=1
\def\s@csym{}
\def\newsec#1{\global\advance\secno by1%
{\toks0{#1}\message{(\the\secno. \the\toks0)}}%
\global\subsecno=0\eqnres@t\let\s@csym\secsym\xdef\secn@m{\the\secno}\noindent
{\bf\hyperdef\hypernoname{section}{\the\secno}{\the\secno.} #1}%
\writetoca{{\string\hyperref{}{section}{\the\secno}{\it\the\secno.}} {{\it #1} }}%
\par\nobreak\medskip\nobreak}
\def\eqnres@t{\xdef\secsym{\the\secno.}\global\meqno=1\bigbreak\bigskip}
\def\sequentialequations{\def\eqnres@t{\bigbreak}}\xdef\secsym{}
\global\newcount\subsecno \global\subsecno=0
\def\subsec#1{\global\advance\subsecno by1%
{\toks0{#1}\message{(\s@csym\the\subsecno. \the\toks0)}}%
\ifnum\lastpenalty>9000\else\bigbreak\fi       \global\subsubsecno=0
\noindent{\it\hyperdef\hypernoname{subsection}{\secn@m.\the\subsecno}%
{\secn@m.\the\subsecno.} #1}\writetoca{\string\quad
{\string\hyperref{}{subsection}{\secn@m.\the\subsecno}{\secn@m.\the\subsecno.}}
{#1}}\par\nobreak\medskip\nobreak}
\def\appendix#1#2{\global\meqno=1\global\subsecno=0\xdef\secsym{\hbox{#1.}}%
\bigbreak\bigskip\noindent{\bf Appendix \hyperdef\hypernoname{appendix}{#1}%
{#1.} #2}{\toks0{(#1. #2)}\message{\the\toks0}}%
\xdef\s@csym{#1.}\xdef\secn@m{#1}%
\writetoca{\string\hyperref{}{appendix}{#1}{{\it Appendix} {\it #1.}} {\it #2}}%
\par\nobreak\medskip\nobreak}
%
%
\def\checkm@de#1#2{\ifmmode{\def\f@rst##1{##1}\hyperdef\hypernoname{equation}%
{#1}{#2}}\else\hyperref{}{equation}{#1}{#2}\fi}
\def\eqnn#1{\DefWarn#1\xdef #1{(\noexpand\relax\noexpand\checkm@de%
{\s@csym\the\meqno}{\secsym\the\meqno})}%
\wrlabeL#1\writedef{#1\leftbracket#1}\global\advance\meqno by1}
\def\f@rst#1{\c@t#1a\em@ark}\def\c@t#1#2\em@ark{#1}
\def\eqna#1{\DefWarn#1\wrlabeL{#1$\{\}$}%
\xdef #1##1{(\noexpand\relax\noexpand\checkm@de%
{\s@csym\the\meqno\noexpand\f@rst{##1}}{\hbox{$\secsym\the\meqno##1$}})}
\writedef{#1\numbersign1\leftbracket#1{\numbersign1}}\global\advance\meqno by1}
\def\eqn#1#2{\DefWarn#1%
\xdef #1{(\noexpand\hyperref{}{equation}{\s@csym\the\meqno}%
{\secsym\the\meqno})}$$#2\eqno(\hyperdef\hypernoname{equation}%
{\s@csym\the\meqno}{\secsym\the\meqno})\eqlabeL#1$$%
\writedef{#1\leftbracket#1}\global\advance\meqno by1}
\def\xeqn{\expandafter\xe@n}\def\xe@n(#1){#1}
\def\xeqna#1{\expandafter\xe@n#1}
\def\eqns#1{(\e@ns #1{\hbox{}})}
\def\e@ns#1{\ifx\UNd@FiNeD#1\message{eqnlabel \string#1 is undefined.}%
\xdef#1{(?.?)}\fi{\let\hyperref=\relax\xdef\next{#1}}%
\ifx\next\em@rk\def\next{}\else%
\ifx\next#1\xeqn#1\else\def\n@xt{#1}\ifx\n@xt\next#1\else\xeqna#1\fi
\fi\let\next=\e@ns\fi\next}

\def\DefWarn#1{\ifx\UNd@FiNeD#1\else
\immediate\write16{*** WARNING: the label \string#1 is already defined ***}\fi}
%
\newskip\footskip\footskip14pt plus 1pt minus 1pt 
\def\footnotefont{\ninepoint}\def\f@t#1{\footnotefont #1\@foot}
\def\f@@t{\baselineskip\footskip\bgroup\footnotefont\aftergroup\@foot\let\next}
\setbox\strutbox=\hbox{\vrule height9.5pt depth4.5pt width0pt}
\global\newcount\ftno \global\ftno=0
\def\foot{\global\advance\ftno by1\def\foot@rg{\hyperref{}{footnote}%
{\the\ftno}{\the\ftno}\xdef\foot@rg{\noexpand\hyperdef\noexpand\hypernoname%
{footnote}{\the\ftno}{\the\ftno}}}\footnote{$^{\foot@rg}$}}
%
\newwrite\ftfile
\def\footend{\def\foot{\global\advance\ftno by1\chardef\wfile=\ftfile
\hyperref{}{footnote}{\the\ftno}{$^{\the\ftno}$}%
\ifnum\ftno=1\immediate\openout\ftfile=\jobname.fts\fi%
\immediate\write\ftfile{\noexpand\smallskip%
\noexpand\item{\noexpand\hyperdef\noexpand\hypernoname{footnote}
{\the\ftno}{f\the\ftno}:\ }\pctsign}\findarg}%
\def\footatend{\vfill\eject\immediate\closeout\ftfile{\parindent=20pt
\centerline{\bf Footnotes}\nobreak\bigskip\input \jobname.fts }}}
\def\footatend{}
%
%
\global\newcount\refno \global\refno=1
\newwrite\rfile
\def\ref{[\hyperref{}{reference}{\the\refno}{\the\refno}]\nref}
\def\nref#1{\DefWarn#1%
\xdef#1{[\noexpand\hyperref{}{reference}{\the\refno}{\the\refno}]}%
\writedef{#1\leftbracket#1}%
\ifnum\refno=1\immediate\openout\rfile=\jobname.refs\fi
\chardef\wfile=\rfile\immediate\write\rfile{\noexpand\item{[\noexpand\hyperdef%
\noexpand\hypernoname{reference}{\the\refno}{\the\refno}]\ }%
\reflabeL{#1\hskip.31in}\pctsign}\global\advance\refno by1\findarg}
\def\findarg#1#{\begingroup\obeylines\newlinechar=`\^^M\pass@rg}
{\obeylines\gdef\pass@rg#1{\writ@line\relax #1^^M\hbox{}^^M}%
\gdef\writ@line#1^^M{\expandafter\toks0\expandafter{\striprel@x #1}%
\edef\next{\the\toks0}\ifx\next\em@rk\let\next=\endgroup\else\ifx\next\empty%
\else\immediate\write\wfile{\the\toks0}\fi\let\next=\writ@line\fi\next\relax}}
\def\striprel@x#1{} \def\em@rk{\hbox{}}
\def\lref{\begingroup\obeylines\lr@f}
\def\lr@f#1#2{\DefWarn#1\gdef#1{\let#1=\UNd@FiNeD\ref#1{#2}}\endgroup\unskip}

\def\addref#1{\immediate\write\rfile{\noexpand\item{}#1}} 
\def\listrefs{\footatend\vfill\supereject\immediate\closeout\rfile\writestoppt
\baselineskip=\footskip\centerline{{\bf References}}\bigskip{\parindent=20pt%
\frenchspacing\escapechar=` \input \jobname.refs\vfill\eject}\nonfrenchspacing}
\def\startrefs#1{\immediate\openout\rfile=\jobname.refs\refno=#1}
\def\xref{\expandafter\xr@f}\def\xr@f[#1]{#1}
\def\refs#1{\count255=1[\r@fs #1{\hbox{}}]}
\def\r@fs#1{\ifx\UNd@FiNeD#1\message{reflabel \string#1 is undefined.}%
\nref#1{need to supply reference \string#1.}\fi%
\vphantom{\hphantom{#1}}{\let\hyperref=\relax\xdef\next{#1}}%
\ifx\next\em@rk\def\next{}%
\else\ifx\next#1\ifodd\count255\relax\xref#1\count255=0\fi%
\else#1\count255=1\fi\let\next=\r@fs\fi\next}
%

%
\newwrite\ffile\global\newcount\figno \global\figno=1
\def\fig{fig.~\hyperref{}{figure}{\the\figno}{\the\figno}\nfig}
\def\nfig#1{\DefWarn#1%
\xdef#1{fig.~\noexpand\hyperref{}{figure}{\the\figno}{\the\figno}}%
\writedef{#1\leftbracket fig.\noexpand~\xfig#1}%
\ifnum\figno=1\immediate\openout\ffile=\jobname.figs\fi\chardef\wfile=\ffile%
{\let\hyperref=\relax
\immediate\write\ffile{\noexpand\medskip\noexpand\item{Fig.\ %
\noexpand\hyperdef\noexpand\hypernoname{figure}{\the\figno}{\the\figno}. }
\reflabeL{#1\hskip.55in}\pctsign}}\global\advance\figno by1\findarg}
\def\listfigs{\vfill\eject\immediate\closeout\ffile{\parindent40pt
\baselineskip14pt\centerline{{\bf Figure Captions}}\nobreak\medskip
\escapechar=` \input \jobname.figs\vfill\eject}}
\def\xfig{\expandafter\xf@g}\def\xf@g fig.\penalty\@M\ {}
\def\figs#1{figs.~\f@gs #1{\hbox{}}}
\def\f@gs#1{{\let\hyperref=\relax\xdef\next{#1}}\ifx\next\em@rk\def\next{}\else
\ifx\next#1\xfig #1\else#1\fi\let\next=\f@gs\fi\next}
\def\figin{\epsfcheck\figin}\def\figins{\epsfcheck\figins}
\def\epsfcheck{\ifx\epsfbox\UNd@FiNeD
\message{(NO epsf.tex, FIGURES WILL BE IGNORED)}
\gdef\figin##1{\vskip2in}\gdef\figins##1{\hskip.5in}
\else\message{(FIGURES WILL BE INCLUDED)}%
\gdef\figin##1{##1}\gdef\figins##1{##1}\fi}
\def\DefWarn#1{}
\def\figinsert{\goodbreak\midinsert}
\def\ifig#1#2#3{\DefWarn#1\xdef#1{Fig.~\noexpand\hyperref{}{figure}%
{\the\figno}{\the\figno}}\writedef{#1\leftbracket fig.\noexpand~\xfig#1}%
\figinsert\figin{\centerline{#3}}\medskip\centerline{\vbox{\baselineskip12pt
\advance\hsize by -1truein\noindent\wrlabeL{#1=#1}\footnotefont%
{\bf Fig.~\hyperdef\hypernoname{figure}{\the\figno}{\the\figno}:} #2}}
\bigskip\endinsert\global\advance\figno by1}
\newwrite\lfile
{\escapechar-1\xdef\pctsign{\string\%}\xdef\leftbracket{\string\{}
\xdef\rightbracket{\string\}}\xdef\numbersign{\string\#}}
\def\writedefs{\immediate\openout\lfile=\jobname.defs \def\writedef##1{%
{\let\hyperref=\relax\let\hyperdef=\relax\let\hypernoname=\relax
 \immediate\write\lfile{\string\def\string##1\rightbracket}}}}%
\def\writestop{\def\writestoppt{\immediate\write\lfile{\string\pageno
 \the\pageno\string\startrefs\leftbracket\the\refno\rightbracket
 \string\def\string\secsym\leftbracket\secsym\rightbracket
 \string\secno\the\secno\string\meqno\the\meqno}\immediate\closeout\lfile}}
\def\writestoppt{}\def\writedef#1{}
\def\seclab#1{\DefWarn#1%
\xdef #1{\noexpand\hyperref{}{section}{\the\secno}{\the\secno}}%
\writedef{#1\leftbracket#1}\wrlabeL{#1=#1}}
\def\subseclab#1{\DefWarn#1%
\xdef #1{\noexpand\hyperref{}{subsection}{\secn@m.\the\subsecno}%
{\secn@m.\the\subsecno}}\writedef{#1\leftbracket#1}\wrlabeL{#1=#1}}
\def\applab#1{\DefWarn#1%
\xdef #1{\noexpand\hyperref{}{appendix}{\secn@m}{\secn@m}}%
\writedef{#1\leftbracket#1}\wrlabeL{#1=#1}}
\newwrite\tfile \def\writetoca#1{}
\def\leaderfill{\leaders\hbox to 1em{\hss.\hss}\hfill}
\def\writetoc{\immediate\openout\tfile=\jobname.toc
   \def\writetoca##1{{\edef\next{\write\tfile{\noindent ##1
   \string\leaderfill {\string\hyperref{}{page}{\noexpand\number\pageno}%
                       {\noexpand\number\pageno}} \par}}\next}}}
\newread\ch@ckfile
\def\listtoc{\immediate\closeout\tfile\immediate\openin\ch@ckfile=\jobname.toc
\ifeof\ch@ckfile\message{no file \jobname.toc, no table of contents this pass}%
\else\closein\ch@ckfile\centerline{\bf Contents}\nobreak\medskip%
{\baselineskip=18.5pt  \footnotefont
\parskip=2pt\catcode`\@=12\input\jobname.toc
\catcode`\@=12\bigbreak\bigskip}\fi}
\catcode`\@=12 
%
\edef\tfontsize{\ifx\answ\bigans scaled\magstep3\else scaled\magstep4\fi}
\font\titlerm=cmr10 \tfontsize \font\titlerms=cmr7 \tfontsize
\font\titlermss=cmr5 \tfontsize \font\titlei=cmmi10 \tfontsize
\font\titleis=cmmi7 \tfontsize \font\titleiss=cmmi5 \tfontsize
\font\titlesy=cmsy10 \tfontsize \font\titlesys=cmsy7 \tfontsize
\font\titlesyss=cmsy5 \tfontsize \font\titleit=cmti10 \tfontsize
\skewchar\titlei='177 \skewchar\titleis='177 \skewchar\titleiss='177
\skewchar\titlesy='60 \skewchar\titlesys='60 \skewchar\titlesyss='60
\def\titlefont{\def\rm{\fam0\titlerm}
\textfont0=\titlerm \scriptfont0=\titlerms \scriptscriptfont0=\titlermss
\textfont1=\titlei \scriptfont1=\titleis \scriptscriptfont1=\titleiss
\textfont2=\titlesy \scriptfont2=\titlesys \scriptscriptfont2=\titlesyss
\textfont\itfam=\titleit \def\it{\fam\itfam\titleit}\rm}
 \ifx\answ\bigans\else scaled\magstep1\fi
\ifx\answ\bigans\def\abstractfont{\tenpoint}\else
\font\absit=cmti10 scaled \magstep1
\font\abssl=cmsl10 scaled \magstep1
\font\absrm=cmr10 scaled\magstep1 \font\absrms=cmr7 scaled\magstep1
\font\absrmss=cmr5 scaled\magstep1 \font\absi=cmmi10 scaled\magstep1
\font\absis=cmmi7 scaled\magstep1 \font\absiss=cmmi5 scaled\magstep1
\font\abssy=cmsy10 scaled\magstep1 \font\abssys=cmsy7 scaled\magstep1
\font\abssyss=cmsy5 scaled\magstep1 \font\absbf=cmbx10 scaled\magstep1
\skewchar\absi='177 \skewchar\absis='177 \skewchar\absiss='177
\skewchar\abssy='60 \skewchar\abssys='60 \skewchar\abssyss='60
\def\abstractfont{\def\rm{\fam0\absrm}
\textfont0=\absrm \scriptfont0=\absrms \scriptscriptfont0=\absrmss
\textfont1=\absi \scriptfont1=\absis \scriptscriptfont1=\absiss
\textfont2=\abssy \scriptfont2=\abssys \scriptscriptfont2=\abssyss
\textfont\itfam=\absit \def\it{\fam\itfam\absit}\def\footnotefont{\tenpoint}%
\textfont\slfam=\abssl \def\sl{\fam\slfam\abssl}%
\textfont\bffam=\absbf \def\bf{\fam\bffam\absbf}\rm}\fi
\def\tenpoint{\def\rm{\fam0\tenrm}
\textfont0=\tenrm \scriptfont0=\sevenrm \scriptscriptfont0=\fiverm
\textfont1=\teni  \scriptfont1=\seveni  \scriptscriptfont1=\fivei
\textfont2=\tensy \scriptfont2=\sevensy \scriptscriptfont2=\fivesy
\textfont\itfam=\tenit \def\it{\fam\itfam\tenit}\def\footnotefont{\ninepoint}%
\textfont\bffam=\tenbf \def\bf{\fam\bffam\tenbf}\def\sl{\fam\slfam\tensl}\rm}
\font\ninerm=cmr9 \font\sixrm=cmr6 \font\ninei=cmmi9 \font\sixi=cmmi6
\font\ninesy=cmsy9 \font\sixsy=cmsy6 \font\ninebf=cmbx9
\font\nineit=cmti9 \font\ninesl=cmsl9 \skewchar\ninei='177
\skewchar\sixi='177 \skewchar\ninesy='60 \skewchar\sixsy='60
\def\ninepoint{\def\rm{\fam0\ninerm}
\textfont0=\ninerm \scriptfont0=\sixrm \scriptscriptfont0=\fiverm
\textfont1=\ninei \scriptfont1=\sixi \scriptscriptfont1=\fivei
\textfont2=\ninesy \scriptfont2=\sixsy \scriptscriptfont2=\fivesy
\textfont\itfam=\ninei \def\it{\fam\itfam\nineit}\def\sl{\fam\slfam\ninesl}%
\textfont\bffam=\ninebf \def\bf{\fam\bffam\ninebf}\rm}
%
%
\def\noblackbox{\overfullrule=0pt}
\hyphenation{anom-aly anom-alies coun-ter-term coun-ter-terms}
\def\inv{^{\raise.15ex\hbox{${\scriptscriptstyle -}$}\kern-.05em 1}}

\def\Dsl{\,\raise.15ex\hbox{/}\mkern-13.5mu D} 
\def\dsl{\raise.15ex\hbox{/}\kern-.57em\partial}

\def\lspace{\ifx\answ\bigans{}\else\qquad\fi}
\def\lbspace{\ifx\answ\bigans{}\else\hskip-.2in\fi} 
\def\boxeqn#1{\vcenter{\vbox{\hrule\hbox{\vrule\kern3pt\vbox{\kern3pt
	\hbox{${\displaystyle #1}$}\kern3pt}\kern3pt\vrule}\hrule}}}
\def\mbox#1#2{\vcenter{\hrule \hbox{\vrule height#2in
		\kern#1in \vrule} \hrule}}  
%

\def\vev#1{\langle #1 \rangle}

\def\darr#1{\raise1.5ex\hbox{$\leftrightarrow$}\mkern-16.5mu #1}

\def\roughly#1{\raise.3ex\hbox{$#1$\kern-.75em\lower1ex\hbox{$\sim$}}}

\global\newcount\subsubsecno \global\subsubsecno=0
\def\subsubsec#1{\global\advance\subsubsecno by1%
{\toks0{#1}\message{(\the\secno\the\subsecno\the\subsubsecno. \the\toks0)}}%
\ifnum\lastpenalty>9000\else\bigbreak\fi
\noindent{\it\hyperdef\hypernoname{subsubsection}{\the\secno.\the\subsecno\the\subsubsecno}%
{\the\secno.\the\subsecno.\the\subsubsecno.} #1}
\par\nobreak\medskip\nobreak}
\def\boxit#1{\vbox{\hrule\hbox{\vrule\kern8pt
\vbox{\hbox{\kern8pt}\hbox{\vbox{#1}}\hbox{\kern8pt}}
\kern8pt\vrule}\hrule}}
\def\mathboxit#1{\vbox{\hrule\hbox{\vrule\kern8pt\vbox{\kern8pt
\hbox{$\displaystyle #1$}\kern8pt}\kern8pt\vrule}\hrule}}
\def\slashchar#1{\setbox0=\hbox{$#1$}           
   \dimen0=\wd0                                 
   \setbox1=\hbox{/} \dimen1=\wd1               
   \ifdim\dimen0>\dimen1                        
      \rlap{\hbox to \dimen0{\hfil/\hfil}}      
      #1                                        
   \else                                        
      \rlap{\hbox to \dimen1{\hfil$#1$\hfil}}   
      /                                         
   \fi}
\def\sqr#1#2{{\vcenter{\vbox{\hrule height.#2pt
         \hbox{\vrule width.#2pt height#1pt \kern#1pt
            \vrule width.#2pt}
         \hrule height.#2pt}}}}


\input amssym.def
\input amssym.tex
\noblackbox
\baselineskip=14.5pt

\def\comment#1{{}}

\def\ap{\alpha'}

\def\cf{{\it cf.\ }}

\def\al{\alpha}

\def\si{\sigma}\def\Si{{\Sigma}}

\def\bet{\beta}

\newif\ifnref

\def\doubref#1#2{\refs{{#1},{#2} }}
\def\threeref#1#2#3{\refs{{#1},{#2},{#3} }}

\nreffalse

\input epsf

\def\figin{\epsfcheck\figin}\def\figins{\epsfcheck\figins}
\def\epsfcheck{\ifx\epsfbox\UnDeFiNeD
\message{(NO epsf.tex, FIGURES WILL BE IGNORED)}
\gdef\figin##1{\vskip2in}\gdef\figins##1{\hskip.5in}
\else\message{(FIGURES WILL BE INCLUDED)}%
\gdef\figin##1{##1}\gdef\figins##1{##1}\fi}
\def\DefWarn#1{}
\def\figinsert{\goodbreak\midinsert}  
\def\ifig#1#2#3{\DefWarn#1\def#1{Fig.~\the\figno}
\writedef{#1\leftbracket fig.\noexpand~\the\figno}%
\figinsert\figin{\centerline{#3}}\medskip\centerline{\vbox{\baselineskip12pt
\advance\hsize by -1truein\noindent\footnotefont\centerline{{\bf
Fig.~\the\figno}\ \sl #2}}}
\bigskip\endinsert\global\advance\figno by1}

\def\iifig#1#2#3#4{\DefWarn#1\xdef#1{Fig.~\the\figno}
\writedef{#1\leftbracket fig.\noexpand~\the\figno}%
\figinsert\figin{\centerline{#4}}\medskip\centerline{\vbox{\baselineskip12pt
\advance\hsize by -1truein\noindent\footnotefont\centerline{{\bf
Fig.~\the\figno}\ \ \sl #2}}}\smallskip\centerline{\vbox{\baselineskip12pt
\advance\hsize by -1truein\noindent\footnotefont\centerline{\ \ \ \sl #3}}}
\bigskip\endinsert\global\advance\figno by1}


\def\appA{A}
\def\appB{B}

\def\tilde{\widetilde}

\def\h {{1\over 2}}

\def\ov {\overline}
\def\o {\over}
\def\fc#1#2{{#1 \o #2}}

\def\IZ{ {\bf Z}}\def\IC{{\bf C}}\def\IR{ {\bf R}}

\def\fs{{\frak s}}


\def\br{\hfill\break}

\def\det {{\rm det}}

\def\lf {\left}
\def\ri {\right}
\def\ra {\rightarrow}
\def\lra {\longrightarrow}

\def\im {{\rm Im}}
\def\p {\partial}

\def\Dc{{\cal D}}

\def\Fc {{\cal F}} 
\def\Cc {{\cal C}} \def\Oc {{\cal O}}
\def\Lc {{\cal L}} \def\Sc {{\cal S}}
\def\Mc {{\cal M}} \def\Ac {{\cal A}}
 \def\Tc {{\cal T}}
\def\Rc {{\cal R}} 
\def\Ic {{\cal I}} \def\Jc {{\cal J}}

\def\Hc{{\cal H}}






\def\lf{\left}
\def\ri{\right}

\def\ra{{\rightarrow}}
\def\lra{{\longrightarrow}}



\lref\Dotsenko{
V.S.~Dotsenko and V.A.~Fateev,
``Conformal Algebra and Multipoint Correlation Functions in Two-Dimensional Statistical Models,''
Nucl. Phys. B {\bf 240} (1984), 312;
``Four Point Correlation Functions and the Operator Algebra in the Two-Dimensional Conformal Invariant Theories with the Central Charge $c<1$,''
Nucl. Phys. B {\bf 251} (1985), 691-734;\br
Vl.S. Dotsenko, S\'erie de Cours sur la Th\'eorie Conforme; Partie I: Th\'eorie Conforme Minimale, 2006.}

\lref\BM{C.P.~Burgess and T.R.~Morris,
``Open and Unoriented Strings \`A La Polyakov,''
Nucl. Phys. B {\bf 291} (1987), 256-284;\br
S.K.~Blau, M.~Clements, S.~Della Pietra, S.~Carlip and V.~Della Pietra,
``The String Amplitude on Surfaces With Boundaries and Crosscaps,''
Nucl. Phys. B {\bf 301} (1988), 285-303. 
}

\lref\StiebergerDAA{
  S.~Stieberger,
``Open \& Closed vs. Pure Open String One--Loop Amplitudes,''
[arXiv:2105.06888 [hep-th]].
}

\lref\BernQJ{
  Z.~Bern, J.J.M.~Carrasco and H.~Johansson,
``New Relations for Gauge-Theory Amplitudes,''
Phys.\ Rev.\ D {\bf 78}, 085011 (2008).
[arXiv:0805.3993 [hep-ph]].
}

\lref\BernREPT{
Z.~Bern, J.J.~Carrasco, M.~Chiodaroli, H.~Johansson and R.~Roiban,
``The Duality Between Color and Kinematics and its Applications,''
[arXiv:1909.01358 [hep-th]].
}

\lref\DHokerPDL{
  E.D'Hoker and D.H.~Phong,
  ``The Geometry of String Perturbation Theory,''
Rev.\ Mod.\ Phys.\  {\bf 60}, 917 (1988).
}

\lref\BernGRAV{
Z.~Bern, J.J.M.~Carrasco and H.~Johansson,
``Perturbative Quantum Gravity as a Double Copy of Gauge Theory,''
Phys. Rev. Lett. {\bf 105}, 061602 (2010)
[arXiv:1004.0476 [hep-th]].
}

\lref\MOS{W. Magnus, F. Oberhettinger, and R.P. Soni,
{\it Formulas and Theorems for the Special Functions of Mathematical Physics}, Springer 1966;\br
A. Erd\'elyi, W. Magnus, F. Oberhettinger, and F.G. Tricomi,
{\it Higher transcendental functions}, Volume II, McGraw-Hill Book Company 1953. }

\lref\StiebergerVYA{
  S.~Stieberger and T.R.~Taylor,
``Disk Scattering of Open and Closed Strings (I),''
Nucl.\ Phys.\ B {\bf 903}, 104 (2016).
[arXiv:1510.01774 [hep-th]].
}

\lref\StiebergerLNG{
  S.~Stieberger and T.R.~Taylor,
``New relations for Einstein--Yang--Mills amplitudes,''
Nucl.\ Phys.\ B {\bf 913}, 151 (2016).
[arXiv:1606.09616 [hep-th]].
}

\lref\EllisDC{
  J.R.~Ellis, P.~Jetzer and L.~Mizrachi,
 ``One Loop String Corrections to the Effective Field Theory,''
Nucl.\ Phys.\ B {\bf 303}, 1 (1988).
}
\lref\StiebergerWK{
  S.~Stieberger and T.R.~Taylor,
 ``Non--Abelian Born-Infeld action and type I - heterotic duality (2): Nonrenormalization theorems,''
Nucl.\ Phys.\ B {\bf 648}, 3 (2003).
[hep-th/0209064].
}

\lref\TsuchiyaVA{
  A.~Tsuchiya,
``More on One Loop Massless Amplitudes of Superstring Theories,''
Phys.\ Rev.\ D {\bf 39}, 1626 (1989).
}

\lref\HoheneggerKQY{
  S.~Hohenegger and S.~Stieberger,
``Monodromy Relations in Higher-Loop String Amplitudes,''
Nucl.\ Phys.\ B {\bf 925}, 63 (2017).
[arXiv:1702.04963 [hep-th]].
}

\lref\TourkineBAK{
  P.~Tourkine and P.~Vanhove,
``Higher-loop amplitude monodromy relations in string and gauge theory,''
Phys.\ Rev.\ Lett.\  {\bf 117}, no. 21, 211601 (2016).
[arXiv:1608.01665 [hep-th]].
}

\lref\CasaliIHM{
  E.~Casali, S.~Mizera and P.~Tourkine,
``Monodromy relations from twisted homology,''
JHEP {\bf 1912}, 087 (2019).
[arXiv:1910.08514 [hep-th]].
}

\lref\KawaiXQ{
  H.~Kawai, D.C.~Lewellen and S.H.H.~Tye,
``A Relation Between Tree Amplitudes of Closed and Open Strings,''
Nucl.\ Phys.\ B {\bf 269}, 1 (1986).
}

\lref\BernSV{
  Z.~Bern, L.J.~Dixon, M.~Perelstein and J.S.~Rozowsky,
``Multileg one loop gravity amplitudes from gauge theory,''
Nucl.\ Phys.\ B {\bf 546}, 423 (1999).
[hep-th/9811140].
}

\lref\BjerrumBohrHN{
  N.E.J.~Bjerrum-Bohr, P.H.~Damgaard, T.~Sondergaard and P.~Vanhove,
 ``The Momentum Kernel of Gauge and Gravity Theories,''
JHEP {\bf 1101}, 001 (2011).
[arXiv:1010.3933 [hep-th]].
}
\lref\BernUG{
  Z.~Bern, L.J.~Dixon, D.C.~Dunbar, M.~Perelstein and J.S.~Rozowsky,
 ``On the relationship between Yang-Mills theory and gravity and its implication for ultraviolet divergences,''
Nucl.\ Phys.\ B {\bf 530}, 401 (1998).
[hep-th/9802162].
}

\lref\StiebergerCEA{
  S.~Stieberger and T.R.~Taylor,
``Graviton as a Pair of Collinear Gauge Bosons,''
Phys.\ Lett.\ B {\bf 739}, 457 (2014).
[arXiv:1409.4771 [hep-th]];
``Graviton Amplitudes from Collinear Limits of Gauge Amplitudes,''
Phys.\ Lett.\ B {\bf 744}, 160 (2015).
[arXiv:1502.00655 [hep-th]].
``Subleading terms in the collinear limit of Yang--Mills amplitudes,''
Phys.\ Lett.\ B {\bf 750}, 587 (2015).
[arXiv:1508.01116 [hep-th]].
}

\lref\StiebergerLNG{
  S.~Stieberger and T.R.~Taylor,
``New relations for Einstein--Yang--Mills amplitudes,''
Nucl.\ Phys.\ B {\bf 913}, 151 (2016).
[arXiv:1606.09616 [hep-th]].
}

\lref\Mazloumi{
P.~Mazloumi and S.~Stieberger,
``Einstein Yang-Mills amplitudes from intersections of twisted forms,''
JHEP {\bf 06}, 125 (2022) [arXiv:2201.00837 [hep-th]].
}

\lref\CohenPV{
  A.G.~Cohen, G.W.~Moore, P.C.~Nelson and J.~Polchinski,
``Semi Off-shell String Amplitudes,''
Nucl.\ Phys.\ B {\bf 281}, 127 (1987).
}

\lref\MizeraCQS{
  S.~Mizera,
``Combinatorics and Topology of Kawai-Lewellen-Tye Relations,''
JHEP {\bf 1708}, 097 (2017).
[arXiv:1706.08527 [hep-th]].
}

\lref\BjerrumBohrRD{
  N.E.J.~Bjerrum-Bohr, P.H.~Damgaard and P.~Vanhove,
``Minimal Basis for Gauge Theory Amplitudes,''
Phys.\ Rev.\ Lett.\  {\bf 103}, 161602 (2009).
[arXiv:0907.1425 [hep-th]].
}
\lref\StiebergerHQ{
  S.~Stieberger,
 ``Open \& Closed vs. Pure Open String Disk Amplitudes,''
[arXiv:0907.2211 [hep-th]].
}

\lref\StiebergerWEA{
  S.~Stieberger,
``Closed superstring amplitudes, single-valued multiple zeta values and the Deligne associator,''
J.\ Phys.\ A {\bf 47}, 155401 (2014).
[arXiv:1310.3259 [hep-th]].
}
\lref\StiebergerHBA{
  S.~Stieberger and T.R.~Taylor,
``Closed String Amplitudes as Single-Valued Open String Amplitudes,''
Nucl.\ Phys.\ B {\bf 881}, 269 (2014).
[arXiv:1401.1218 [hep-th]].
}

\lref\BrownGIA{
  F.~Brown,
 ``Single-valued Motivic Periods and Multiple Zeta Values,''
SIGMA {\bf 2}, e25 (2014).
[arXiv:1309.5309 [math.NT]].
}

\lref\BrownOMK{
F.~Brown and C.~Dupont,
``Single-valued integration and double copy,''
J. Reine Angew. Math. {\bf 775} (2021), 145--196
[arXiv:1810.07682 [math.NT]];
``Single-valued integration and superstring amplitudes in genus zero,''
Commun. Math. Phys. {\bf 382} (2021) no.2, 815-874
[arXiv:1910.01107 [math.NT]].
}

\lref\StiebergerXHS{
S.~Stieberger,
``Periods and Superstring Amplitudes,''
Springer Proceedings in Mathematics \& Statistics, vol {\bf 314}, Springer, Cham (2020).
[arXiv:1605.03630 [hep-th]].
}

\lref\Hsue{
C.S.~Hsue, B.~Sakita and M.A.~Virasoro,
``Formulation of dual theory in terms of functional integrations,''
Phys. Rev. D {\bf 2}, 2857-2868 (1970).
}

\lref\GeyerBJA{
  Y.~Geyer, L.~Mason, R.~Monteiro and P.~Tourkine,
``Loop Integrands for Scattering Amplitudes from the Riemann Sphere,''
Phys.\ Rev.\ Lett.\  {\bf 115}, no. 12, 121603 (2015).
[arXiv:1507.00321 [hep-th]].
}

\lref\GeyerJCH{
  Y.~Geyer, L.~Mason, R.~Monteiro and P.~Tourkine,
``One-loop amplitudes on the Riemann sphere,''
JHEP {\bf 1603}, 114 (2016).
[arXiv:1511.06315 [hep-th]].
}

\lref\SongHe{S.~He and O.~Schlotterer,
``New Relations for Gauge-Theory and Gravity Amplitudes at Loop Level,''
Phys. Rev. Lett. {\bf 118} (2017) no.16, 161601 [arXiv:1612.00417 [hep-th]];\br
A.~Edison, M.~Guillen, H.~Johansson, O.~Schlotterer and F.~Teng,
``One-loop matrix elements of effective superstring interactions: $\ap$--expanding loop integrands,''
JHEP {\bf 12}, 007 (2021) [arXiv:2107.08009 [hep-th]].}

\lref\BrownGraph{
E.D'Hoker, M.~B.~Green, \"O.~G\"urdogan and P.~Vanhove,
``Modular Graph Functions,''
Commun. Num. Theor. Phys. {\bf 11} (2017), 165-218
[arXiv:1512.06779 [hep-th]];\br
F.~Brown,
``A class of non-holomorphic modular forms I,''
[arXiv:1707.01230 [math.NT]];
``A Class of Nonholomorphic Modular Forms II: Equivariant Iterated Eisenstein Integrals,''
Forum Math. Sigma {\bf 8} (2020), e31.
}

\lref\Kaidi{
E.D'Hoker and J.~Kaidi,
``Lectures on modular forms and strings,''
[arXiv:2208.07242 [hep-th]].}

\lref\Cohen{
A.G.~Cohen, G.W.~Moore, P.C.~Nelson and J.~Polchinski,
``An Off-Shell Propagator for String Theory,''
Nucl. Phys. {\bf B 267}, 143-157 (1986)}

\lref\Broedel{
J.~Broedel and A.~Kaderli,
``Amplitude recursions with an extra marked point,''
Commun. Num. Theor. Phys. {\bf 16}, no.1, 75-158 (2022)
[arXiv:1912.09927 [hep-th]].}
\Title{\vbox{\rightline{MPP--2021--162}
}}
{\vbox{
\centerline{A Relation between One--Loop Amplitudes}\vskip2mm
\centerline{of Closed and Open Strings}\vskip5mm
\centerline{\it ( One--Loop KLT Relation )}
}}
\medskip
\centerline{S. Stieberger}
\bigskip

\medskip
\centerline{\it Max--Planck--Institut  f\"ur Physik}
\centerline{\it Werner--Heisenberg--Institut, 80805 M\"unchen, Germany}
\medskip

\vskip15pt

\vskip15pt

\medskip
\bigskip\bigskip\bigskip
\centerline{\bf Abstract}
\vskip .2in
\noindent
We express one--loop closed string amplitudes as weighted sums over squares of open string one--loop subamplitudes. 
These findings generalize -- subject to final complex structure modulus integration -- the celebrated tree--level relationships  known as Kawai--Lewellen--Tye (KLT)  relations to higher loops and can be applied for  both the massless and massive case -- with or without supersymmetry. As a consequence in the field--theory limit our relations capitalize  solid one--loop gauge--gravity relations including loop--level color kinematics duality.
In particular, in gravitational  one--loop amplitudes a graviton is traded for two gluons
just like at tree--level.
Our results are derived on the underlying string world--sheet torus by splitting each complex  string coordinate into a pair of two real cylinder coordinates by means of an analytic continuation on the elliptic curve. This manipulation  involves   the loop momentum, which mediates between the two one--loop open string cylinder amplitudes.

\noindent

\Date{}
\noindent
\goodbreak
\listtoc
\writetoc
\break
\newsec{Introduction}

The first indication of an amplitude relation between gauge theory and gravity was incepted by the Kawai--Lewellen--Tye (KLT) relations at the perturbative tree--level\foot{Actually, the mathematical  foundation  of these relations had been laid   by Dotsenko and Fateev in their seminal work~\Dotsenko.} \KawaiXQ.
This linkage was then reformulated as double--copy structure  by Bern, Carrasco and Johansson (BCJ) based on  a duality between color and kinematics derived from the (conjectural) existence of relations between partial gluon subamplitudes  \BernQJ. The general proof of such tree--level BCJ amplitude identities  was subsequently  presented by using string theory and the power of world--sheet monodromy properties \doubref\StiebergerHQ\BjerrumBohrRD. The formulation of the color--kinematics duality has since then seen  a large number of further extensions, we refer to \BernREPT\ for a comprehensive review. 
An  immediate extension  to loop amplitudes via generalized unitarity was conjectured in \BernGRAV. However, so far formal proofs have been constructed for only tree--level scattering amplitudes in these theories and yet  a loop--level generalization of the KLT relations is still missing. Therefore finding the one--loop generalization of the tree--level KLT relations is most pressing. 
They will allow elevating the various gauge--gravity (double--copy) relations at tree--level to loop--level and beyond. In particular, a loop--level proof of color kinematics duality is capitalized.

In fact, it is string theory where gauge--gravity connections become manifest -- at least in the most direct and obvious way.
Certainly, one of the most well--known example is the KLT relation.
Note, that this relation is derived from pure string world--sheet properties without any assumptions on the underlying string background (except that there exists an underlying conformal field
theory description) nor on the amount of space--time supersymmetries.
In string theory a massless graviton state  appears as lowest closed superstring mode whilst a gluon is the lowest massless open superstring state. The closed string modes can be described by a tensor product of left--movers and right--movers with each one describing open strings. On the  string world--sheets, which describe perturbative amplitudes, the left-- and right--movers are linked by monodromies, which are specified by  some kernel or intersection matrix.
It is this hybrid construction which entails the connection between gravitational and gauge amplitudes  in perturbative string theory. In fact, many gauge--gravity relations derive from string theory albeit they give already rise to unexpected relations in field theory. 

In tackling any closed string amplitude calculation it useful to first consider the corresponding open string computation. At field--theory level this means that for graviton amplitudes one should borrow results from gauge amplitudes.
Kawai, Lewellen and Tye have given expressions for closed string tree amplitudes as weighted sum over squares of open string tree amplitudes \KawaiXQ. This relation holds to all orders in the inverse string tension $\ap$, with the latter  related to the string mass as follows $\ap\!\sim\! M_{\rm string}^{-2}$.
Furthermore, in \StiebergerHQ\ it has been shown that any tree--level string amplitude involving 
$n_c$ closed strings and $n_o$ open strings can be written as linear combination of pure open string amplitudes involving only $2n_c\!+\!n_o$ open strings. 
These structures have been further explored and extended in \StiebergerCEA\ leading to new relations \StiebergerLNG\ and formulations \Mazloumi\ for Einstein--Yang--Mills (EYM) amplitudes.
At any rate in the field--theory limit, these findings imply that at least at tree--level properties of gravitational amplitudes  are inherited from gauge amplitudes \BernSV.
 Therefore, it is natural to ask for a one--loop generalization thereof. In fact, recently
this task has been accomplished on the cylinder world--sheet allowing to express any one--loop
amplitudes of open and closed strings in terms of pure open string one--loop amplitudes \StiebergerDAA, which gives rise  to the one--loop analog of \StiebergerHQ\ and \StiebergerCEA.
 In fact, we shall see that the  approach developed  in \StiebergerDAA\ paves the way for a one--loop generalization of the KLT relations.

In this work we study one--loop string amplitudes involving closed oriented strings. The latter may be both massless or massive and we do not need any  space--time supersymmetries.
We show that on the torus world--sheet these amplitudes can be written as pure open string amplitudes with each closed string replaced by two open strings (subject to final complex structure modulus integration). 
More precisely, by means of an analytic continuation each complex torus coordinate accounting  for a closed string insertion is split into a pair of two real cylinder coordinates. The latter naturally give rise to  open string vertex positions located at the boundaries of two cylinders.
These steps involve  a one--loop kernel depending on the loop momentum, which mediates between the two one--loop open string amplitudes.
Although, our work is essentially a one--loop generalization of the tree--level case \KawaiXQ,  various additional problems arise at one--loop. These issues can be addressed  
to the lack of holomorphic double periodic functions on the elliptic curve. As a consequence we have to deal with  quasi--periodic functions with non--harmonic contributions. The latter can be traded for introducing the loop momentum and working with loop integrands.
The non--harmonic contributions hamper the (full) one--loop integrand to factorize into a product of holomorphic and anti--holomorphic functions. Eventually, we find a way to split each complex torus string coordinate into a pair of two real cylinder coordinates by means of an analytic continuation on the elliptic curve. Open string monodromy relations on doubled surfaces will play an important role. 

The organization of the present work is as follows.
In Section 2 we introduce the relevant objects describing generic $n$--point  one--loop closed string amplitudes. In Section~3 we deform the closed string position integrations on the torus along a closed contour thereby disentangling holomorphic and anti--holomorphic coordinates and converting them into real coordinates describing open string positions on two separate  cylinders. 
In Section 4 we summarize our results by stating the one--loop KLT relations. The one--loop closed string amplitude is written as a weighted sum over squares of (color ordered) one--loop open string amplitudes. To familiarize with this result we shall discuss the explicit result for the one--loop four closed string amplitude (four--graviton amplitude in type II superstring theory).
Furthermore, we discuss the role of the loop momentum flowing between the two cylinder diagrams. Besides, we extract the field--theory limit from our one--loop KLT result. 
In Section 5 we put our results into context and give some concluding remarks. Finally, in the Appendix we shall discuss uplifting one--loop open string monodromy relations onto doubled surfaces. This leads to  the notion of doubles of counters which elevate open string monodromy relations on surfaces with boundaries to relations on surfaces without boundaries.


\goodbreak

\newsec{One--loop torus amplitudes with closed strings}

We consider the scattering of $n$ closed oriented strings at one--loop.
The on--shell string amplitude $\Ac_{n}^{(1)}$ is described by the topology of a world--sheet torus without boundaries.
The underlying string world--sheet is described by a torus $\Tc$ with $n$ on--shell open strings attached to the world--sheet at points $z_r,\ r=1,\ldots,n$.
The torus has a complex structure modulus $\tau$ (modular parameter). Closed strings are inserted at points $z_r\in\Tc$. 
The bosonic closed string Green's function on the torus  is given by
\eqn\BOSG{
G^{(1)}(z_1,z_2)=\ln\lf|\fc{\theta_1(z_1-z_2,\tau)}{\theta_1'(0,\tau)}\ri|^2-\fc{2\pi}{\Im\tau}[\Im(z_2-z_1)]^2\ ,}
and $q=e^{2\pi i\tau}$. The second term of \BOSG\ renders 
the function to be single--valued on the torus under the shift $z\ra z+\tau$ \DHokerPDL.

A generic (on--shell) plane wave closed string state is represented by
\eqn\plane{
|q\rangle=\lim_{\rho,\bar \rho\ra0} :e^{iq_L^\mu X_\mu(\rho)}e^{iq_R^\mu X_\mu(\bar \rho)}:|0\rangle\ ,}
with complex plane coordinates $\rho,\bar\rho\in\IC$ and the total momentum $q\!=\!q_L+q_R$. The interaction between left-- and right--moving fields can conveniently be described  by correlators on the torus. 
In particular,  we choose the left-- and right--moving  momenta $q_L,q_R$ as
\eqn\Split{
q_{L}=\h q\ \ \ ,\ \ \ q_{R}=\h q\ .}
Massless closed  states enjoy $q_{L}^2=q_{R}^2=0$. 
Not only its plane wave part \plane\ but the full closed string vertex operator $V_{closed}(\bar z,z)$ splits into a direct
product of left-- and right--moving open string vertex operators:
\eqn\Vertexsplit{
V_{closed}(q,z,\bar z)=V_{open}(q_L,z)\otimes V_{open}(q_R,\bar z)\ .}
This product structure, which holds both for matter and ghost parts, is then reflected in any perturbative string amplitudes which are
derived from correlation functions of vertex operators. In fact, it is this product structure
\Vertexsplit\ why we expect closed string amplitudes to factorize into products of open string amplitudes.

The amplitude  of $n$ closed oriented strings at one--loop in $d$ space--time dimensions 
\eqn\Gerk{
{\cal A}_{n}^{(1)}(q_1,\ldots,q_n)=g_{c}^n\int_{\Mc_{1,n}}  d\mu\ \vev{V_1(q_1,z_1,\bar z_1)\ldots V_n(q_n,z_n,\bar z_n)}\ ,}
involves a correlator of $n$ closed string vertex operators \Vertexsplit\ integrated over the moduli space $\Mc_{1,n}$  of a $n$ punctured torus $\Tc$ with measure~$d\mu$. The latter can be written as integral over the fundamental region of the torus $\Fc_1=\{ \tau=\tau_1+i\tau_2\in\IC \; |\;  \tau_2>0,\; -\h\leq\tau_1\leq\h,\; |\tau|\geq1\}$. Above we have introduced 
the closed string coupling constant $g_{c}$.
The corresponding world--sheet torus with closed string insertions $z_r$ is depicted in  Fig.~1.
\ifig\SigmaWorldsheet{String world--sheet torus with $n$ closed string insertions $z_r$.}{\epsfxsize=0.45\hsize\epsfbox{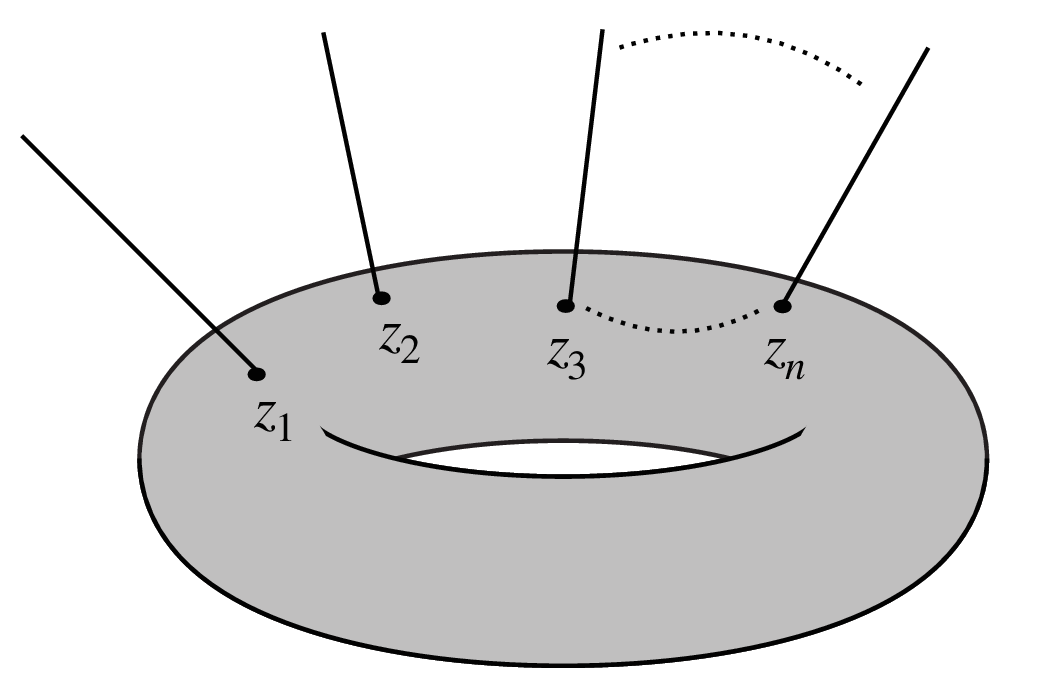}}
\noindent  The amplitude \Gerk\ assumes the generic form 
\eqn\Start{\eqalign{
{\cal A}_{n}^{(1)}(q_1,\ldots,q_n)&=\h\ g_{c}^n\;\delta^{(d)}\lf(\sum_{r=1}^{n}q_r\ri)\int_{\Fc_1} \!\!\fc{d^2\tau}{\tau_2}\; V_{CKG}^{-1}(\Tc) \cr
&\times\left(\!\int_{\Tc}\prod_{s=1}^{n}d^2z_s\!\right)I(\{z_s,\bar z_s\})\ Q(\{z_s,\bar z_s\})\ ,}}
with $\Tc=\{z=x+\tau y\; |\; x,y\in \IR,\ 0\leq x,y\leq 1\}$ and the momentum conservation condition:
\eqn\conserv{
\sum_{r=1}^{n}q_r=0\ .}
The conformal Killing group consists of the translations of the torus. The volume of the conformal Killing group $V_{CKG}(\Tc)\!=\!\tau_2$
can be cancelled by taking into account the translation invariance  and fixing one  vertex position $z_s\in\Tc$ subject to $\int_\Tc d^2z_s =\tau_2$. The factor of $1/2$ accounts for the discrete transformation $z\ra-z$ of the closed (oriented) string world--sheet, which leaves the torus metric invariant.
More precisely, this symmetry can be fixed by restricting a second vertex operator to half the torus or likewise integrate it over the full torus by including a factor of $1/2$.   Furthermore, we have
\eqn\KN{
I(\{z_s,\bar z_s\})=\prod_{1\leq r<s\leq n}e^{\h\ap q_rq_s G^{(1)}(z_s,z_r)}}
accounting for the insertion of an exponential factor universal to all vertex 
operators~\Vertexsplit. The doubly--periodic modular function $Q$ of weight $w$, i.e. 
$Q(\{\fc{z_s}{\tau},\fc{\bar z_s}{\bar\tau}\})\!=\!|\tau|^{2w}Q(\{z_s,\bar z_s\})$ 
(with $w\!=\!n$ for the massless case), depends on the set of positions $\{z_s,\bar z_s\}$ and describes the remaining contractions of vertex operators including
ghost and matter contributions and the partition function.   The simplest cases represent $n$ tachyon scattering in $d\!=\!26$ bosonic string theory, with $Q\!=\!\tau_2^{-12}|\eta(\tau)|^{-48}$ or four--graviton scattering in $d\!=\!10$ type II superstring theory, with $Q\!=\!\tau_2^{-4}$. Other explicit expressions for $Q$ for five-- and six--point (heterotic) closed string one--loop amplitudes can be found in \StiebergerWK, cf. also \doubref\EllisDC\TsuchiyaVA.

As a consequence of \Vertexsplit\ for any closed string amplitude the function $Q$  can be split into a product of holomorphic and anti--holomorphic modular functions of weight $w$ as:
\eqn\splitQ{
Q(\{z_s,\bar z_s\})=\tau_2^{1-d/2}\ Q_L(\tau,\{z_s\})\cdot Q_R(\ov\tau,\{\bar z_s\})\ .}
On the other hand, for \KN\ we explicitly have
\eqn\Generali{
I(\{z_s,\bar z_s\})=
\Theta_L(\{z_s\})\ 
\Theta_R(\{\bar z_s\})\ I_{nh}(\{z_s,\bar z_s\})\ ,}
with the functions
\eqn\thetaLR{\eqalign{
\Theta_L(\{z_s\})&=\prod_{1\leq r<s\leq n}\lf[\fc{\theta_1(z_s-z_r,\tau)}{\theta_1'(0,\tau)}\ri]^{\h\ap q_sq_r} \ ,\cr
\Theta_R(\{\bar z_s\})&= \prod_{1\leq r< s\leq n}\lf[\fc{\bar \theta_1(\bar z_s-\bar z_r,\bar \tau)}{\bar\theta_1'(0,\bar \tau)}\ri]^{\h\ap q_sq_r}\ ,}}
and the non--holomorphic factor only depending on the imaginary parts $\Im z_r$ of the closed string positions $z_r$:
\eqn\nonholo{
I_{nh}(\{z_s,\bar z_s\})=\prod_{r,s=1\atop r<s}^{n} e^{-\fc{\pi\ap}{\tau_2}q_{r}q_{s}\Im(z_r-z_s)^2}=\exp\lf\{\fc{\pi\ap}{\tau_2}\lf(\sum_{r=1}^n q_r\Im z_r \ri)^2\ri\}\ .}
There exists a formulation of the non--holomorphic factor \nonholo\ in terms the loop momentum $\ell$, which allows\foot{Note, that this recast is also valid if some of the external closed strings are not massless,~i.e. $q_i^2\neq0$.} to rewrite $I_{nh}$ as follows \DHokerPDL:
\eqn\looprepr{(\ap \tau_2)^{-d/2}\   I_{nh}(\{z_s,\bar z_s\})=\int_{-\infty}^\infty d^{d}\ell\ \exp\lf\{-\pi\ap \tau_2\ell^2-\pi i\ap\ell\sum_{r=1}^{n}
q_r(z_r-\bar z_r)\ri\}\ .}
With this information in \KN\ the holomorphic and anti--holomorphic parts can be factorized as:
\eqn\LoopE{
(\ap \tau_2)^{-d/2}\ I(\{z_s,\bar z_s\})=\int\limits_{-\infty}^\infty d^d\ell\ 
\lf|\exp\lf\{\fc{1}{2} i\pi\ap \tau\ell^2-\pi i\ap\ell\sum_{r=1}^{n}
q_rz_r\ri\}\ri|^2 \ \Big|\Theta_L(\{z_s\})\Big|^2\ .}
The integrand of \LoopE\ is single--valued as function in the complex coordinates $z_s$.
The hermitian pairing furnished in \LoopE\ is familiar from two--dimensional conformal field theory where loop momenta  label conformal blocks.

In the following we shall concentrate on a rectangular torus, i.e. $\tau=i\tau_2:=it$, with $t\in (0,\infty)$. Hereby, we shall make profit from the identities (B.13). Generalizations will be discussed elsewhere. The torus coordinate $z=\sigma^1+i\sigma^2$ is then parameterized by 
the intervals $(\sigma^1,\sigma^2)\in [0,1]\times[0,\tau_2]$.

\newsec{Closed string torus coordinates  as  pairs of open string cylinder coordinates}

In this section we perform the necessary steps to convert complex closed string positions  on the torus into pairs  of open string cylinder coordinates. We shall perform an analytic continuation of the closed string vertex operator positions  $z_t\in \Tc$ on the torus (with $t=1,\ldots,n$) by considering some contour integral on the torus.  
As a result for each  complex coordinate $z_i$ we obtain a pair of two real coordinates $\xi_t,\eta_t$  describing two real open string positions.

To discuss the dependence of the relevant integrand on the closed string coordinates $z_r$ given in \Generali\ we define
\eqn\relevant{\eqalign{
I(\{z_s,\bar z_s\};\ell)&:=\exp\lf\{-\pi\ap \tau_2\ell^2-\pi i\ap\ell\sum_{r=1}^{n}
q_r(z_r-\bar z_r)\ri\}\cr 
&\times\prod_{1\leq r<s\leq n}\theta_1(z_s-z_r,\tau)^{\h\ap q_sq_r}\ 
\theta_1(\bar z_s-\bar z_r,\tau)^{\h\ap q_sq_r} \ ,}}
depending on the  loop momentum $\ell$. 
Clearly the expression \relevant\ is invariant under shifts $z_t\ra z_t+1$. On the other hand, shifting one coordinate $z_t$ by an integer period $\pm\tau$ amounts to:
\eqn\Amountshift{
I(z_t\pm\tau,\bar z_t\pm\bar\tau;\ell\pm q_t)=I(z_t,\bar z_t;\ell)\ .}
Shifts in the loop momentum may be compensated in the loop integral. Therefore, thanks to \Amountshift\ for convenience in the following  we shall  parameterize the torus coordinates $z_s\in\Tc$ by the rectangle 
$\Tc=\{z=x+\tau y\; |\; x,y\in \IR,\ 0\leq x\leq 1,\ -\h\leq y\leq \h\}$.
Eventually, after introducing the parameterization
\eqn\sig{
z_s=\sigma_s^1+i\sigma_s^2\ \ \ ,\ \ \ \sigma_s^1\in(0,1)\ ,\ \sigma_s^2\in\lf(-\fc{t}{2},\fc{t}{2}\ri)\ ,\ s=1,\ldots,n\ ,}
the expression \relevant\ becomes:
\eqnn\Relevant{
$$\eqalignno{
I(\{z_s,\bar z_s\};&\ell)\equiv I(\{\sigma^1_s,\sigma^2_s\};\ell):=\Lc(\{\si^2_s\})&\Relevant\cr
&\times\prod_{r,s=1\atop r<s}^n\theta_1(\sigma^1_s+i\sigma^2_s-\sigma_r^1-i\sigma_r^2,\tau)^{\h\ap q_sq_r}\ 
 \theta_1(\sigma^1_s-i\sigma^2_s-\sigma_r^1+i\sigma_r^2, \tau)^{\h\ap q_sq_r} \ ,}$$}
 with the loop momentum dependent factor:
\eqn\Loopf{
\Lc(\{\si^2_s\}):=\exp\lf\{-\pi\ap \tau_2\ell^2+2\pi\ap \ell \sum_{r=1}^{n} q_r\sigma_r^2\ri\}\ .}
Actually, in addition to \relevant\ we shall also need the "non--planar" object
\eqnn\Object{
$$\eqalignno{
I_{A_1,A_2}(\{z_s,\bar z_s\};\ell)&:=\exp\lf\{-\pi\ap \tau_2\ell^2-\pi i\ap\ell\sum_{r=1}^{n}
q_r(z_r-\bar z_r)\ri\}\cr 
&\times\prod_{r,s\in A_1 \atop r,s\in A_2}\prod_{r<s}\theta_1(z_s-z_r,\tau)^{\h\ap q_sq_r}\ 
\theta_1(\bar z_s-\bar z_r,\tau)^{\h\ap q_sq_r}\cr
&\times\prod_{r\in A_1\atop s\in A_2}\theta_4(z_s-z_r,\tau)^{\h\ap q_sq_r}\ 
\theta_4(\bar z_s-\bar z_r,\tau)^{\h\ap q_sq_r}\ ,&\Object}$$}
with $a_1=|A_1|,\;a_2=|A_2|$. For the parameterization \sig\ we obtain:
\eqnn\Relevantnp{
$$\eqalignno{
I_{A_1,A_2}&(\{z_s,\bar z_s\};\ell)\equiv I_{A_1,A_2}(\{\sigma^1_s,\sigma^2_s\};\ell):=\Lc(\{\si^2_s\})&\Relevantnp\cr
&\times\prod_{r,s\in A_1 \atop r,s\in A_2}\prod_{r<s}\theta_1(\sigma^1_s+i\sigma^2_s-\sigma_r^1-i\sigma_r^2,\tau)^{\h\ap q_sq_r}\ 
 \theta_1(\sigma^1_s-i\sigma^2_s-\sigma_r^1+i\sigma_r^2, \tau)^{\h\ap q_sq_r} \cr
 &\times\prod_{r\in A_1\atop s\in A_2}\theta_4(\sigma^1_s+i\sigma^2_s-\sigma_r^1-i\sigma_r^2,\tau)^{\h\ap q_sq_r}\ \theta_4(\sigma^1_s-i\sigma^2_s-\sigma_r^1+i\sigma_r^2, \tau)^{\h\ap q_sq_r}\ .}$$}

\subsec{One--dimensional complex torus integration}

To substantiate the concept of analytic continuation on the torus for the moment let us single out one particular coordinate $\sigma^2_t$ with $t\in\{1,\ldots,n\}$ and 
investigate the $\sigma_t^2$--dependence of the integrands \Relevant\ and \Relevantnp. More precisely, we shall be concerned with the local system
\eqn\Local{\eqalign{
I_{A_1,A_2}(\{z_s,\bar z_s\};\ell)\hookrightarrow I_{A_1,A_2}(z_t,\bar z_t;\ell)\equiv I_{A_1,A_2}(\sigma^1_t,\sigma^2_t;\ell)\ ,}}
which is holomorphic in the coordinate $\sigma_t^2$. Despite the expression \Local\ is the same as \Relevant\
or \Relevantnp, in this subsection we only display its  dependence on $\sigma^1_t$ and $\sigma^2_t$ and consider the remaining coordinates as spectators. We want to discuss the dependence of \Local\ in  the complex $\sigma^2_t$--plane and consider the closed cycle (polygon) $\Gamma$ defined by the 
four edges $C_1,C_2,C_1'$ and~$C_2'$
\eqn\edges{
\Gamma=C_1\cup C_2 \cup C_1'\cup C_2'\ ,}
and depicted in Fig.~2. \ifig\SigmaWorldsheet{Closed contours in the complex $\sigma^2_t$--plane and branch points.}{\epsfxsize=0.6\hsize\epsfbox{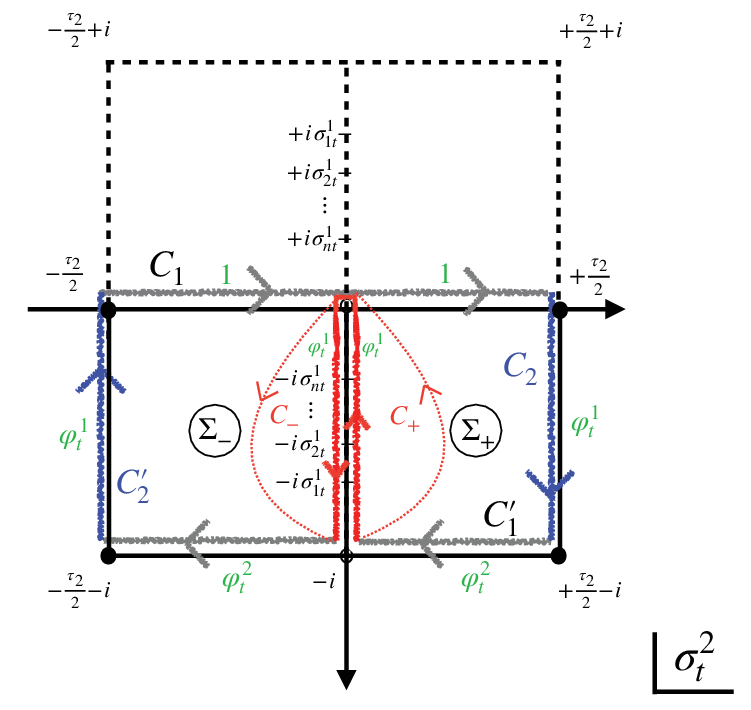}}
\noindent  
The cycle $\Gamma$ enfolds the doubled surface $\si^2_t\in\Tc$ which is divided into the two 
cylinders $\Si_-: -\fc{\tau_2}{2}\leq\Re\si^2_t\leq0$ and $\Si_+: 0\leq\Re\si^2_t\leq+\fc{\tau_2}{2}$, respectively with $-1\leq \Im \si^2_t\leq 0$.
The expression \Local\ as a holomorphic function in $\sigma^2_t$  has $2n-2$ branch points\foot{Note, that generically we have
$\sigma_{st}^1\in (-1,1)$. Hence, only half of $n\!-\!1$ branch points are located inside the cycle $\si^2_t\in \Rc$. However,  by appropriately  shifting  the arguments of the theta--functions by unity    we can always probe all branch points inside the cycle $\si^2_t\in \Rc$.} in the complex $\si^2_t$--plane   at $\sigma^2_t=-i \sigma_{st}^1+ \sigma^2_s$ and $\sigma^2_t=+i \sigma_{st}^1+ \sigma^2_s$ ($s\neq t$), where the arguments of the theta--functions become  zero. Additional phases need to be introduced when passing their corresponding branch cuts. 
When we shall discuss (in the next subsection) analytic continuation in all coordinates $\sigma^2_r$ 
all branch points are located along the imaginary axes $\Re(\si_t^2)=0$ as illustrated in Fig.~2. 
Then, their effect will be taken into account  after introducing  new coordinates parameterizing the edges $C_\pm$. 
In addition, the holomorphic loop--dependent factor  of \Relevant\ exhibits branch points in the complex $\si^2_t$--plane along the  lines $\si_t^2=\pm \fc{\tau_2}{2}$ for $\Im(\si_t^2)\in\{0,-1\}$. This can be substantiated by
looking at the different theta--function representations
\comment{Alternatively, for $\si^2_r\in\Si_+$ 
this branch cut structure can also be examined by noting:
\eqn\Regalpwand{
\Lc(\{\si^2_s\})=e^{-\pi\ap \tau_2\ell^2}\;\prod_{r=1}^n
\lf(\fc{\theta_1(i\si^2_r-\fc{i\tau_2}{2})}
{\theta_4(i\si^2_r)}\ri)^{-\ap\ell q_r}\ 
\lf(\fc{\theta_1(i\si^2_r-\fc{i\tau_2}{2}-1)}{\theta_4(i\si^2_r-1)}
\ri)^{-\ap\ell q_r}.}
Then, we associate branch points at $\si_t^2=\fc{\tau_2}{2}$ and $\si_t^2=\fc{\tau_2}{2}-i$  each with
monodromy phase $\varphi_t$.}
of the loop--momentum dependent factor \Loopf\ as
\eqn\Regalpwand{\eqalign{
\Lc(\{\si^2_s\})&=e^{-\pi\ap \tau_2\ell^2}\;
\prod_{r=1}^n\lf(\fc{\theta_1(i\si^2_r-\fc{i\tau_2}{2})}{\theta_4(i\si^2_r)}\ri)^{-2\ap\ell q_r}
\!\!\!\!\!=e^{-\pi\ap \tau_2\ell^2}\;
\prod_{r=1}^n\lf(\fc{\theta_1(i\si^2_r-\fc{i\tau_2}{2}-1)}{\theta_4(i\si^2_r-1)}\ri)^{-2\ap\ell q_r}\cr
&=e^{-\pi\ap \tau_2\ell^2}\;
\prod_{r=1}^n\lf(\fc{\theta_1(i\si^2_r+\fc{i\tau_2}{2})}{\theta_4(i\si^2_r)}\ri)^{2\ap\ell q_r}
\!\!\!\!\!=e^{-\pi\ap \tau_2\ell^2}\;
\prod_{r=1}^n\lf(\fc{\theta_1(i\si^2_r+\fc{i\tau_2}{2}-1)}{\theta_4(i\si^2_r-1)}\ri)^{2\ap\ell q_r},}}
which are appropriate  to discuss the monodromies in the $\si^2_t$--plane and extract the relevant branchings when crossing these points. As a consequence, while moving along $C_2$ and passing  the pair of points $\sigma_t^2=+\fc{\tau_2}{2}$ and $\sigma_t^2=+\fc{\tau_2}{2}-i$ in clockwise (negative) orientation by a quarter arc we encounter  the phase 
\eqn\orientation{
\varphi_t:=e^{-\pi i\ap\ell q_t}\ ,} 
respectively. Similarly, the phase   $\varphi_t^{-1}$ is picked up along $C_2'$ at the pair of points $\sigma_t^2=-\fc{\tau_2}{2}$ and $\sigma_t^2=-\fc{\tau_2}{2}-i$, respectively leading to a $\varphi_t$ branch on $C_2'.$
The function \Local\  along $C_1'$ differs from that along $C_1$ by a phase:
\eqn\differ{
I_{A_1,A_2}(\sigma^1_t,\sigma^2_t-i;\ell)=\varphi_t^2\ I_{A_1,A_2}(\sigma^1_t,\sigma^2_t;\ell)\ .}
The phase factor in \differ\ agrees with the total phase stemming from moving along $C_2$ and encircling by a quarter arc in the negative direction (clockwise) the two branch points  $\sigma_t^2=+\fc{\tau_2}{2}$ and $\sigma_t^2=+\fc{\tau_2}{2}-i$.
Following the discussion on contours on doubled surfaces in Appendix \appA\ we shall also consider the path along
\eqn\edgeR{
\Rc=C_+\cup C_-\ ,}
spanned by the two contours $C_+$ and $C_-$ at $\Re(\sigma_t^2)\!=\!0$. The cycle $\Rc$ avoids all the  $n\!-\!1$ branch points located within  the interval  $-1\leq\im(\si^2)\leq0$   by counter--clockwise encircling them, cf. Fig.~2. The branchings of the latter become relevant later when we shall introduce new coordinates and choose an integration path to  take into account their effects. 
On the other hand, the branch cut structure of the loop--dependent factor \Loopf\ can be inferred  along $\Re(\sigma_t^2)\!=\!0$ 
by rewriting \Regalpwand\ as:
\eqn\Antoni{
\Lc(\{\si^2_s\})=e^{-\pi\ap \tau_2\ell^2}\;\prod_{r=1}^n\lf(\fc{\theta_1(i\si^2_r)}{\theta_4(i\si^2_r-\fc{i\tau_2}{2})}\ri)^{2\ap\ell q_r}=e^{-\pi\ap \tau_2\ell^2}\;\prod_{r=1}^n\lf(\fc{\theta_1(i\si^2_r-1)}{\theta_4(i\si^2_r-1-\fc{i\tau_2}{2})}\ri)^{2\ap\ell q_r}.}
From \Antoni\ we associate branch points at $\si_t^2=-i$ and $\si_t^2=0$, respectively  each with
monodromy phase $\varphi_t^2$, respectively. 

As a consequence of \Amountshift\ shifting a coordinate $\si^2_t$ by a full period 
$\tau_2$ may  be compensated by a shift in the loop momentum $\ell$:
\eqn\amountshift{
I_{A_1,A_2}(\sigma^1_t,\sigma^2_t+\tau_2;\ell+q_t)=I_{A_1,A_2}(\sigma^1_t,\sigma^2_t;\ell)\ .}
On the other hand, along the half--periods $C_2,C_2'$ the value of the function \Local\ differs from that along $C_\pm$ by a shift in the loop momentum in the following way
 \eqn\differi{
I_{A_1,A_2}(\sigma^1_t,\sigma^2_t\pm\fc{\tau_2}{2};\ell)=I_{A_1',A_2'}(\sigma^1_t,\sigma^2_t;\ell\mp\fc{q_t}{2})\ ,}
respectively with $A_1'=A_1\slash\{t\}$ and $A_2'=A_2\cup\{t\}$.
Eventually we need to integrate the coordinate $z_t$  over the full torus $\Tc$:
\eqn\need{
\int_{\Tc}d^2z_t\ I(z_t,\bar z_t;\ell)=\int_0^1d\sigma^1_t\int\limits_{-\tau_2/2}^{\tau_2/2}d\sigma^2_t\ I(\sigma^1_t,\sigma^2_t;\ell)\ .}
Thus in \need\ we are supposed to integrate $\sigma_t^2$ along the (grey) edge $C_1$ in Fig.~2. 
We seek to convert this integration to integrations along the imaginary axis  by considering the closed cycle  $\Gamma$  and using Cauchy's integral theorem stating that  for given $\si^1_t\in(0,1)$:
\eqn\Cauchy{
\oint_{\Gamma}d\sigma_t^2\  \hat I(\sigma^1_t,\sigma^2_t;\ell)=-\oint_\Rc\ d\sigma_t^2\ \hat I(\sigma^1_t,\sigma^2_t;\ell)\ .}
The integrand $\hat I$ differs from $I$  by choosing a correct integration branch  when moving in the complex $\sigma_t^2$--plane and passing the branch cuts. This will be achieved by appending  to the integrand \Local\ appropriate  phase factors. The relation \Cauchy\ corresponds to  Eq. (A.2) describing
open string monodromy relations on the doubled surface $\Tc$. Note, that the union of two cycles $\Gamma$ and $\Rc$   resembles the cycle which is used for a  Sommerfeld--Watson transformation. 
By using Cauchy's theorem \Cauchy\ and taking into account the branch cuts when moving along the paths $C_2$ and $C_2'$ thanks to the relations \differ\ and \differi\ the  contour integral along $\Gamma$ can be expressed  for given $\si^1_t\in(0,1)$ as:
\eqnn\Pyramidenspitze{
$$\eqalignno{
\oint_{\Gamma}d\sigma_t^2\  \hat I(\sigma^1_t,&\sigma^2_t;\ell)=\lf(1-\varphi_t^2\ri) \ \int\limits_{-\tau_2/2}^{\tau_2/2}\ d\sigma_t^2\ I(\sigma_t^1,\sigma_t^2;\ell)&\Pyramidenspitze\cr
&+i\varphi_t\int_0^1\!d\tilde\sigma_t^2\ \lf[\; I_{A_1',A_2'}(\sigma^1_t,-i\tilde\sigma^2_t;\ell-\fc{q_t}{2})- I_{A_1',A_2'}(\sigma^1_t,-i\tilde\sigma^2_t;\ell+\fc{q_t}{2})\;\ri].}$$}
At $C_2,C_2'$ we have  the respective non--planar configuration $A_1',A_2'$ following from \differi. Furthermore, the relevant theta--functions in the object \Relevantnp\ have even characteristics  and thus   do not produce branch points along  $\sigma^2_t\in C_2,C_2'$. Altogether, thanks to \differi\ the last line of \Pyramidenspitze\ gives a vanishing contribution (after integration over the loop momentum $\ell$).
On the other hand, for the contour $\Rc$   we have:
\eqn\Peterskoepfl{
\oint_\Rc\ d\sigma_t^2\ \hat I(\sigma^1_t,\sigma^2_t;\ell)= i\; \varphi_t\;(\varphi_t^{-1}-\varphi_t)\ \int_0^1\! d\tilde\sigma_t^2\  \hat I(\sigma_t^1,-i\tilde\sigma_t^2;\ell)\ .}
Above, we have introduced the parameterization $\tilde\sigma^2_t:=i\sigma^2_t$ subject to $\sigma^2_t\in+ i(-1,0)$ for $C_+$ and $\sigma^2_t\in +i(0,-1)$ for   $C_-$, respectively. The integrands $I$ for  the contributions along $C_+$ and $C_-$ are identical subject to \amountshift.
The appropriate  phase factors accounting for the branch cut structure along $C_\pm$ are accommodated   
by accordingly choosing  the integration paths along $\tilde\sigma_t^2\in(0,1)$ and passing the branch points in suitable direction. Compared to $C_+$ at $C_-$ the opposite integration direction is realized and the branch points are passed in opposite orientation. The branch point at $\si^2_t=0$ prevents the contributions of $C_+$ and $C_-$ from cancelling out. This situation  is similar to the monodromy structure of the two contours $C_\pm$ in the tree--level case discussed in Appendix A.2. As a consequence in \Peterskoepfl\ we are dealing with  the integrand $\hat I$ that accounts for the two paths of integrations along $\tilde\sigma_t^2\in(0,1)$, respectively.
Inserting \Pyramidenspitze\ and \Peterskoepfl\ into \Cauchy\ yields for given $\si^1_t\in(0,1)$:
\eqn\Contour{
 \int\limits_{-\tau_2/2}^{\tau_2/2}\ d\sigma_t^2\ I(\sigma_t^1,\sigma_t^2;\ell)= -i \int_0^1\! d\tilde\sigma_t^2\ \hat I(\sigma_t^1,-i\tilde\sigma_t^2;\ell)\ .}
Notice, that the combinations of cycles \edges\ and \edgeR\ resembles a version of Hankel contour  -- a path in the complex plane. An alternative derivation of \Contour\ will be presented in Appendix A.4.

For $\tilde\sigma_t^2\in(0,1)$ we can introduce   the following new real coordinates\foot{Note, that due to $\tilde\si_t^2\in(0,1)$ this map implies $\xi_t\!>\!\eta_t$. Below we shall relax this constraint by performing appropriate shifts in the $\xi_t,\eta_t$--plane while 
retaining the cut structure of the integrand of \Peterskoepfl\ along $\tilde\si_t^2\in(0,1)$.
\comment{at the cost of introducing some loop--momentum dependent phase factors}}
\eqn\newcoords{\eqalign{
\xi_t&=\sigma^1_t+\tilde\sigma^2_t\ ,\cr
\eta_t&=\sigma^1_t-\tilde\sigma^2_t\ ,}}
with $\det\lf(\fc{\p(\xi_t,\eta_t)}{\p(\sigma^1_t,\tilde\sigma^2_t)}\ri)=-2$. 
With this  parameterization  the integrand in \Contour\ can be specified as 
some function depending on $\xi_t$ and $\eta_t$ as:
\eqn\integrands{
\hat I(\sigma^1_t,-i\tilde\sigma^2_t;\ell)\equiv \hat I(\xi_t,\eta_t;\ell)\ .}
Actually, the precise form of the latter with the correct phases will be displayed in the next subsection. 
At any rate, the function \integrands\  entering the integrand \Contour\ is holomorphic and single--valued w.r.t. $\xi_t,\eta_t$ after imposing the correct branch cut structure.
In fact, a manifestly single--valued expression for the momentum dependent factor \Loopf\ is obtained from \Regalpwand\ by considering  the following representation:
\eqn\Breakthrough{\eqalign{
\Lc(\{\si^2_s\})&=e^{-\pi\ap \tau_2\ell^2}\ \prod_{r=1}^{n} \lf|\theta_1(\si^1_r+i\si^2_r+\fc{\tau}{2})\ri|^{\ap\ell q_r}\ \lf|\theta_1(\si^1_r+i\si^2_r-\fc{\tau}{2})\ri|^{-\ap\ell q_r}\cr
&=e^{-\pi\ap \tau_2\ell^2}\ \prod_{r=1}^{n} \lf(\fc{\theta_1(\si^1_r+i\si^2_r+\fc{\tau}{2})}{\theta_1(\si^1_r+i\si^2_r-\fc{\tau}{2})}\ri)^{\fc{\ap}{2}\ell q_r}\ \lf(\fc{\theta_1(\si^1_r-i\si^2_r-\fc{\tau}{2})}{\theta_1(\si^1_r-i\si^2_r+\fc{\tau}{2})}\ri)^{\fc{\ap}{2}\ell q_r}\ .}}
This object furnishes the dependence on $\si^1_t$ and is suitable to discuss and impose the branch cut structure along $\si^2_t\in C_\pm$.  It exhibits a pairwise branch cut structure in the complex $\si^2_t$--plane referring to the 
cuts  $\si_t^2=\pm i\si_t^1-\h t$ with branching $\h\ap \ell q_t$. Likewise, we have  cuts with branching $-\h\ap \ell q_t$ at  $\si_t^2=\pm i\si_t^1+\h t$. The local effect of these branchings has  already been exhibited by \Regalpwand. 
The expression \Breakthrough\  will be used hereinafter when discussing the analytic continuation of \Relevant\ and \integrands\ along $C_\pm$ and imposing the correct phase convention.

Due to \newcoords\ the region $(\sigma^1_t,\tilde\sigma^2_t)\in[0,1]^2$ is (biholomorphically) mapped to a rhombus  accounting for the $\xi_t,\eta_t$--dependence of \integrands, cf. Fig.~3 and  \StiebergerDAA. 
This diamond can be divided into the four triangular domains
\eqnn\domains{
$$\eqalignno{
\Dc_{I}&=\{(\xi_t,\eta_t)\ |\ 0<\xi_t<1\ \wedge\ 0\leq\eta_t\leq\xi_t\} \ ,\cr
\Dc_{II}&=\{(\xi_t,\eta_t)\ |\ 1<\xi_t<2\ \wedge\ 0\leq\eta_t\leq2-\xi_t\} \ ,\cr
\Dc_{III}&=\{(\xi_t,\eta_t)\ |\ 1<\xi_t<2\ \wedge\ \xi_t-2\leq\eta_t\leq0\}\ ,\cr
\Dc_{IV}&=\{(\xi_t,\eta_t)\ |\ 0<\xi_t<1\ \wedge\ -\xi_t\leq\eta_t\leq0\}  \ ,&\domains}$$}
depicted in Fig.~3.
\ifig\Region{Bijective map from the rectangle to the rhombus divided into the four domains $\Dc_i$}{\epsfxsize=0.85\hsize\epsfbox{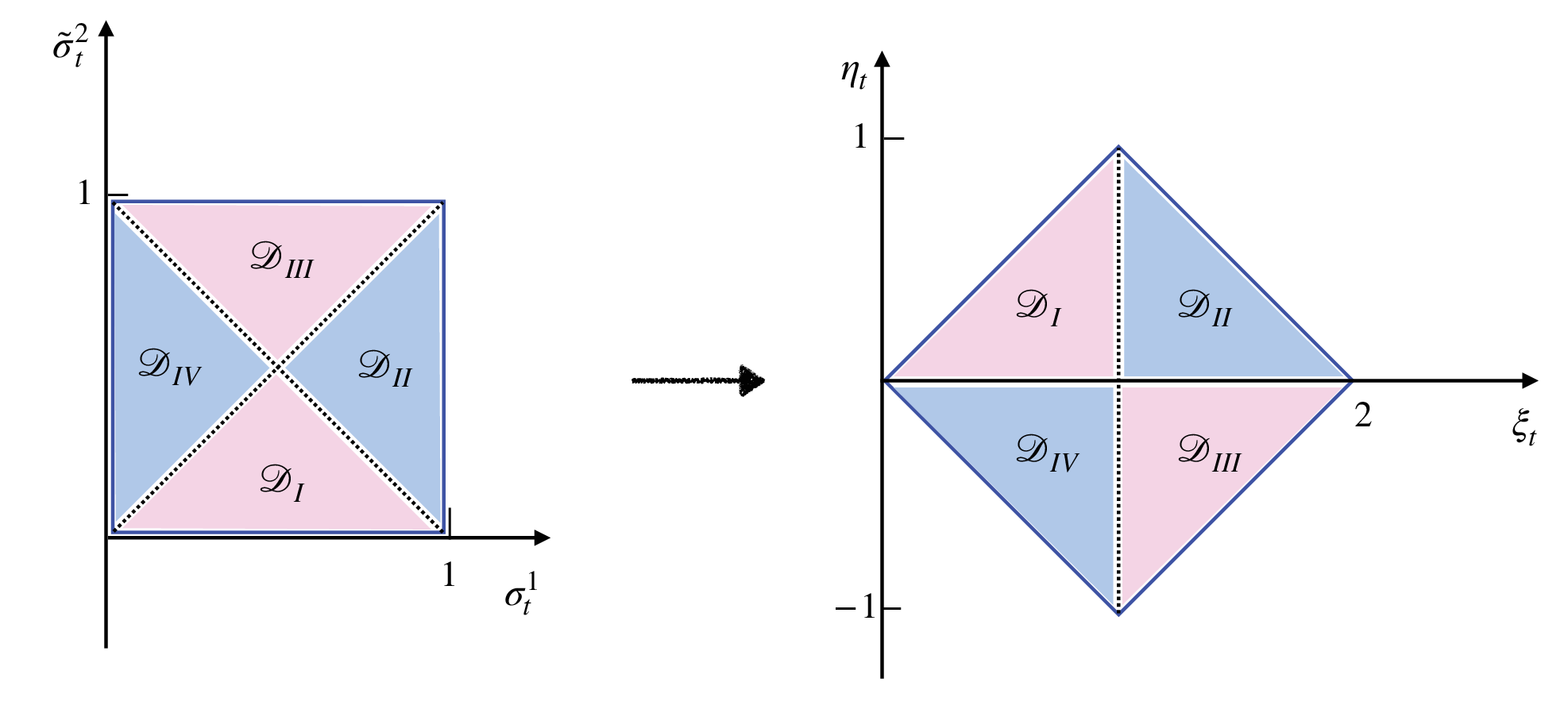}}
\noindent 
For  these four domains for \Breakthrough\ we have
\eqn\Seeberg{
\Lc(\{\si^2_s\})=e^{-\pi\ap \tau_2\ell^2}\ \prod_{r=1}^{n} \lf(\fc{\theta_1(\xi_r+\fc{\tau}{2})}{\theta_1(\xi_r-\fc{\tau}{2})}\ri)^{\fc{\ap}{2}\ell q_r}\ \lf(\fc{\theta_1(\eta_r-\fc{\tau}{2})}{\theta_1(\eta_r+\fc{\tau}{2})}\ri)^{\fc{\ap}{2}\ell q_r}\ ,}
and we can study the local system
\eqn\local{\eqalign{
\hat I(\xi_t,\eta_t;\ell)&\sim \exp\lf\{-\pi\ap \tau_2\ell^2-\pi i \ell\ap \sum_{r=1}^n q_r(\xi_r-\eta_r)\ri\}\cr 
&\times\prod_{s\neq t}^n|\theta_1(\xi_s-\xi_t,\tau)|^{\fc{\ap}{2}q_tq_s}\; |\theta_1(\eta_s-\eta_t,\tau)|^{\fc{\ap}{2}q_tq_s}\ \Pi_q(t,s)\ ,}}
with the phase factors $\Pi_q$ to be specified below. The latter account for the correct branch when moving in the $(\xi_t,\eta_t)$--plane and crossing the branch cuts (cf. also the comment in  Fn.~4.). In the end, the phase factors  
$\Pi_q$ can be taken into account by appropriately  integrating along $C^\pm$, respectively. After a careful inspection of the dependence of the local system \local\ on the regions \domains\ and performing changes  of integration variables  we find that by shifts in $\xi_t$ and $\eta_t$ the two blue triangles $\Dc_{II}$ and $\Dc_{IV}$ can be  combined to give the unit square $[0,1]^2$:
\eqn\together{
\int_{\Dc_{II}}\hat I(\xi_t,\eta_t;\ell)+\int_{\Dc_{IV}}\hat I(\xi_t,\eta_t;\ell)=e^{-\pi i\ap\ell q_t}\ \int_0^1d\xi_t\int_0^1d\eta_t\ \hat I(\xi_t,\eta_t;\ell)\ .}
Likewise, for the red triangles $\Dc_{I}$ and $\Dc_{III}$ we find:
\eqn\togetheri{\eqalign{
\int_{\Dc_{I}}\hat I(\xi_t,\eta_t;\ell)&= \int_0^1d\xi_t    \int_0^{\xi_t} d\eta_t\ 
\hat I(\xi_t,\eta_t;\ell)\ ,\cr
\int_{\Dc_{III}}\hat I(\xi_t,\eta_t;\ell)&=  e^{-2\pi i\ap\ell q_t}\ \int_0^1d\xi_t   \int_{\xi_t}^1 d\eta_t\ \hat I(\xi_t,\eta_t;\ell)\ .}}
Altogether, we may assemble the four contributions of the regions $\Dc_i$ into:
\eqn\totalcontri{
\int_{\bigcup_{i}\Dc_{i}} \hat I(\xi_t,\eta_t;\ell)= \lf(1+e^{-\pi i\ap\ell q_t}\ri)\ 
\int_0^1d\xi_t   \int_0^1 d\eta_t\ e^{-\pi i\ap\ell q_t\; \theta(\eta_t-\xi_t)}\ \hat I(\xi_t,\eta_t;\ell)\ .}
Eventually, with \Contour\ and  \totalcontri\  the complex integral \need\ over the closed string vertex position $z_t$ becomes:
\eqn\final{\eqalign{
\int_{\Tc}d^2z_t\ I(z_t,\bar z_t;\ell)&=\fc{i}{2}\lf(1+e^{-\pi i\ap\ell q_t}\ri)\ 
e^{-\h\pi i\ap\ell q_t} \cr
&\times\int_0^1d\xi_t\int_0^1d\eta_t\ e^{\h\pi i\ap\ell q_t\;{\rm sgn}(\xi_t-\eta_t)}\;
 \hat I(\xi_t,\eta_t;\ell)\ .}}

\comment{
In these four domains we can study the local system \integrands:
\eqnn\local{
$$\eqalignno{
\hat I(\xi_t,\eta_t;\ell)&\sim e^{-\pi\ap \tau_2\ell^2}\ \prod_{r=1}^{n} \lf(\fc{\theta_1(\xi_r+\fc{\tau}{2})}{\theta_1(\xi_r-\fc{\tau}{2})}\ri)^{\fc{\ap}{2}\ell q_r}\ \lf(\fc{\theta_1(\eta_r-\fc{\tau}{2})}{\theta_1(\eta_r+\fc{\tau}{2})}\ri)^{\fc{\ap}{2}\ell q_r}\cr
&\times\prod_{s\neq t}^n|\theta_1(\xi_s-\xi_t,\tau)|^{\fc{\ap}{2}q_tq_s}\; |\theta_1(\eta_s-\eta_t,\tau)|^{\fc{\ap}{2}q_tq_s}\ \Pi_q(t,s)\ .&\local}$$}
\comment{with two pairs of unintegrated (real) points
These real numbers \Unintegrated\ can be analytically continued to the four complex values
\Additional}
In \local\  we have the  phase factors $\Pi_q$ to be specified below. The latter  account for the correct branch when moving in the $(\xi_t,\eta_t)$--plane and crossing the branch cuts (cf. also the comment in  Fn.~4.). In the end, the phase factors 
$\Pi_q$ can be taken into account by appropriately  integrating along $C_\pm$, respectively. After a careful inspection of the dependence of the local system \local\ on the regions \domains\ and performing changes  of integration variables  we find that by (integer) shifts in $\xi_t$ and $\eta_t$ the two (blue) triangles $\Dc_{II}$ and $\Dc_{IV}$ can be  
combined to give the unit square $[0,1]^2$:
\eqn\together{
\int_{\Dc_{II}}\hat I(\xi_t,\eta_t;\ell)+\int_{\Dc_{IV}}\hat I(\xi_t,\eta_t;\ell)= \int_0^1d\xi_t\int_0^1d\eta_t\ \hat I(\xi_t,\eta_t;\ell)\ .}
Similarly, for the (red) triangles $\Dc_{I}$ and $\Dc_{III}$ we find:
\eqn\togetheri{
\int_{\Dc_{I}}\hat I(\xi_t,\eta_t;\ell)+\int_{\Dc_{III}}\hat I(\xi_t,\eta_t;\ell)= \int_0^1d\xi_t   \int_0^1 d\eta_t\ \hat I(\xi_t,\eta_t;\ell)\ .}
Altogether, we may assemble the four contributions of the regions $\Dc_i$ into:
\eqn\totalcontri{
\int_{\bigcup_{i}\Dc_{i}} \hat I(\xi_t,\eta_t;\ell)=2\   \int_0^1d\xi_t   \int_0^1 d\eta_t\ \hat I(\xi_t,\eta_t;\ell)\ .}    }

\comment{Above we have used the identity:
$$-i\ e^{\h\pi i\ap\ell q_t}\ \fc{1-e^{-2\pi i\ap\ell q_t}}{1+e^{-\pi i\ap\ell q_t}}=2\ \sin\lf(\fc{\pi\ap\ell q_t}{2}\ri)\ .$$}
\comment{For given orderings $\sigma,\rho\in S_{n-1}$ of the positions $\xi_s,\eta_s$, respectively  along $C_\pm$ the phase factor $\Pi^\pm$ of \local\ is a constant $e^{\pm\pi i\ap\Phi_{\sigma\rho}}$
and contribute the terms
\eqnn\display{
$$\eqalignno{
e^{-\pi\ap \tau_2\ell^2}\ &e^{\pm i\pi\ap\Phi_{\sigma\rho}}\int_{\xi_t\in \Ic_\sigma} d\xi_t\  e^{\h\pi i\ap\ell q_t\;{\rm sgn}(\xi_t-\eta_t)}\ e^{-\pi i \ell\ap \sum\limits_{r=1}^n q_r\xi_r}
\prod_{s\neq t}^n|\theta_1(\xi_s-\xi_t,\tau)|^{\fc{\ap}{2}q_tq_s}\cr
&\enspace\qquad\times\int_{\eta_t\in \Ic_\rho} d\eta_t \;  e^{\pi i \ell\ap \sum\limits_{r=1}^n q_r\eta_r}
\prod_{s\neq t}^n|\theta_1(\eta_s-\eta_t,\tau)|^{\fc{\ap}{2}q_tq_s}\ .&\display}$$}
to the  r.h.s. of \final, respectively.}  

Let us summarize our results so far. Eq. \final\ enables us to express the complex torus integral \need\ over $z_t$ in terms two real  integrations each along one cylinder boundary. This promotion can be illustrated by representing the torus $\Tc$  as doubled surface  $\Tc=\Si_-\cup\Si_+$ with the two cylinders $\Si_-,\Si_+$, cf. Fig.~4.
\iifig\Soistes{Deforming the $\pm \Im(z_t)$ axes to the $\Re(z_t)$ axis}{in the torus represented as doubled surface  $\Tc=\Si_-\cup\Si_+$.}{\epsfxsize=0.6\hsize\epsfbox{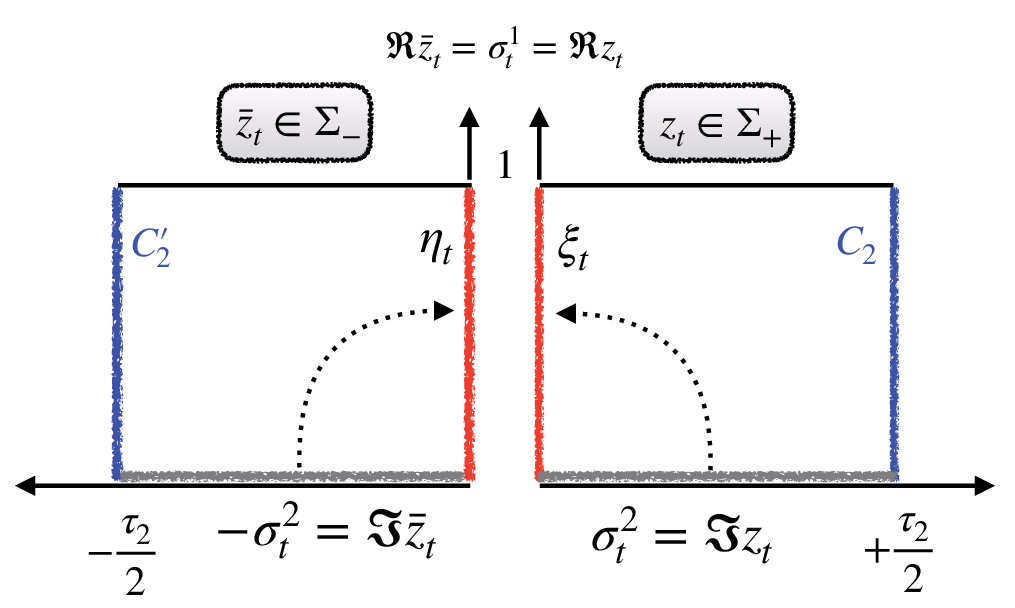}}
\noindent 
After splitting the complex torus coordinate as \sig\ the $\sigma^2_t=\Im(z_t)$--integration 
is deformed to be aligned to $\sigma^1_t=\Re(z_t)$ yielding the real integration variable 
$\xi_t$. Simultaneously, the $\Im(\bar z_t)$--integration 
is deformed to be aligned to $\Re(z_t)$ yielding the real integration variable $\eta_t$.  
For both edges $C_+$ and $C_-$ the real integrations $\xi_t$ and $\eta_t$ describe (iterated) integrals along two  distinct cylinder  boundaries, respectively, cf. also Fig.~6.
Hence, the dependence on the complex closed string coordinate $z_t$ has been converted into contributions from a pair of open strings. 
Eq. \final\ derives from \Contour\ which in turn follows by considering the contour \Cauchy\ in the complex $\sigma^2_t$--plane.

\subsec{Multi--dimensional complex torus integrations}

The expression \Relevant\  is holomorphic in all $n$ coordinates $\sigma_t^2,\; t\!=\!1,\ldots,n$.
In the previous subsection we have exhibited how one complex closed string coordinate $z_t$ is 
converted into a pair of two real open string positions by means of considering a closed 
contour in the complex $\sigma^2_t$--plane and applying Cauchy's theorem  resulting in the relation~\final. 
Now the task is to integrate all closed string coordinates $z_t$ over the full torus~$\Tc$, i.e.
to perform
\eqn\Need{
\lf(\prod_{t=1}^{n}\int_{\Tc}d^2z_t\ri)\ I(\{z_s,\bar z_s\};\ell)=\lf(\prod_{t=1}^{n}\int_0^1d\sigma^1_t\int\limits_{-\tau_2/2}^{\tau_2/2}d\sigma^2_t\ri)\ 
I(\{\sigma^1_s,\sigma^2_s\},\ell)\ ,}
with  the integrands \relevant\ and \Relevant, respectively.

For each of the $n$ closed string coordinates $z_t$ we want to repeat the steps from the previous subsection and perform an analytic continuation of all closed string coordinates $\sigma_t^2,\ t\!=\!1,\ldots,n$, 
defined in \sig. As before we shall discuss the dependence of the integrand \Relevant\ in  the complex $\sigma^2_t$--planes, consider in each complex $\sigma^2_t$--plane closed cycles (polygons) $\Gamma$ defined by the four edges  \edges\ and apply \Cauchy. We simultaneously deform all of the $\sigma^2_t$--integration contours from the real axis to the pure imaginary axes described by the contours $\Rc$ and obtain:
\eqn\VordereKesselschneid{\eqalign{ 
\lf[\prod_{t=1}^n(1-e^{-2\pi i\ap\ell q_t})\ri] &\lf(\prod_{t=1}^n\int\limits_{-\tau_2/2}^{\tau_2/2} d\sigma_t^2\ri) I(\{\sigma^1_s,\sigma^2_s\},\ell)\cr
&=(-1)^n\oint_\Rc\; d\sigma_1^2\ldots\oint_\Rc\; 
d\sigma_n^2\; \hat I(\{\sigma^1_s,\sigma^2_s\},\ell)\ .}}
The multi--dimensional complex contours describing \VordereKesselschneid\ are  emblematized as a multi--page book in Fig.~5. 
 Interestingly, the structure of the right hand side of \VordereKesselschneid\ can be compared with the tree--level analogs (A.17) and (A.19). In particular, the pairs of contours $C_+$ and $C_-$ can be thought as doubles of contours. 
 As in the case of one integration w.r.t. $\si^2_t$, discussed in the previous subsection, the contributions from $C_2$ and $C_2'$ cancel each other.

In the multi--dimensional complex  space $\sigma^2\in\IC^n$ there are now many branch points, all of them lying along the imaginary axis given by $\sigma^2_{st}=\pm i \sigma^1_{st}$. These conveniently divide themselves into pairs (given by $z_r-z_s=0$ and $\bar z_r-\bar z_s=0$), which may be examined essentially independently in the same way as in the previous subsection. On the other hand,  not any cuts appear 
alongside $\Re(\si^2_t)= \fc{\tau_2}{2}$ with $\Im(\si^2_t)\in(-1,0)$ since in the  object \Relevantnp\ the relevant  theta--functions appear with even characteristics. 
\ifig\Buch{Multi--dimensional complex contours for the analytic continuation on the torus.}{\epsfxsize=0.6\hsize\epsfbox{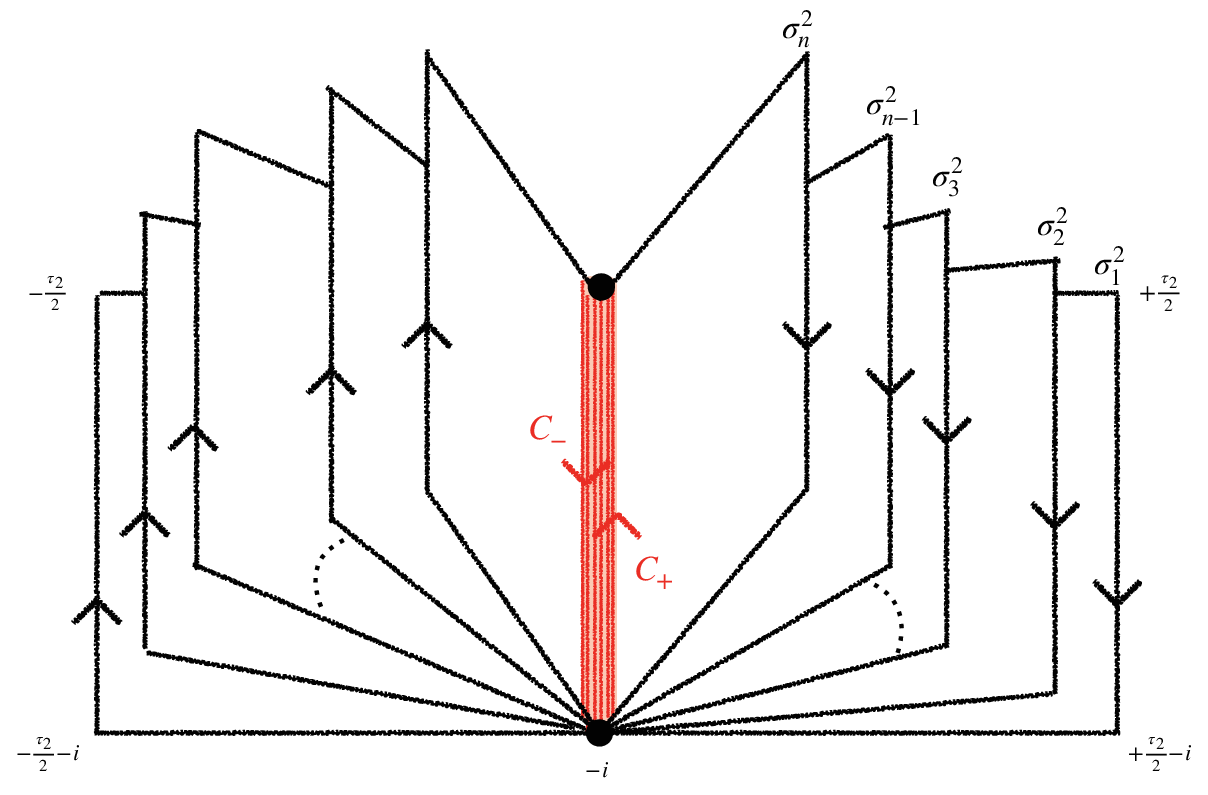}}

After having deformed all $\si^2_t$--integrations to the contours $\Rc$ along the latter we may introduce the real coordinates $\xi_t,\eta_t$ given by \newcoords. As a result,  combining \Cauchy\ and \Pyramidenspitze\ allows to trade in each $\sigma_t^2$ integration by means of deforming its integration region $C_1$ to $C_\pm$.
For $\sigma^2_t\in C_\pm$ we can now define the single--valued function generalizing  the local system \local
\eqnn\Integrands{
$$\eqalignno{
\hat I(\{\xi_t,\eta_t\};\ell)&=e^{-\pi\ap \tau_2\ell^2}\ \exp\lf\{-\pi\ap \tau_2\ell^2-\pi i \ell\ap \sum_{r=1}^n q_r(\xi_r-\eta_r)\ri\}&\Integrands\cr
& \times\prod_{r,s=1\atop r<s}^{n} |\theta_1(\xi_s-\xi_r,\tau)|^{\h\ap q_sq_r} 
 |\theta_1(\eta_s-\eta_r, \tau)|^{\h\ap q_sq_r}\; \Pi_q(r,s)\ ,\ \si^2_t\in C_\pm.}$$}
For a generic configuration of $n$ closed  string vertex positions  the phase factors $\Pi_q$ render  the correct branch of the integrand \Integrands\  after having deformed the contour $C_1$ to the integration paths $C_-$ and $C_+$:
\eqn\Phases{
\Pi_q(r,s):=\Pi(\xi_s,\xi_r,\eta_s,\eta_r; q_rq_s)=e^{\h\pi i\ap q_rq_s (1-\theta[(\xi_r-\xi_s)(\eta_r-\eta_s)])}\ ,\ \si^2_t\in C_\pm\ .}
In other words, the phase factors $\Pi_q$ make sure, that we stay in the correct branch when the coordinates $\xi_t,\eta_t$ are varied along $C_\pm$, respectively. On the other hand, the first line of \Integrands\ corresponds to just one particular selection of cuts and thus a particular choice of analytic analytic continuation of \Loopf.  In Subsection 4.5 we shall employ a different representation  thereof. Now,  in \Integrands\ all the theta--functions  entering the integrand \VordereKesselschneid\ are analytic and single--valued w.r.t. $\xi_s,\eta_s\in\IR$. For each given orderings of $\xi_s$ and $\eta_s$  the phase factors \Phases\ are constants and they can be taken into account by appropriately encircling the branch points along $C_\pm$ (cf. also Appendix A.2). Eq. \final\  can be applied for the full integrand  \Integrands\ as
\eqn\Final{
\int_{\Tc}d^2z_t\ I(\{z_s,\bar z_s\};\ell)=\fc{i}{2}\ \int_0^1d\xi_t\int_0^1d\eta_t\ \Psi(\xi_t,\eta_t;\ell)\; \hat I(\{\xi_s,\eta_s\};\ell)\ ,}
with:
\eqn\EllisJetzer{
\Psi(\xi_t,\eta_t;\ell)=\lf(1+e^{-\pi i\ap\ell q_t}\ri)\ 
 e^{-\pi i\ap\ell q_t\;\theta(\eta_t-\xi_t)}  \ .}
 The splitting function $\Psi$ essentially subjects level matching conditions to the left-- and 
 right--movers.
\comment{
On the other hand, the analytic continuation from the points \Additional\ to the real numbers \Unintegrated\ is  described by the choice:}
\comment{with  the additional unintegrated  points \Additional\ and the corresponding momenta: 
In \Integrands\ the prime at the product excludes the case $r\!=\!0,s\!=\!n+1$.}

To convert all $n$ complex world--sheet torus integrals into pairs of real integrals
we start with a closed string position (e.g. $z_t$ with $t\!=\!1$) and apply \Final.
Then, we successively  apply the formula \Final\ for all other complex coordinates $z_r,r=2,\ldots,n$. 
After reinstalling the correct branch cut structure we shall end up with $2n$ ($2n-2$ after cancelling the conformal Killing factor $V_{CKG}$) real integrations $\xi_r,\eta_r$ along the cylinder  boundaries $C_+,C_-$:
\eqn\Gruttenhuette{
\lf(\prod_{r=1}^{n}\int_{\Tc}d^2z_r\ri)\ I(\{z_s,\bar z_s\};\ell)
= \lf(\fc{i}{2}\ri)^n\;\lf(\prod_{t=1}^{n}\int_0^1d\xi_t\int_0^1d\eta_t\  \Psi(\xi_t,\eta_t;\ell)\ri)\  
\hat I(\{\xi_s,\eta_s\};\ell) \ .}
\comment{Actually, in \Gruttenhuette\ we may  incorporate  the following relation: 
$$\lf(\prod_{r=1}^{n}\fs\lf(q_t,\ell\ri)\ri)\;(-1)^n\;I^-(\{\xi_s,\eta_s\};\ell)=
\lf\{\;\lf(\prod_{r=0}^{n+1}\fs\lf(q_t,-\ell\ri)\ri) I^+(\{\xi_s,\eta_s\};-\ell)\;\ri\}^\ast\ .$$
With this information and}
After  putting everything together  Eq. \Need\ amounts to
\eqnn\Gaudeamushuette{
$$\eqalignno{
\int_{-\infty}^\infty d^d\ell\ &\lf(\prod_{r=1}^{n}\int_{\Tc}d^2z_r\ri)\ I(\{z_s,\bar z_s\}
=\lf(\fc{i}{2}\ri)^n\;\int_{-\infty}^\infty\!\! d^d\ell\  e^{-\pi\ap \tau_2\ell^2}\cr
&\times\lf(\prod_{r=1}^{n}\int_0^1d\xi_r\int_0^1d\eta_r\ \Psi(\xi_t,\eta_t;\ell) \ri) \exp\lf\{-i\pi\ap \ell\sum\limits_{r=1}^{n}q_r(\xi_r-\eta_r)\ri\} \cr
&\times    \prod_{r,s=1\atop r<s}^{n}|\theta_1(\xi_s-\xi_r,\tau)|^{\h\ap q_sq_r}\; 
 |\theta_1(\eta_s-\eta_r, \tau)|^{\h\ap q_sq_r}\ \Pi_q(r,s)\ ,&\Gaudeamushuette}$$}
with  the phase factor $\Pi_q$ defined in \Phases. 
A closed string sector (akin two additional open strings) is furnished by the loop momentum dependent phase factor in the second line of \Gaudeamushuette, cf. Section 4. The particular monodromy structure of the aforementioned  matches the correct large complex structure limit $\tau\ra i\infty$, cf. Section 4. 

Lastly, before introducing the new coordinates \newcoords\ and phases \Phases\ the integral over the loop momentum $\ell$ can be performed as Hubbard--Stratonovich  transformation
\eqn\Prama{
\int_{-\infty}^\infty d^d\ell\ e^{-\pi\ap \tau_2\ell^2}\ 
e^{-2\pi i\ap \ell\sum\limits_{r=1}^nq_r\tilde\si^2_r}\!\!=(\ap\tau_2)^{-d/2}\exp\lf\{-\fc{\pi\ap}{\tau_2}\lf(\sum_{r=1}^n
q_r\tilde\si_r^2\ri)^2\ri\}\ ,}
which however obscures the direct product structure of our one--loop KLT result \Gaudeamushuette\
after introducing the coordinates \newcoords.

\comment{Above we have reinstalled the prefactor $(\ap\tau_2)^{d/2}$ of \looprepr.}

\comment{referring to the the pair of points \Additional\ }
\comment{The latter takes into account the analytic continuation described by the pair of points \Additional\ and the phases \PhaseEYM.}

\newsec{The one--loop KLT relation}

In this section we both summarize and discuss our main results from the previous section. The latter give rise to the one--loop KLT relation expressing (subject to final complex structure modulus integration) one--loop closed string amplitudes \Start\  as a (weighted) sum over squares of certain one--loop open string subamplitudes. The latter are expressed in the so--called closed string channel.
With the result \Gaudeamushuette\ the one--loop $n$ closed string torus amplitude \Start\ in $d$ space--time dimensions becomes
\eqnn\Scheffau{
$$\eqalignno{
\Ac_n^{(1)}(q_1,\ldots,q_n)&=\h\; g_c^n\ \delta^{(d)}\lf(\sum_{i=1}^n q_i\ri)  
\int d\tau_2\  V_{CKG}^{-1}(\Tc) \int_{-\infty}^\infty d^d\ell\ e^{-\pi\ap \tau_2\ell^2}\cr
&\times\lf(\fc{i}{2}\ri)^n \lf(\prod_{r=1}^{n}\int_0^1d\xi_r\int_0^1d\eta_r\ \Psi(\xi_t,\eta_t;\ell)\ri)\prod_{r,s=1\atop r<s}^{n}\!\!\Pi_q(r,s)\cr
&\times  \lf(\prod_{r,s=1\atop r<s}^{n} |\theta_1(\xi_s-\xi_r,\tau)|^{\h\ap q_sq_r}\ri)\ e^{-i\pi\ap \ell\sum\limits_{r=1}^{n}q_r\xi_r}\;Q_L(\tau,\{\xi_s\}) \cr
&\times \lf(\prod_{r,s=1\atop r<s}^{n} 
 |\theta_1(\eta_s-\eta_r, \tau)|^{\h\ap q_sq_r}\ri)\ e^{i\pi\ap \ell\sum\limits_{r=1}^{n}q_r\eta_r}\; Q_R(\ov\tau,\{\eta_s\}),&\Scheffau}$$}
with  the phases specified in \Phases\ and \EllisJetzer, respectively.
Furthermore, we have the  modular functions $Q_L,Q_R$ (accounting for the left-- and right--moving modes with weights  $n+1-d/2$ in the massless case) defined in \splitQ\  and the complex structure modulus $\tau\!=\!i\tau_2$.
In addition, in \Scheffau\ the volume $V_{CKG}(\Tc)$ of the conformal Killing group can be cancelled by fixing one complex closed string vertex operator position e.g. $z_n=\bar z_n=1$, which in turn results in fixing  the two real coordinates $\xi_n,\eta_n=1$ on the two cylinders, respectively:
\eqn\fix{
\xi_n=1\ \ \ ,\ \ \ \eta_n=1\ .}

\subsec{Geometric picture of the one--loop KLT relation}

The underlying world--sheet of the expression \Scheffau\ can be interpreted as a non--planar cylinder   with a closed string insertion, cf. Fig.~6.
More precisely, the one--loop torus is sliced along the $A$--cycle with $n$ pairs of open string positions $\xi_i$ and $\eta_j$ located along the two boundaries, respectively resulting in a non--planar one--loop cylinder 
configuration. The details of the cutting procedure is governed by the splitting function $\Psi$ defined in \EllisJetzer\ and originating  from the change of coordinates \newcoords\ along the two boundaries.
\iifig\BasicCut{Slicing the torus into one cylinder with}{a closed string insertion of momentum  $\pm\ell$.}{\epsfxsize=0.5\hsize\epsfbox{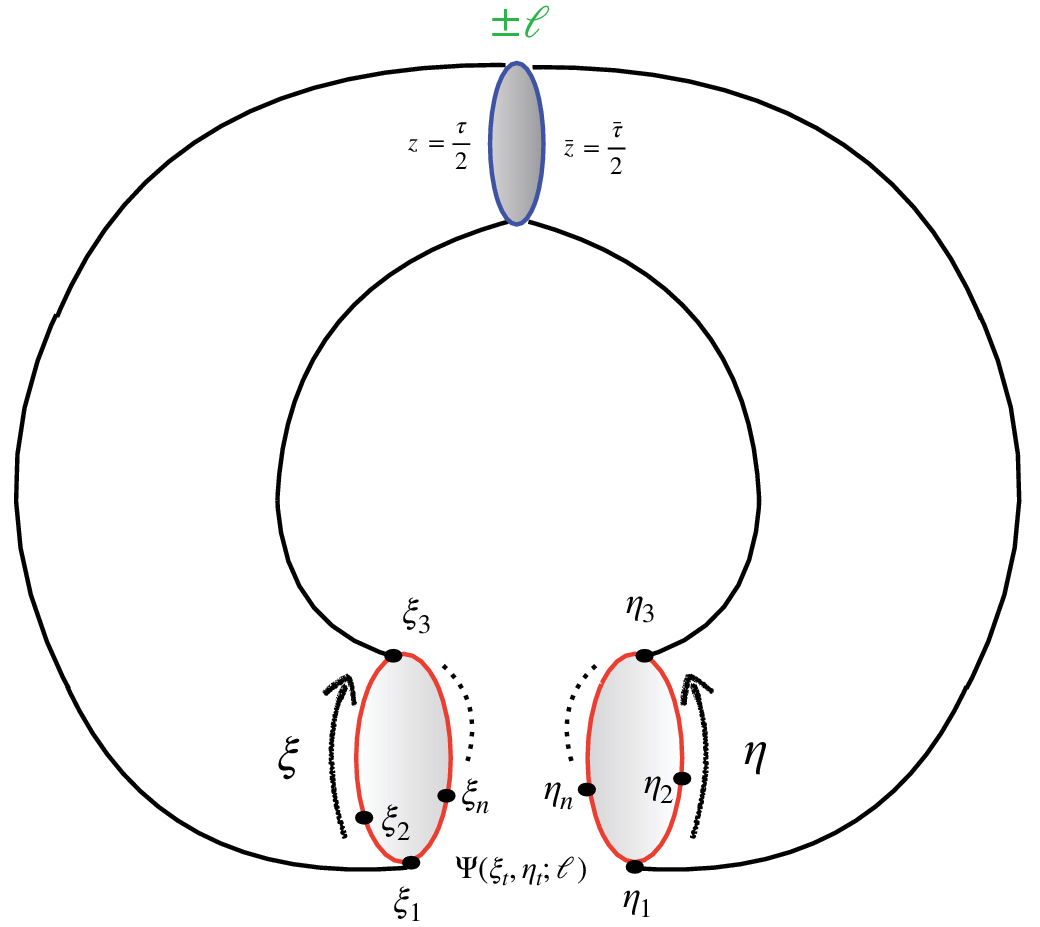}}
\noindent
Note, that in \Scheffau\  the theta--function expressions are given in the closed string channel, i.e. 
open string vertex positions are located at the boundary $\xi_r,\eta_r\in(0,1)$ of the 
cylinder  corresponding to the real unit segment. 
The loop momentum dependent phase factors of \Scheffau\ 
can be written in the following way
\eqn\HintereKesselschneid{\eqalign{
e^{-i\pi\ap \ell\sum\limits_{r=1}^{n}q_r\xi_r}&=\prod_{r=1}^n\theta_1\lf(\xi_r-z,\tau\ri)^{-\h\ap \ell q_r}\ \theta_1\lf(\xi_r-\ov z,\tau\ri)^{\h\ap \ell q_r}\ ,\cr
e^{i\pi\ap \ell\sum\limits_{r=1}^{n}q_r\eta_r}&=\prod_{r=1}^n\theta_1\lf(\eta_r-z,\tau\ri)^{\h\ap \ell q_r}\ \theta_1\lf(\eta_r-\ov z,\tau\ri)^{-\h\ap \ell q_r}\ ,}}
respectively.
Then, there is an unintegrated  (off--shell) closed string insertion with Dirichlet boundary conditions at
\eqn\Ebbs{
z=\fc{\tau}{2}+a\ \ \ ,\ \ \ \bar z=\fc{\ov\tau}{2}+a\ ,}
closed string momenta 
\eqn\Schanz{
q_L=-\h\ell\ \ \ ,\ \ \ q_R=+\h\ell\ ,}
and some free parameter $a\in\IR$ along the $A$--cycle of the torus.

World--sheet geometries with smooth boundary components describe (semi) off--shell string amplitudes \CohenPV. More concretely,  while the two boundary circles  (depicted  in Fig.~6  in red) account for closed (semi) off--shell strings, the circle (depicted  in Fig.~6  in blue)  parameterized by \Ebbs\  accounts for the insertion of a single closed  string state with momenta  \Schanz\ and Dirichlet boundary  conditions. It can be shown, that the net effect is described by the integrand of  \Scheffau. In particular, together with the modulus integration the factor $\int_0^\infty d\tau_2\;e^{-\pi\ap \tau_2\ell^2}$ 
constitutes the semi off--shell propagator in momentum space \Cohen.
Incidentally, the splitting function defined in \EllisJetzer\  may be understood as result from sewing the two red boundaries with insertions $\xi_i,\eta_j$ along the gluing boundary.

The internal loop momentum $\ell$ describes the momentum flowing through a specific torus cycle and is defined as average \DHokerPDL. Concretely, for the torus $B$--cycle  we have:
\eqn\loopflow{
\ell^\mu=\oint_{B}\fc{dz}{2\pi}\ \p_z X^\mu(z)\ .}
In the large complex structure limit $\tau\ra i\infty$ the closed string \Ebbs\ becomes a node connecting
two degenerating cylinders, cf. Fig.~7.
\iifig\BasicCut{Slicing the torus into two cylinders connected}{by a closed string node exchanging the loop momentum  $\pm\ell$.}{\epsfxsize=0.5\hsize\epsfbox{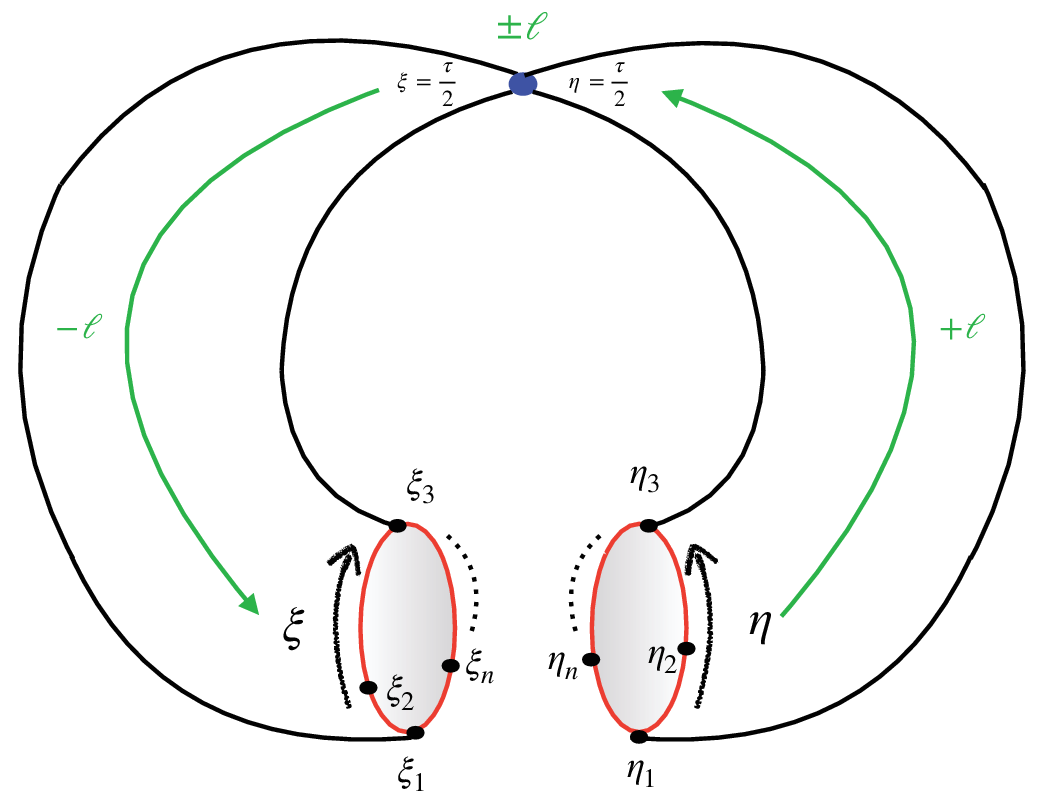}}
\noindent
In this limit the torus is pinched to a node along the $B$--cycle and  the diagram Fig.~6 turns into Fig.~7 displaying a product of two disk diagrams each with a single closed string insertion at the node. 
In Fig.~7 the loop momentum \loopflow\ flowing along the $B$--cycle can now be exchanged between the two surfaces through the node at infinity, while in Fig.~6 the loop momentum endows the (red) $A$--cycles with the splitting function $\Psi$.
The large complex structure limit $\tau\ra i\infty$  will be throughly  discussed in Subsection 4.5.

\subsec{One--loop closed string amplitude as  sum over squares of open string amplitudes}
\def\If{\frak I}

At tree--level the string  KLT relation \KawaiXQ\ is triggered by the so--called KLT kernel \doubref\BernSV\BjerrumBohrHN, which is an intersection matrix combining different (color) orderings from  left-- and right--movers. Actually,  the tree--level KLT kernel derives from \Phases\ by applying open string monodromy relations in both sectors. 
Therefore, we expect the phase factor $\Pi_q$  to also conspire with one--loop open string monodromy relations into  a similar pattern. 
However, now the splitting function $\Psi$ intertwines the real (open string) integrations with the phase factor $\Pi_q$.

According to the specific structure of the phase factors \Phases\ in the result \Scheffau\ the real integrations of the coordinates $\xi_r,\eta_r\in(0,1)$ can be divided into segments subject to their orderings along the unit interval.
More precisely, for given orderings $\sigma,\rho\in S_{n-1}$ of the real coordinates $\xi_r,\eta_r\in(0,1)$ within the unit interval  (with \fix, i.e. $\xi_n=1,\eta_n=1$) the domain of iterated integration is given by
\eqn\domain{
\Ic_\si=\{(\xi_1,\ldots,\xi_{n-1})\in \IR^{n-1}\ |\ 0\leq \xi_{\si(1)}<\ldots<\xi_{\si(n-1)}< 1\}\ ,}
subject to the   ordering $\si$ of open string vertex operator insertions along the unit interval and similarly for $\Ic_\rho$.
Then, for these orderings $\sigma,\rho$ our product of phase factors $\Pi_q$ entering  \Scheffau\ reduces  to a constant $e^{i\pi \ap\Phi(\sigma,\rho)}$ depending only on the external momenta $q_s$. In addition,  for given  orderings $\sigma,\rho$ the splitting function \EllisJetzer\ subjects the real (iterated) integrations $\xi_t\in\Ic_\sigma$ and $\eta_t\in\Ic_\rho$.
Initially, we may define the one--loop intersection matrix~as
\comment{is furnished by appropriately choosing arcs with phases $e^{i\pi\phi_{\sigma,\rho}(r)}$ around the $n$ points $\eta_r$ such that $\Phi_{\sigma,\rho}=\sum_{r=1}^{n}\phi_{\sigma,\rho}(r)$ and connect the latter by performing iterated integrations subject to the orderings $\sigma,\rho$ and orientations of $C_\pm$.
In this setup the difference between integrating along $C_+$ or $C_-$ results in opposite phases 
$\phi_{\sigma,\rho}(r)$ and integration directions between the points.
As a result we can write
\eqn\Steingrubenwand{\eqalign{
V_{CKG}^{-1}(\Tc) &\lf(\prod_{r=1}^{n}\int_0^1d\xi_r\ri)   \lf(\prod_{r=1}^{n}\int_0^1d\eta_r \ri)\ \Theta(\ell;\{\xi_s,\eta_s\})
 \prod_{t=1}^n \fs\lf(q_t,\ell\ri) \prod_{r,s=1\atop r<s}^n\Pi(r,s)\cr
&=\sum_{\si,\rho\in S_{n-1}}  \prod_{t=1}^{n-1}\lf(e^{i\pi\phi_{\sigma,\rho}(t)}-e^{-i\pi\phi_{\sigma,\rho}(t)}\ri)
  \int_{\Ic_\sigma} 
\!\!\!\lf(\prod_{r=1}^{n-1}d\xi_r\ri)   \int_{\Ic_\rho} 
\!\!\!\lf(\prod_{r=1}^{n-1}d\eta_r\ri) \Theta(\ell;\{\xi_s,\eta_s\}),}}
with $\Theta$ accounting for the remaining functions in the integrand \Scheffau.}
\eqn\KERNEL{
S^{(1)}[\sigma|\rho]:=\lf(\fc{i}{2}\ri)^{n-1}\ e^{i\pi \ap\Phi(\sigma,\rho)}\ ,}
with $\Phi(\sigma,\rho)$ accounting for the total phase stemming from \Phases\ for the  orderings $\sigma,\rho$ of the $\xi$ and $\eta$ subject to \fix.
For any orderings $\sigma,\rho\in S_{n-1}$ the matrix \KERNEL\ boils down to a (trigonometric) expression depending only on the external momenta $q_s$ and the  loop momentum $\ell$. The dependence on the latter enters after applying monodromy relations.
In practice, for a given ordering $\si$ of the $n\!-\!1$ coordinates $\xi_s$ the total phase $e^{i\pi \ap\Phi(\sigma,\rho)}$ originating from \Phases\ is furnished by appropriately choosing  paths for the $\eta$--coordinates along the interval $\eta_s\in(0,1)$, cf. also Appendix A.2.
We relegate further simplifications of \KERNEL\ to the end of this subsection. At any rate, the final result \Scheffau\ turns out to be a real number expression.

\comment{In order to produce  the phase factor $e^{- \pi i\ap\ell q_t\delta^t_{\sigma,\rho}}$ the path for $\eta_t$ passes the coordinate $\xi_t$ in clock--wise (negative) direction.
As a consequence we may always deform  the $\eta_t$ cycle to the left or right of $\eta_t=\xi_t$. This manipulation removes the  disentanglement $\delta^t_{\sigma,\rho}$ of the factor $e^{- \pi i\ap\ell q_t\delta^t_{\sigma,\rho}}$ by replacing the latter by some $\sin(\pi\ap\ell q_t)$ factor.}

As a consequence, for given orderings $\sigma,\rho\in S_{n-1}$ of the coordinates $\xi_r,\eta_r\in(0,1)$ with \fix\ the amplitude \Scheffau\  represents a sum over squares of certain one--loop  open string cylinder integrands, each with  $n$ open string insertions $\xi_i$ and $\eta_i$ ordered  along one of the two boundaries of the two cylinders, respectively.
In fact, let us define the following functions
\eqn\OpenW{\eqalign{
\If_{n+2}(\ell)&=g_o^n \  \Bigg(\!\prod_{r,s=1\atop r<s}^n 
|\theta_1(\xi_s-\xi_r,\tau)|^{\fc{\ap}{2} q_sq_r}\!\!\Bigg)
\;e^{-\pi i\ap \ell\sum\limits_{r=1}^{n}q_r\xi_r}\ Q_L(\tau,\{\xi_s\}),\cr
\tilde \If_{n+2}(\ell)&=g_o^n \   \Bigg(\!\prod_{r,s=1\atop r<s}^n 
 |\theta_1(\eta_s-\eta_r, \tau)|^{\fc{\ap}{2} q_sq_r}\!\!\Bigg)
\;e^{\pi i\ap \ell\sum\limits_{r=1}^{n}q_r\eta_r} Q_R(\ov\tau,\{\eta_s\}),}}
relating to specific integrands of (planar) one--loop open string amplitudes.
The $n$ open string momenta $p_i$ entering the open string integrands \OpenW\  are given by the closed string momenta as:
\eqn\OpenMom{
p_i=\h\ q_i\ \ \ ,\ \ \ i=1,\ldots,n\ .}
With 
\eqn\KLTCKG{
 g_c=g_o^2\ ,}
and \OpenW\ we can express \Scheffau\ in the following way:
\eqnn\EllmauW{
$$\eqalignno{
\qquad\Ac_n^{(1)}(q_1,\ldots,q_n)&=\h\  \delta^{(d)}\lf(\sum_{i=1}^n q_i\ri)\;\int d\tau_2\int_{-\infty}^\infty d^d\ell\ e^{-\pi\ap \tau_2\ell^2}\cr
&\times \!\!
\sum_{\sigma,\rho\in S_{n-1}} \int_{\Ic_\sigma}\Bigg(\prod_{r=1}^{n-1}d\xi_r\Bigg)\int_{\Ic_\rho}\Bigg(\prod_{r=1}^{n-1}d\eta_r\Bigg)\lf(\prod_{t=1}^{n-1}\Psi(\xi_t,\eta_t;\ell)\ri)\cr
&\times  \If_{n+2}(\ell)\ 
S^{(1)}[\sigma|\rho]\ \tilde \If_{n+2}(\ell)\ .&\EllmauW }$$}
By cutting the torus world--sheet into one cylinder we have accomplished to express (subject to final complex structure modulus integration) the torus amplitude involving $n$ closed strings \Start\ as sum over  squares of certain cylinder integrands \OpenW\ involving $2n$ open strings. While the foundation of this cutting procedure originates from the monodromy discussions in  Fig.~4, the final result \Scheffau\ is illustrated in Fig.~6.

The position dependent phase factors \HintereKesselschneid\ appearing in \OpenW\  are familiar from discussing open string one--loop monodromy relations \HoheneggerKQY. For \HintereKesselschneid\ we can write down the following identities
\eqn\loopoff{\eqalign{
e^{-i\pi\ap \ell\sum\limits_{r=1}^{n}q_r\xi_r}& =\prod_{r=1}^n\lf(\fc{\theta_1(\xi_r-\xi_0,\tau)}{\theta_1(\xi_r-\xi_{n+1},\tau)}\ri)^{\h\ap \ell q_r}\ ,\cr
e^{i\pi\ap \ell\sum\limits_{r=1}^{n}q_r\eta_r}&=\prod_{r=1}^n\lf(\fc{\theta_1(\eta_r-\eta_0,\tau)}{\theta_1(\eta_r-\eta_{n+1},\tau)}\ri)^{\h\ap \ell q_r}\ ,}}
subject to the four marked points
\eqn\Additionala{\eqalign{
\xi_0&=-\fc{\tau}{2}+a\ \ \ ,\ \ \ \eta_0=+\fc{\tau}{2}+a\ ,\cr
\xi_{n+1}&=+\fc{\tau}{2}+a\ \ \ ,\ \ \  \eta_{n+1}=-\fc{\tau}{2}+a\ ,}}
with some complex number $a\in\IC$ with $0<\Re a<1$ and $0<\Im a<\fc{t}{2}$. The latter account for two additional pairs of (unintegrated) auxiliary open strings inserted at  \Additionala, respectively with corresponding (open string) momenta:
\eqn\newpoints{
p_0=\h\;\ell\ \ \ ,\ \ \ p_{n+1}=-\h\;\ell\ .}
Thus, we may interpret \OpenW\ as specific one--loop (planar) cylinder integrands 
\eqn\OpennW{\eqalign{
\If_{n+2}(\ell)&=g_o^n   \Bigg(\prod_{r,s=1\atop r<s}^{n} 
|\theta_1(\xi_s-\xi_r,\tau)|^{2\ap p_sp_r}\Bigg) \Bigg(\prod_{r=1\atop i=0,n+1}^n\theta_1(\xi_r-\xi_i)^{2\ap p_rp_i}\Bigg)\; Q_L(\tau,\{\xi_s\})\ ,\cr
\tilde \If_{n+2}(\ell)&=g_o^n    \Bigg(\prod_{r,s=1\atop r<s}^{n} 
 |\theta_1(\eta_s-\eta_r, \tau)|^{2\ap p_sp_r}\Bigg) \Bigg(\prod_{r=1\atop i=0,n+1}^n\theta_1(\eta_r-\eta_i)^{2\ap p_rp_i}\Bigg)\; Q_R(\bar\tau,\{\eta_s\})\ ,}}
with $n\!+\!2$ open string insertions and three marked points corresponding to the fixing \fix\ and \Additionala, cf. also Subsection 4.4. 
Altogether, that choice  corresponds\foot{It would be interesting to clarify whether 
the extra marked points \Additionala\ can be related to the auxiliary point, which allows the recursive calculation of cylinder integrals using an extra marked point in a differential equation \Broedel. Accordingly, $n$--point open string amplitudes at genus one can be obtained from $n\!+\!2$--point open--string amplitudes at tree level.} to tree--level (off--shell) scattering of $n\!+\!2$ strings involving $n$ massless open strings with momenta \OpenMom\ and two massive strings states with  appropriate momenta \newpoints, respectively, cf. also Subsection~4.5. 
Note, that in the expressions \OpennW\ all $n\!+\!2$ pairs of (open string) positions $\xi_s,\eta_s$ and $s=0,\ldots,n\!+\!1$ are put on the same footing.

Finally, we have already proposed that in  \KERNEL\ the entanglements w.r.t. the open string orderings $\si,\rho$  can be further simplified. The aforementioned  refer to certain 
iterated integrations  of the coordinates $\xi_r,\eta_r\in(0,1)$ alongside the unit interval, with  $\xi_n\!=\!\eta_n=1$. Subject to \EllisJetzer\ one--loop open string monodromy relations \refs{\HoheneggerKQY,\TourkineBAK} may be invoked to further simplify \KERNEL. In fact, in \EllmauW\ for a given loop--momentum $\ell$ one--loop open string monodromy relations  may be applied to express each open string sector \OpennW\   in terms of a basis of open string subamplitudes. Note, that this manipulation generically implicates a boundary term\foot{This boundary term has been first found, explicitly furnished and demonstrated to  be crucial to close the monodromy relations and to match the correct $\alpha'$--expansion  in Ref. \HoheneggerKQY. Subsequently, the existence  of this term has  been reconfirmed and its importance has been restated once again in \CasaliIHM, while this boundary term  (among other issues) had been overlooked in \TourkineBAK.}  entering the  one--loop open string monodromy relations \HoheneggerKQY.

\subsec{Example: One--loop four closed string amplitude}

To familiarize with our one--loop KLT formulae \Scheffau\ and \EllmauW\ here we shall discuss one example. We consider the one--loop four closed string amplitude. More specifically, we look at  the four--graviton scattering in $d\!=\!10$ type II superstring theory, with $Q\!=\!\tau_2^{-4}$. According to \fix\ after choosing the gauge choice $z_4,\bar z_4=1$, i.e. $\xi_4\!=\!\eta_4\!=\!1$ Eq. \Scheffau\ becomes\foot{Up to some normalizations we have included the overall kinematical factor $\Rc^4$ accounting for the unique scalar contraction of four powers of the Riemann tensor $\Rc$ compatible with maximal supersymmetry. Furthermore, we have Newton's constant $\kappa_{10}$ in $d\!=\!10$.}
\eqnn\scheffau{
$$\eqalignno{
\Ac^{(1)}_4(q_1,q_2,q_3,q_4)&=\h\ \delta^{(10)}\lf(q_1+q_2+q_3+q_4\ri)  \kappa_{10}^2 \Rc^4\int d\tau_2\ \int_{-\infty}^\infty d^{10}\ell\ e^{-\pi\ap \tau_2\ell^2}\cr
&\times\sum_{\sigma,\rho\in S_3}S^{(1)}[\sigma|\rho]\ \int_{\Ic_\sigma}\!\!\lf(\prod_{r=1}^{3}d\xi_r\ri)\;\int_{\Ic_\rho}\!\!\lf(\prod_{r=1}^{3}d\eta_r \ri)\lf(\prod_{t=1}^{3}\Psi(\xi_t,\eta_t;\ell)\ri)\cr
&\times e^{-i\pi\ap \ell\sum\limits_{r=1}^4q_r\xi_r}\prod_{r,s=1\atop r<s}^4 |\theta_1(\xi_s-\xi_r,\tau)|^{\fc{\ap}{2} q_sq_r}\cr
&\times e^{i\pi\ap \ell\sum\limits_{r=1}^4q_r\eta_r}\prod_{r,s=1\atop r<s}^4 |\theta_1(\eta_s-\eta_r, \tau)|^{\fc{\ap}{2} q_sq_r}\ ,&\scheffau}$$}
with the one--loop kernel \KERNEL\ and the splitting function \EllisJetzer. 



\subsec{Closed string exchange and loop momentum flow between two cylinder amplitudes}

Let us now discuss the two one--loop cylinder  integrands \OpenW\ and \OpennW\ entering the KLT formula \EllmauW\ and give an interpretation of their additional loop momentum dependent factors \HintereKesselschneid. Note, that both integrands \OpenW\ are given in the closed string channel, i.e. 
open string vertex positions are located at the boundary $\xi_r,\eta_r\in(0,1)$ of the 
cylinder $\Cc$ corresponding to the real unit segment. On the other hand, closed strings are inserted 
in the bulk $\xi\in\Cc$ or $\eta\in\Cc$ of the two cylinders, respectively.
The two expressions \HintereKesselschneid\ may be interpreted as those parts of two cylinder amplitudes which describe an (unintegrated)  single closed string  state coupled to $n$ open strings, cf.~Fig.~6.
The closed strings are inserted  at the node (cf. also \Additionala)
\eqn\Newpointsi{\eqalign{
\xi&=\fc{i\tau_2}{2}+a\equiv \xi_{n+1}\ \ \ ,\ \ \ \bar \xi=-\fc{i\tau_2}{2}+a\equiv \xi_{0}\ ,\cr
\eta&=\fc{i\tau_2}{2}+a\equiv \eta_0\ \ \ ,\ \ \ \bar \eta=-\fc{i\tau_2}{2}+a\equiv \eta_{n+1}\ ,}}
with (auxiliary) closed string vertex operators 
\eqn\Auxil{\eqalign{
U(\xi,\ov \xi)=:e^{-\fc{i}{2}\ell^\mu X_\mu(\xi)}e^{\fc{i}{2}\ell^\mu X_\mu(\ov \xi)}:\ ,\cr
V(\eta,\ov \eta)=:e^{\fc{i}{2}\ell^\mu X_\mu(\eta)}e^{-\fc{i}{2}\ell^\mu X_\mu(\ov \eta)}:\ ,}}
respectively.
Such one--loop cylinder amplitudes  have been thoroughly  studied in \StiebergerHQ\ and denoted by $A^{(1)}(n;1)$. In this setup the closed string has left-- and right--moving momenta 
\eqn\leftrightmomenta{
q_{L,R}=\mp \h\ell\ \ \ ,\ \ \ q_{L,R}=\pm \h\ell\ ,}
respectively corresponding to Dirichlet boundary conditions~\StiebergerHQ.
This interpretation entails two cylinder world--sheets each with $n$ open string insertions and one additional closed string insertion with (transverse) momentum $q^\perp=\mp\ell$, respectively.
Altogether, the two cylinder diagrams are connected by a single closed string exchange with the momentum~$q^\perp$.

Likewise, according to \StiebergerHQ\ each cylinder amplitude involving $n$ open string and one closed string may be written as pure open string amplitude involving $n+2$ open strings 
with the corresponding (open string) momenta \Additionala, cf. Subsection 4.2.


\subsec{Large complex structure limit $\tau\ra i\infty$}

In this subsection we shall derive the large complex structure limit $\tau\ra i\infty$ of our result \Scheffau. In this limit  the splitting function $\Psi$ along the $A$--cycle becomes trivial $\Psi\simeq 1$ as the two boundaries can be connected through infinity along the $B$--cycle, cf. also Fig.~7. 
Therefore,  it is convenient to rewrite the integrand \Integrands\ in the following way
\eqnn\Integrandsnew{
$$\eqalignno{
\hat I(\{\xi_t,\eta_t\};\ell)&=e^{-\pi\ap \tau_2\ell^2}\ \prod_{r=1}^{n} \lf(\fc{\theta_1(\xi_r-\xi_0)}{\theta_1(\xi_r-\xi_{n+1})}\ri)^{\fc{\ap}{2}\ell q_r}\ \lf(\fc{\theta_1(\eta_r-\eta_0)}{\theta_1(\eta_r-\eta_{n+1})}\ri)^{\fc{\ap}{2}\ell q_r}&\Integrandsnew\cr
& \times\prod_{r,s=1\atop r<s}^{n} |\theta_1(\xi_s-\xi_r,\tau)|^{\h\ap q_sq_r} 
 |\theta_1(\eta_s-\eta_r, \tau)|^{\h\ap q_sq_r}\; \Pi_q(r,s)\ ,\ \si^2_t\in C_\pm,}$$}
with the two pairs of complex points (cf. \Additionala) 
\eqn\Additional{\eqalign{
\xi_0&=-\fc{\tau}{2}\ \ \ ,\ \ \ \eta_0=+\fc{\tau}{2}\ ,\cr
\xi_{n+1}&=+\fc{\tau}{2}\ \ \ ,\ \ \  \eta_{n+1}=-\fc{\tau}{2}\ ,}}
with $\xi_0=\bar\eta_0$ and $\xi_{n+1}=\bar\eta_{n+1}$.
The particular branch cut structure of \Integrandsnew\ and  the analytic continuation \VordereKesselschneid\ also lead to \Scheffau\ (albeit now with $\Psi\simeq 1$) and are appropriate to study the  correct large complex structure limit $\tau\ra i\infty$. We shall substantiate the expression  \Integrandsnew, which leads to \Scheffau\  with $\Psi\simeq 1$ to furnish the correct large complex structure  limit $\tau\ra i\infty$.
Note, that in the expressions \Integrandsnew\ all $n\!+\!2$ pairs of (open string) positions $\xi_s,\eta_s$ and $s=0,\ldots,n\!+\!1$ are put on the same footing.
Since the $\xi_r$ and $\eta_r$ integrations along the $A$--cycle boundaries  can be performed separately with the integrands \OpenW\ or \OpennW\ we may define the following open string cylinder amplitudes
\eqn\Open{\eqalign{
A^{(1)}(\sigma(1,\ldots,n-1),n;-\ell)&=\int_{\Ic_\sigma} 
\lf(\prod_{r=1}^{n-1}d\xi_r\ri)\If_{n+2}(\ell)\ ,\cr
\tilde A^{(1)}(\rho(1,\ldots,n-1),n;\ell)&=\int_{\Ic_\rho} 
\lf(\prod_{r=1}^{n-1}d\eta_r\ri)\tilde \If_{n+2}(\ell)\ ,}}
with the domain of integration \domain\ and similarly for $\Ic_\rho$ with $n\!+\!2$ open string insertions and three marked points corresponding to the fixing \fix\ and \Additional.
With these preparations, our one--loop KLT result \EllmauW\ can be cast into the following form
\eqnn\Ellmau{
$$\eqalignno{
\qquad\Ac_n^{(1)}(q_1,\ldots,q_n)&=\h\  \delta^{(d)}\lf(\sum_{i=1}^n q_i\ri)\;\int d\tau_2\int_{-\infty}^\infty d^d\ell\ e^{-\pi\ap \tau_2\ell^2}\ &\Ellmau \cr
&\kern-3em\times \!\!
\sum_{\sigma,\rho\in S_{n-1}}\!\!\! A^{(1)}(\si(1,\ldots,n-1),n;-\ell)\ 
S^{(1)}[\sigma|\rho]\ 
\tilde A^{(1)}(\rho(1,\ldots,n-1),n;\ell)\ ,}$$}
which is the appropriate expression to analyze  the large complex structure limit.

One--loop supergravity amplitudes are reproduced from closed string amplitudes \Start\ in the large complex structure limit $\tau\ra i\infty$ where one of the two torus cycles is pinched and the torus degenerates to a nodal sphere. Specifically, the field--theory limit $\ap\ra0$ is obtained by keeping the quantity $\ap\tau_2$ finite. Likewise, the one--loop cylinder amplitudes \Open\ give rise to specific gauge amplitudes in the 
large $\tau_2$--limit and the cylinder degenerates to a disk.
In this limit  the one--loop subamplitudes \Open\ can be approximated~as
\eqn\diskamplitude{\eqalign{
A^{(1)}(\sigma(1,\ldots,n-1),n;+\ell)&\simeq g_o^n \!\!\int_{\Ic_{\sigma}} 
\!\!\!\lf(\prod_{r=1}^{n-1}\fc{d\rho_r}{2\pi i\rho_r}\!\ri)\!\!\prod_{r,s=1\atop r<s}^n 
|\rho_s-\rho_r|^{\fc{\ap}{2} q_sq_r}\! \prod_{r=1}^n (\rho_r-\rho_{n+1})^{-\fc{\ap}{2} \ell q_r}c_L,\cr
\tilde A^{(1)}(\rho(1,\ldots,n-1),n;-\ell)&\simeq g_o^n \!\!\int_{\Ic_{\rho}^t} 
\!\!\!\lf(\prod_{r=1}^{n-1}\fc{d\si_r}{2\pi i\si_r}\!\ri)\!\!\prod_{r,s=1\atop r<s}^n 
|\sigma_s-\si_r|^{\fc{\ap}{2} q_sq_r}\! \prod_{r=1}^n (\si_r-\si_{n+1})^{-\fc{\ap}{2} \ell q_r}c_R,}}
subject to the choice \fix, with $c_{L,R}:=\lim\limits_{\tau\ra i \infty} Q_{L,R}$. Note, that the latter may include (negative) integer powers of $\rho_{sr}$ and $\si_{sr}$, respectively accounting for the specific  amplitude \Start\ under consideration.
Above we have introduced the following $n\!+\!2$ disk coordinates
\eqn\diskcoords{\eqalign{
\rho_i&=e^{2\pi i \xi_i}\ ,\ i=1,\ldots,n\ ,\cr
\rho_n&=1\ \ \ ,\ \ \ \rho_{n+1}=0\ \ \ ,\ \ \ \rho_0=\infty\ ,}}
and similarly 
\eqn\diskcoordsi{\eqalign{
\si_i&=e^{-2\pi i \eta_i}\ ,\ i=1,\ldots,n\ ,\cr
\si_n&=1\ \ \ ,\ \ \ \si_{n+1}=0\ \ \ ,\ \ \ \si_0=\infty\ ,}}
used momentum conservation \conserv\  and the relation
\eqn\thetalimit{\eqalign{
\ln\theta_1(\xi_s-\xi_r,\tau)&=\ln2+\fc{\pi i\tau}{4}+\ln\sin(\pi \xi_{sr})+\Oc(q)\ ,\cr
&=-\ln i+\fc{\pi i\tau}{4}-\h \ln(\rho_s\rho_r)+\ln(\rho_s-\rho_r)+\Oc(q)\ ,}}
following from (B.3). The $n$ open strings are located along the boundary $\p \Dc$ of the unit disk $\Dc$, i.e. $|\rho_i|=1$ and $|\si_i|=1,i=1,\ldots,n$ subject to the parameterizations\foot{Alternatively, with $x_i=\cot(\pi\xi_i)$ the three unintegrated coordinates  \diskcoords\ may be identified with the points  $x_n=\infty, x_{n+1}=-i$ and $x_0=i$ on the upper half--plane $\Hc_+$, and similarly $y_i=-\cot(\pi\eta_i)\in \Hc_+$ for \diskcoordsi.} \diskcoords\ and \diskcoordsi, respectively (cf. also \Hsue).
Furthermore, in \diskcoordsi\ there is a reflection of the $n$ coordinates $\si_i$ along the boundary of the disk reverting the integration direction to $\Ic_\rho^t$. The latter can be rescinded by implying an additional factor of $(-1)^n$. Thanks to  momentum conservation \conserv\  the additional factors of $\prod_{r=1}^n(\rho_r-\rho_0)^{\fc{\ap}{2} \ell q_r}$ and $\prod_{r=1}^n(\si_r-\si_0)^{\fc{\ap}{2} \ell q_r}$ may be included in \diskamplitude, respectively. 
With this extension,  each expression  of \diskamplitude\ represents a disk amplitude involving 
$n$ open strings and one closed string located at $\rho\!=\!0$ and $\si\!=\!0$, respectively. Note, that this fact directly follows\foot{Furthermore, the factor $e^{-\pi\ap\tau_2\ell^2}$ of \Ellmau\ conspires the correct closed string propagator factors $(\rho_{n+1}-\rho_0)^{\h\ap\ell^2}(\si_{n+1}-\si_0)^{\h\ap\ell^2}$, with $q^2=(q_L-q_R)^2=\ell^2$, \cf also \StiebergerHQ.} from \HintereKesselschneid\ subject to \Newpointsi, with $\rho_{n+1},\si_{n+1}=e^{2\pi i\lf(\fc{i\tau_2}{2}\ri)}\ra0$ and $\rho_0,\si_0=e^{2\pi i\lf(\fc{-i\tau_2}{2}\ri)}\ra\infty$. In \diskcoords\ we are left with the three fixed positions $\rho_n=1,\rho_{n+1}=0,\rho_{0}=\infty$ describing a choice of $SL(2,\IR)$ frame on the   world--sheet disk and similarly in \diskcoordsi.
Alternatively, each of the two expressions  of \diskamplitude\ can be interpreted as disk amplitude involving $n+2$ open strings. This interpretation becomes clear after studying the monodromy properties of a single closed string on the disk. 

The computation of disk amplitudes involving both open and closed strings has been throughly developed  in \doubref\StiebergerHQ\StiebergerVYA\ and allows to write the  disk amplitudes \diskamplitude\  of one closed string and $n$ open strings in terms of pure open string amplitudes involving $n\!+\!2$ open strings. Thus, in the $\tau\ra i\infty$ limit  the amplitudes \diskamplitude\  boil down to linear combinations of $(n\!+\!2)$--point tree subamplitudes (A.5)
\eqn\forwardAmp{\eqalign{
\lim_{\tau\ra i\infty}A^{(1)}(\si(1,\ldots,n-1),n;+\ell)&=g_o^n\sum_{\tau\in OP(\al,\bet)} e^{\pi i \ap\varphi(\tau)}\;A^{(0)}(\tau,n)\ ,\cr
\lim_{\tau\ra i\infty}\tilde A^{(1)}(\rho(1,\ldots,n-1),n;-\ell)&=g_o^n\;(-1)^n\sum_{\upsilon\in OP(\gamma,\delta)} e^{\pi i \ap\vartheta(\upsilon)}\;\tilde A^{(0)}(\upsilon,n)\ ,}}
with coefficients $e^{i\pi \ap\varphi(\tau)}$ and $e^{i\pi \ap\vartheta(\upsilon)}$ to be specified below. Actually, after expanding w.r.t.  $\ap$ the latter start linearly in $\ap$ since their lowest order  sums up to zero \StiebergerVYA. Furthermore, the sum is over the ordered permutations $\tau\in OP(\al,\bet)$, that is, all permutations $\tau$ of $\al\cup\bet$ which maintain the order of the individual elements $\al\!=\!\{\si(1,\ldots,n\!-\!1)\},\bet\!=\!\{0,n\!+\!1\}$ belonging to each set within the joint set $\tau$. Similarly, for $\upsilon\in OP(\gamma,\delta)$ with $\gamma\!=\!\{\rho(1,\ldots,n\!-\!1)\},\delta\!=\!\{0,n\!+\!1\}$.
Concretely, discussing the monodromy properties before introducing the disk coordinates \diskcoords\ and \diskcoordsi\ we follow \StiebergerVYA\ and analytically  continue the closed string coordinates \Newpointsi\ to real values 
\eqn\Unintegrated{
\tilde \xi_0,\tilde \eta_0\in\IR\ \ \ ,\ \ \ \tilde\eta_{n+1},\tilde\eta_{n+1}\in\IR\ ,}
describing two additional open strings,  respectively.  For this we  introduce  the additional phase factors 
\eqn\PhaseEYMi{
\Pi_\ell(r)=e^{-\h\pi i \ap q_r\ell\theta(\xi_r-\tilde \xi_{n+1})}\ e^{\h\pi i\ap q_r\ell\theta(\xi_r-\tilde\xi_0)}\ e^{-\h\pi i \ap q_r\ell\theta(\eta_r-\tilde \eta_{n+1})}\ 
e^{\h\pi i\ap q_r\ell\theta(\eta_r-\tilde \eta_0)}}
\comment{ (albeit in the following  labelled by the pairs of real coordinates $\xi_0,\xi_{n+1}$ and $ \eta_0,\eta_{n+1}$)}
to impose for each pair of open string coordinates $\xi_r,\eta_r, r\in\{1,\ldots,n\}$ the correct branch cut structure on the disk. The phases  in \forwardAmp\ directly follow 
from \PhaseEYMi, i.e.
\eqn\Dictionary{\eqalign{
\varphi(\tau)&\simeq \h\sum_{r=1}^nq_r\ell\lf\{\theta(\xi_r-\tilde\xi_0)- \theta(\xi_r-\tilde\xi_{n+1}) \ri\}\ ,\ \tau\in OP(\al,\bet)\ ,\cr
\vartheta(\upsilon)&\simeq \h\sum_{r=1}^nq_r\ell\lf\{\theta(\eta_r-\tilde\eta_0)- \theta(\eta_r-\tilde\eta_{n+1}) \ri\}\ ,\ \upsilon\in OP(\gamma,\delta)\ ,}}
which  encompasses  a dictionary between the permutations $\al,\bet$ and the orderings of positions 
$\{\bigcup\limits_{r=1}^{n-1}\xi_r\},\{\tilde\xi_0,\tilde\xi_{n+1}\}$ and similarly for 
$\gamma,\delta$ with $\{\bigcup\limits_{r=1}^{n-1}\eta_r\},\{\tilde\eta_0,\tilde\eta_{n+1}\}$. Now after applying the map \diskcoords\ all the $n\!+\!2$ open string coordinates 
$\rho_r\simeq(e^{2\pi i \tilde\xi_0},e^{2\pi i \xi_i},e^{2\pi i \tilde\xi_{n+1}})$ are located along the boundary $\p \Dc$ of the disk,  where we may choose e.g. $\rho_n=i,\rho_{n+1}=1,\rho_0=-1$ and similarly for the $n\!+\!2$ coordinates $\si_i\simeq(e^{-2\pi i \tilde\eta_0},e^{-2\pi i \eta_i},e^{-2\pi i \tilde\eta_{n+1}})$.
Interestingly, we may identify  \PhaseEYMi\ with the expression
\eqn\PhaseEYM{
\Pi_\ell(r)=e^{-\h\pi i\ap q_r\ell\{1-\theta[(\xi_r-\tilde\xi_{n+1})(\eta_r-\tilde\eta_{n+1})]\}}\ e^{\h\pi i\ap q_r\ell\{1-\theta[(\xi_r-\tilde\xi_0)(\eta_r-\tilde\eta_0)]\}}\ ,} 
which assumes the same structure as the phase factor $\Pi$ introduced in \Phases. 
This observation allows us to combine both factors into the following widened phase factor
\eqn\Steingrube{ 
\prod_{r,s=1\atop r<s}^n\Pi_q(r,s)\ \times\prod_{r=1}^n\Pi_\ell(r)=\prod_{r,s=0\atop r<s}^{n+1}\Pi(r,s)\ ,}
with the additional (closed string) momenta: 
\eqn\Huettling{
q_0=\ell\ \ \ ,\ \ \  q_{n+1}=-\ell\ .}
Note, that the equivalence of \PhaseEYMi\ and \PhaseEYM\ establishes a relationship between our analytic continuation leading to \Integrands\ and the large complex structure limit $\tau\ra i\infty$ giving rise to \Unintegrated.

Eventually, after this excursion into the structure of the disk amplitudes \diskamplitude\ (involving  both open and closed strings) and their analytic continuation \forwardAmp\ we are now ready to  extract the field--theory limit of the expression  \Ellmau. Inserting the EYM amplitudes \forwardAmp\ into the second line of \Ellmau\ we may perform the following reorganization of sums
\eqnn\Durchholzen{$$\eqalignno{
&\quad\lim_{\tau\ra i\infty} \sum_{\si,\rho\in S_{n-1}}A^{(1)}(\si(1,\ldots,n-1),n;-\ell)\; 
S^{(1)}[\sigma|\rho]\; 
\tilde A^{(1)}(\rho(1,\ldots,n-1),n;\ell)&\Durchholzen\cr
&=-g_c^n\!\lf(-\fc{i}{2}\ri)^{n-1}\!\!\!\!\!\sum_{\si,\rho\in S_{n-1}}e^{i\pi \ap\Phi(\sigma,\rho)}\!\!
\sum_{\tau\in OP(\al,\bet)} e^{\pi i \ap\varphi(\tau)} A^{(0)}(\tau,n)
\!\!\sum_{\upsilon\in OP(\gamma,\delta)}e^{\pi i \ap\vartheta(\upsilon)} \tilde A^{(0)}(\upsilon,n)\cr
& \quad=:-g_c^n\!\lf(-\fc{i}{2}\ri)^{n-1}\!\!\!\!\!\sum_{\hat\si,\hat\rho\in S_{n+1}}^\prime e^{i\pi \ap\Phi^\ell(\hat\sigma,\hat\rho)}  \; 
A^{(0)}(\hat\si(0,1,\ldots,n),n+1)\; \tilde A^{(0)}(\hat\rho(0,1,\ldots,n),n+1),}$$}
with the sets $\alpha,\beta,\gamma,\delta$ defined below \forwardAmp. Note, that the relation \Durchholzen\ holds for any order in $\ap$.
Due to our choice of $\rho_{n+1}\!=\!1,\si_{n+1}\!=\!1$ and 
$\rho_n,\si_n\!=\!i$ in \Durchholzen\ the prime at the sum restricts to all 
$\h(n+1)!$ permutations $\hat\si,\hat\rho$ with leg $n$  to the right of  leg $0$ with momentum $+\ell$. In other words, those permutations  preserve the ordering of the sets $\beta,\delta\!=\!\{0,n+1\}$ in \forwardAmp. The phase factor $e^{i\pi \ap\Phi^\ell(\hat\sigma,\hat\rho)}$, which straightforwardly   derives from \Steingrube, gives rise to an $(n\!+\!1)\!\times\! (n\!+\!1)$ intersection matrix and  relates (simplifies) to the tree--level KLT kernel $S^{(0)}[\sigma|\rho]_\ell$, i.e. (subject to  open string monodromy relations)
\eqn\MapKernel{
\lf(-\fc{i}{2}\ri)^{n-1}\; e^{i\pi \ap\Phi^\ell(\hat\sigma,\hat\rho)} \ \simeq\  S^{(0)}[\sigma|\rho]_\ell\ ,}
for a pair of tree--level $(n\!+\!2)$--point subamplitudes $A^{(0)}(0,\si(1,\ldots,n-1),n,n+1)$ and 
$\tilde A^{(0)}(0,\si(1,\ldots,n-1),n+1,n)$. More concretely, we have \doubref\KawaiXQ\BjerrumBohrHN
\eqn\Vogelbad{\eqalign{
&\hskip0.25cm\Big(\fc{i}{2}\Big)^{n-1}\int_{-\infty}^\infty\!\!\! d\tilde x_1\ldots d\tilde x_{n-1}\int_{-\infty}^\infty\!\!\! d\tilde y_1\ldots d\tilde y_{n-1}\!\!\prod_{r,s=0\atop r<s}^{n+1}|\tilde x_s-\tilde x_r|^{\fc{\ap}{2} q_rq_s}|\tilde y_s-\tilde y_r|^{\fc{\ap}{2} q_rq_s}\;\Pi(r,s)\;c_Lc_R\cr
&\qquad=\lf(\fc{i}{2}\ri)^{n-1}\!\!\!\sum_{\hat\si,\hat\rho\in S_{n+1}}^\prime  e^{i\pi \ap\Phi^\ell(\hat\sigma,\hat\rho)} \
A^{(0)}(\hat\si(0,1,\ldots,n),n+1)\ \tilde A^{(0)}(\hat\rho(0,1,\ldots,n),n+1)\cr
&=(-1)^{n-1}\!\!\!\sum_{\si,\rho\in S_{n-1}}\!\!\!  A^{(0)}(0,\si(1,\ldots,n-1),n,n\!+\!1)\; S^{(0)}[\sigma|\rho]_\ell\; 
\tilde A^{(0)}(0,\rho(1,\ldots,n-1),n\!+\!1,n),}}
with a pair of $n\!+\!2$ real coordinates $\tilde x_i,\tilde y_j\in\IR$ along  two disk boundaries $\p \Hc_+$ and a gauge choice $\tilde x_0,\tilde y_0=0, \tilde x_n,\tilde y_n=1,\tilde x_{n+1},\tilde y_{n+1}=\infty$, respectively. The latter relate  to our choice $\rho_0,\si_0=-1,\rho_n,\si_n=i,\rho_{n+1},\si_{n+1}=1$ along our two disk boundaries $\p \Dc$ after the maps $\rho\!=\!\fc{x+i}{x-i},\si\!=\!\fc{y+i}{y-i}$, respectively.
\comment{ and subsequently can be mapped to the disk coordinates $\rho_r\simeq(e^{2\pi i \tilde\xi_0},e^{2\pi i \xi_i},e^{2\pi i \tilde\xi_{n+1}})$ used above.}
Furthermore, the KLT kernel $S^{(0)}[\sigma|\rho]_\ell$  is defined as follows.
For given (cyclic)  orderings $\rho,\sigma\in S_{k}$ of the $k=n-1$ coordinates $\tilde x_r,\tilde y_r$ along the real axis and a reference momentum $\ell$ one defines  the KLT kernel as the $k!\times k!$--matrix \doubref\BernSV\BjerrumBohrHN
\eqn\Bohrkern{\eqalign{
S^{(0)}[\sigma|\rho]_\ell&:=S^{(0)}[\, \si(1,\ldots,k) \, | \, \rho(1,\ldots,k) \, ]_\ell \cr
&=\prod_{t=1}^{k} \sin\lf(\pi  \ap \!\lf[\ell q_{t_\si}+\sum_{r<t}q_{r_\si}q_ {t_\si} \theta(r_\si,t_\si)\ri]\ri)\ ,}}
with $j_\si=\si(j)$ and  $\theta(r_\si,t_\si)=1$
if the ordering of the legs $r_\si,t_\si$ is the same in both orderings
$\si(1,\ldots,k)$ and $\rho(1,\ldots,k)$, and zero otherwise.
We emphasize that the second term of the argument of the sinus does not depend on  the reference momentum $\ell$.  
Thus, putting together \Durchholzen\ and \Vogelbad\ yields the ($z_j$--integrated) large complex structure limit  $\tau\ra i\infty$ of the one--loop torus amplitude \Ellmau\ including  the full--fledged dependence on $\ap$:
\eqn\Ackerlhuette{\eqalign{
\quad\lim_{\tau\ra i\infty}&\Ac_n^{(1)}(q_1,\ldots,q_n)\simeq\h\;  g_c^n\;\delta^{(d)}\Big(\sum_{i=1}^n q_i\Big)\ \lim_{\tau\ra i\infty} \sum_{\si,\rho\in S_{n-1}}A^{(1)}(\si(1,\ldots,n-1),n;-\ell)\cr 
&\qquad\times\; S^{(1)}[\sigma|\rho]\; \tilde A^{(1)}(\rho(1,\ldots,n-1),n;\ell)\ =\ 
-\h\;  g_c^n\;\delta^{(d)}\Big(\sum_{i=1}^n q_i\Big)\cr
&\!\!\!\times\!\!\sum_{\si,\rho\in S_{n-1}}\!\!\!  A^{(0)}(0,\si(1,\ldots,n-1),n,n\!+\!1)\; S^{(0)}[\sigma|\rho]_\ell\; 
\tilde A^{(0)}(0,\rho(1,\ldots,n-1),n\!+\!1,n).}}
In the following we shall also need the field--theory limit of \Bohrkern\ given by:
\eqn\Bohrkerni{
S^{(0)}_{FT}[\sigma|\rho]_\ell = \prod_{t=1}^{k} \lf(\ell q_{t_\si}+\sum_{r<t}q_{r_\si}q_ {t_\si} \theta(r_\si,t_\si)\ri)\ .}
Thereafter, as final comment we note that in the field--theory limit $\ap\ra0$ 
the pair of $(n\!+\!2)$--point subamplitudes, which  enter \Vogelbad\ and in the following we shall write as $A^{(0)}_{FT}(+\ell,\si(1,\ldots,n-1),n,-\ell)$ and $\tilde A^{(0)}_{FT}(+\ell,\rho(1,\ldots,n-1),-\ell,n)$, represent tree--level $(n\!+\!2)$--point amplitudes
in the forward limit of two off--shell legs $+\ell$ and $-\ell$. The aforementioned  constitute the decomposition of $n$--point loop amplitudes in the forward limit  (after taking into account all cyclic orderings) \doubref\GeyerBJA\GeyerJCH:
\eqn\Winkelalm{
A^{(1)}_{FT}(1,\ldots,n)=\int\fc{d^d\ell}{\ell^2} \sum_{\gamma\in cyc(1,\ldots,n)} A^{(0)}_{FT}(+\ell,\gamma(1,\ldots,n),-\ell)\ .}
In this formulation the loop momentum $\ell$ is identified with a light--like external momentum of a tree--level amplitude $A_{FT}^{(0)}$.
Finally, with $\int_0^{\infty} d\tau_2\;e^{-\pi\ap\tau_2\ell^2} \ra(\pi\ap\ell^2)^{-1}$ for $\tau\ra i\infty$, the relations \Durchholzen\ and \Vogelbad\ give rise to  the one--loop field--theory limit 
$\tau\ra i\infty$ of \Ellmau\
\eqnn\Ansatz{
$$\eqalignno{
\lim_{\tau\ra i\infty\atop \ap\ra 0}\Ac_n^{(1)}(q_1,\ldots,q_n)&\simeq-\h\; g_c^n\,\delta^{(d)}\Big(\sum_{i=1}^n q_i\Big) \int \fc{d^d\ell}{\ell^2}\!\!\!\sum_{\si,\rho\in S_{n-1}}\!\! A^{(0)}_{FT}(+\ell,\sigma(1,\ldots,n-1),n,-\ell)\cr\cr
&\times\; S^{(0)}_{FT}[\sigma|\rho]_\ell\;\tilde A^{(0)}_{FT}(+\ell,\rho(1,\ldots,n-1),-\ell,n)\ ,&\Ansatz}$$}
which has the form of a $n\!+\!2$--point tree--level KLT relation involving the field--theory KLT kernel \Bohrkerni\ with  the pivot leg $\ell$ and $k=n-1$. For the case of  $n$ gravitons the field--theory limit \Ansatz\ agrees with a conjecture made in \SongHe\ for the one--loop supergravity integrand.

\comment{Note, that  \KERNEL\ starts at $\Oc(\ap^0)$ in the inverse string tension $\ap$.
Therefore, in the $\ap\ra0$ limit in \Ellmau\ the one--loop kernel \KERNEL\  reduces to:
\eqn\Durchholzen{\eqalign{
S^{(1)}[\sigma|\rho]_\ell&=i^{n-1}\ e^{i\pi \Phi_{\sigma,\rho}}\ \prod_{t=1}^{n-1} \cos\lf(\fc{\pi \ap\ell q_t}{2}\ri)\ e^{- \pi i\ap\ell q_t\delta^t_{\sigma,\rho}}\cr
&\lra\; i^{n-1}\ \; e^{i\pi \Phi_{\sigma,\rho}}\ .}}
We have already anticipated after \KERNEL, that this phase factor is related to \Phases, i.e. 
$e^{i\pi \Phi_{\sigma,\rho}}\simeq \prod\limits_{r,s=1\atop r<s}^n\Pi(r,s)$. }
\comment{For the one--loop cylinder amplitudes which enter the integrand of \Ellmau and boil down to \diskamplitude\ in  the limit $\tau\ra i\infty$ the  manipulations described above are performed leading 
to the pair of $(n\!+\!2)$--point amplitudes $A^{(0)}(+\ell,\si(1,\ldots,n-1),n,-\ell)$ and 
$\tilde A^{(0)}(+\ell,\si(1,\ldots,n-1),-\ell,n)$ multiplied by \PhaseEYM.}

\comment{ \ifig\BasicCut{Configuration in the complex $\eta_i$  plane for $0=\xi_0<\ldots<\xi_{n}=1$}{\epsfxsize=0.7\hsize\epsfbox{Kernel.eps}}
\noindent}

\comment{From \GeyerBJA\ it is known, that we may rewrite the momentum independent factors of the integrands of \diskamplitude\  as:
\eqn\knownGeyer{
\lf(\prod_{r=1}^{n-1}\fc{d\rho_r}{\rho_r}\ri)=\lf(\prod_{r=1}^{n-1}\fc{d\rho_r}{\rho_r-\rho_{n+1}}\ri)=V_{CKG}^{-1}\sum_{\pi\in S_{n}}\fc{d\rho_1 \cdot\ldots\cdot d\rho_{n}}{\rho_{n+1,\pi(1)}\rho_{\pi(1),\pi(2)}\cdot\ldots\cdot\rho_{\pi(n-1),\pi(n)}}\ .}}

To summarize our findings, in this subsection we have considered the leading order $\Oc(q^0)$ of the large complex structure limit of \Ellmau. It gives rise to a pair of fully fledged tree--level disk amplitudes \forwardAmp\ each involving  one closed and $n$ open strings. The resulting complete monodromy structure \Steingrube\ can be arranged
into a tree--level KLT kernel \Bohrkern\ intertwining in \Vogelbad\ a pair of tree--level $(n\!+\!2)$--point amplitudes 
in the forward limit of two off--shell legs with momenta \Huettling. We expect the higher orders in $q$ of \Ellmau\ to follow a similar pattern than \Ackerlhuette. In particular, the latter arrange  
into pairs of mixed string amplitudes involving one closed and $n$ open strings dictated by the monodromy 
\PhaseEYMi, cf. also Subsection 4.4.

\newsec{Concluding remarks}

In tackling any closed string amplitude calculation it useful to first consider the corresponding open string computation. At field--theory level this means that for graviton amplitudes one should recycle results from gauge amplitudes.
 Kawai, Lewellen and Tye (KLT) have given expressions for closed string tree amplitudes as weighted sum over squares of open string tree amplitudes \KawaiXQ. In the field--theory limit, this implies that properties of gravity tree  amplitudes should be reflected in gauge theory tree amplitudes. In fact, these characteristics have   been further extended for EYM amplitudes in \StiebergerCEA.
 Therefore, it is natural to ask for a one--loop generalization thereof.

We have found the one--loop  generalization \EllmauW\ of the renowned  KLT relations relating closed  string amplitudes to a weighted sum over squares of certain open string integrands \OpennW. Through this map  (oriented) closed string one--loop amplitudes formulated on a  genus one  world--sheet torus ($g\!=\!1$) are related (subject to final complex structure modulus integration) to a pair of (oriented) open string one--loop amplitudes described by a cylinder world--sheet. In string perturbation theory the loop momentum  $\ell$  mediates between the two open string cylinder integrands \OpennW\ 
 -- both involving $n\!+\!2$ open strings.
Likewise, we have proposed a description  as two cylinder diagrams, which are connected by a single closed string  exchanging the loop momentum $\ell$, cf. Fig.~6. Such one--loop cylinder amplitudes (involving $n$ open strings and one closed string) can  be related to 
the boundary term appearing in the one--loop open string monodromy relations  \StiebergerDAA, cf. also 
the comments before Eq. \loopoff.

String theory serves as important framework to provide non--trivial  gauge/gravity relations for field theory. 
The torus amplitude \EllmauW\ involving $n$ closed strings is written as a sum over
squares of certain cylinder amplitudes connected through a loop momentum $\ell$ and each involving $n$ open strings. If the closed strings represent  
graviton states the amplitude \EllmauW\ computes the $n$--point one--loop  gravitational  amplitude  in terms of a pair of certain $n$--point one--loop gluon amplitudes connected through a loop momentum.
We can compute their field--theory limits to obtain  the
$n$ graviton one--loop amplitudes  in terms of $n$ gluon amplitudes and establish a product structure underlying the gravitational amplitudes -- independent on the amount of supersymmetries.  Hence, 
at the field--theory level  our one--loop KLT relations express the one--loop gravity amplitude in terms of   one--loop gauge amplitudes connected through a loop momentum  $\ell$. This step provides powerful relations between gravitational and gauge amplitudes at one--loop  generalizing the tree--level results \doubref\KawaiXQ\BernSV. Likewise, they provide the ground
to proof the gravity--gauge theory relations imposed from unitarity, which allow (at least in the supersymmetric case) to construct  gravity tree amplitudes out of  gluon amplitudes by applying tree--level KLT relations and  using them as input into cutting rules \BernUG. Besides, from \EllmauW\ we have derived the $n$--point one--loop  gravitational  amplitude  \Ansatz\  (including $\ap$ corrections accounting for higher derivative gravitational interactions) expressed 
as double copy of two $(n\!+\!2)$--point tree--level gluon subamplitudes in the forward limit of two off--shell legs $+\ell$ and $-\ell$.
Finally, at tree--level the  still--enigmatic double--copy construction \BernQJ\ follows from string theory by means of applying KLT relations together with subamplitude relations \doubref\StiebergerHQ\BjerrumBohrRD. On the other hand, at one--loop this construction is still conjectural. Therefore, our one--loop KLT relations play an important role for establishing and proving one--loop double--copy constructions in field--theory. 
To this end, the structure of string amplitudes has deep impact on the form and organization of quantum field theory amplitudes. Phrased differently, properties of scattering amplitudes in both gauge and gravity theories suggest a deeper understanding from string theory.

Our derivation relies on world--sheet properties and does not use any amount of space--time supersymmetries, in particular holds also in the case of no supersymmetry.
Though we have specialised our derivation to massless external string states, it is straightforward  to generalize our results to massive string states. Furthermore, we have restricted to the case of a rectangular torus, i.e. $\tau\!=\!i\tau_2$. A generalization to arbitrary  complex structure $\tau$ is possible and will be presented elsewhere.
After having accomplished the one--loop generalization of the KLT relations \EllmauW\ moving beyond one--loop  in string perturbation theory (arbitrary genus $g\!\geq\! 2$) should now conceptually be similar with a formulation involving $g$ loop momenta $\ell_I, I\!=\!1,\ldots,g$ entering 
the $g$--loop generalization of the integrand~\relevant
\eqnn\highloop{
$$\eqalignno{
I(\{z_s,\bar z_s\};\{\ell_I\})&\sim (\det\Im\Omega)^{d/2}\lf|\exp\lf\{\h i\pi\ap\ell^\mu_I\Omega^{IJ}\ell_J^\mu-\pi i \pi \ap\sum_{I=1}^g\ell_I^\mu\sum_{r=1}^{n}
q^\mu_r\int_P^{z_r}\omega_I\ri\}\ri|^2 \cr
&\times\prod_{1\leq r<s\leq n}E(z_s,z_r)^{\h\ap q_sq_r}\ 
\bar E(\bar z_s,\bar z_r)^{\h\ap q_sq_r}\ ,&\highloop}$$ }
with the genus $g$ prime form $E$ and period matrix~$\Omega$. Again, the $n$ closed strings can holomorphically be split into $n$ pairs of open strings located at boundary curves with their 
 branch cut structure triggered by the one--loop intersection matrix $S^{(1)}[\sigma|\rho]$ 
 introduced in~\KERNEL.

The latter is the one--loop extension of the tree--level KLT kernel \Bohrkern\ defined in
\threeref\KawaiXQ\BernSV\BjerrumBohrHN. Actually, both objects are described by the  same phase structures, cf.~\Phases. In the $\tau\ra i \infty$ limit, i.e. selecting a particular analytic continuation, in \Durchholzen\ the one--loop kernel $S^{(1)}[\sigma|\rho]$ can be reorganized  to match  in \MapKernel\ an  extended tree--level kernel $S^{(0)}[\sigma|\rho]_\ell$ containing also the loop momentum $\ell$.
At tree--level the KLT kernel is related to topological properties of twisted
homology encoded in intersection numbers of twisted cycles \MizeraCQS.
The latter are associated to domain integrations of multi--valued functions. 
The formalism of twisted de Rham cohomology has recently been discussed for one--loop open string amplitudes  \CasaliIHM\ (to revisit and confirm the results  \HoheneggerKQY\ from a different perspective).
Building up on this approach the kernels \KERNEL\ and \MapKernel\ should  also play an important r\^ole in intersection theory and  generalized twisted cohomology theory on the elliptic curve describing one--loop string amplitudes. In this language the KLT factorization of the closed string one--loop amplitude \EllmauW\ should be expressible  as bilinear form of the elements of the underlying basis of twisted cycles
whose intersection numbers are described by  \KERNEL\ and \MapKernel\ .

At genus zero ($g\!=\!0$) there exists an intriguing map (called single--valued map \BrownGIA), which maps full ($z_j$--integrated) tree--level closed string amplitudes to the single--valued projection of  open string disk amplitudes. This map, which  acts on period integrals over the moduli space of Riemann spheres of $n\!-\!3$ marked points has been observed, conjectured  and established in \doubref\StiebergerWEA\StiebergerHBA\  and then later rigorously been proven in \BrownOMK.
Note, that this map acts at the full ($z_j$--integrated) amplitude without the necessity to perform an expansion in $\ap$. With our one--loop KLT relations it could be possible to define such a map at genus one at the level of the fully--fledged   (i.e. $z_j$--integrated) one--loop torus amplitude including the (non--analytic) effects from the massless states or branch cuts of $\ln(s_{ij})$ terms. Note, that the latter are generically omitted when focusing only at a power  series expansion in $\ap$ yielding (after $z_j$--integrations) as coefficients modular graph functions (expanded around the cusp $\tau\ra i\infty$) \BrownGraph. In contrast, in this work we have allowed for the full analytic structure of the amplitude.

\medskip
\noindent
{\it Acknowledgments:} I wish to thank Johannes Broedel for 
interesting and useful discussions over the past several years. Moreover, an insightful, valuable and helpful long discussion with Peter~Mayr is greatly acknowledged.
\comment{Furthermore, I wish to thank The Kolleg Mathematik Physik Berlin of
Humboldt Universit\"at for warm hospitality and financial support.}


\appendix\appA{Open string monodromy relations on  doubled surfaces}

In this section we shall discuss some extension of  open string monodromy relations. 
The latter are formulated on surfaces $\ov\Sigma$ with boundaries.
Surfaces with boundaries are obtained from their doubles $\Sigma$ by involution $\sigma$ \BM.
In the following we shall assume that the manifold $\Sigma$ can be constructed by gluing together two copies $\Si_+,\Si_-$ of the bordered surface $\ov\Sigma$ along its boundaries $\p\bar\Si$. 
At string tree--level open string  monodromy relations are derived from considering closed cycles on the
disk world--sheet $\bar\Sigma=\Hc_+$, which can be represented as upper half plane $\Hc_+$ \doubref\StiebergerHQ\BjerrumBohrRD. Likewise, at one--loop  open string  monodromy relations are derived from studying  closed cycles on the cylinder world--sheet $\ov\Sigma=\Cc$ \refs{\HoheneggerKQY,\TourkineBAK}.

We want to extend the monodromy discussion on the bordered surface $\ov\Si$ to the doubled surface $\Sigma$. 
On $\ov\Si$ the $n$ open string coordinates $\si_i,\; i\!=\!1,\ldots,n$ are parameterized along its boundaries. For finding the open string monodromy relations on the manifold $\ov\Sigma$ one  selects one coordinate $\si_1$ and considers a closed cycle $\gamma$ along the boundary component $\p\ov\Si$ of the surfaces $\ov\Sigma$.  With the local system $ I(\{\si_i\})$ w.r.t. to the coordinate $\si_1$ one considers the following contour integral on $\ov\Sigma$:
\eqn\othercycle{
\oint_{\gamma}d\si^1\  \hat I(\{\si_i\})=0\ .}
Again, $\hat I$ is single--valued and takes into account the correct branch cut structure of the holomorphic function  $I$ with branch cuts. This is achieved by introducing proper monodromy phases, which render the correct branches in the complex $\si^1$--plane.
After deforming  the cycle $\gamma$  to  a  boundary component of $\ov\Sigma$ there are only contributions from this boundary cycles, which  give rise to the usual open string monodromy relations.
Our task is to amplify the monodromy discussion on the bordered surfaces $\ov\Si$ to the doubled surface $\Sigma$ and extend \othercycle\ by a global contour $\Gamma$ on this doubled surface enclosing the gluing regions.
Then, we should also expect potential residua contributions from these gluing regions. Thus \othercycle\ generically amounts to the following relation  on the doubled surface $\Si$:
\eqn\Othercycle{
\oint_{\Gamma}d\si^1\  \hat I(\{\Sigma_i\})=-\oint_{\Rc}d\si^1\  \hat I(\{\Sigma_i\})\ .}

\subsec{Open string tree--level monodromy relations on the sphere}

Let us start with the open string tree--level monodromy relations \doubref\StiebergerHQ\BjerrumBohrRD:
\eqn\treeMono{
 A^{(0)}(1,2,\ldots,n)+ e^{\pi i \ap s_{12}}\ A^{(0)}(2,1,\ldots,n)+\ldots+e^{\pi i \ap  \sum\limits_{j=2}^{n-1}s_{1j}}\ A^{(0)}(2,3,\ldots,n-1,1,n)=0,}
with the kinematic invariants 
\eqn\Mandel{
s_{ij}=(p_i+p_j)^2=2 p_ip_j\ ,\ i,j=1,\ldots,n\ ,}
 for $n$ massless open string momenta $p_i$ and the open string subamplitudes
\eqn\openTreeAmp{
A^{(0)}(\sigma(1,\ldots,n))=V_{CKG}^{-1}(\Dc)\ \int_{\Ic_\si}\!\!\!\lf(\prod_{r=1}^{n}d\xi_r \ri)\prod_{r<s}|\xi_s-\xi_r|^{2\ap p_rp_s}\ ,}
subject to the ordering $\si$ of open string vertex operators along the boundary $\p\Hc_+$ of the disk
$\Dc\simeq \Hc_+$.
For the ordering $\si$ we have the  domain of integration:
\eqn\Domain{
\Ic_\si=\{(\xi_1,\ldots,\xi_{n})\in \IR^{n}\ |\ -\infty\leq \xi_{\sigma(1)}<\ldots<\xi_{\si(n)}<\infty\}\ .}
Furthermore, in \openTreeAmp\ there is the   volume $V_{CKG}(\Dc) $ of the conformal Killing group on the disk, which  can be cancelled  by choosing   $\xi_1=0,\xi_{n-1}=1$ and $\xi_n=\infty$.
The identities \treeMono\ can be derived by considering \othercycle\ for the contour $\gamma\!=\!C\cup C_+$ in the upper half--plane $\Si_+=\Hc_+$ and deforming the half--cycle $C$ to infinity, cf. 
Fig.~8.
\ifig\SigmaWorldsheet{Monodromy cycles on the complex sphere.}{\epsfxsize=0.5\hsize\epsfbox{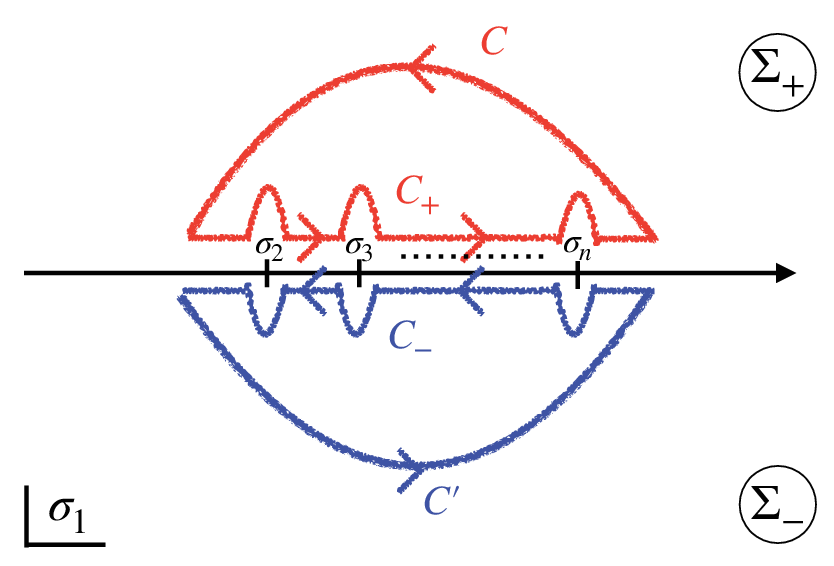}}
\noindent
The counterclockwise (positive) arcs generate the correct branch cut structure.
Alternatively, we may also consider \othercycle\ for the closed cycle $\gamma'=C'\cup C_-$ in the lower half--plane $\Si_-=\Hc_-$ leading to:
\eqn\treeMonoi{
 -A^{(0)}(1,2,\ldots,n)- e^{-\pi i \ap  s_{12}}\ A^{(0)}(2,1,\ldots,n)-\ldots-e^{-\pi i \ap  \sum\limits_{j=2}^{n-1}s_{1j}}\ A^{(0)}(2,3,\ldots,n-1,1,n)=0.}
So far we have only considered closed contours $\gamma,\gamma'$ on the two separate  surfaces $\Si_+,\Si_-$ with their boundaries leading to the two sets of equations \treeMono\ and
\treeMonoi, respectively. 

Let us now consider a closed cycle $\Gamma$ on the doubled surface $\Sigma=
\Si_+\cup \Si_-$. The latter manifold is represented by a complex sphere on which we choose the (closed)  cycle $\Gamma=C\cup C'$  to fulfill    \Othercycle. Again, $C$ and $C'$ can be deformed to infinity. However, there could be some potential residua contributions from encircling the vertex positions $\si_i$ along the real axis. They can be collected by considering the path along $\Rc=C_+\cup C_-$  with closed cycles encircling the points $\si_i$. Thus,  \Othercycle\ yields:
\eqn\sinuscontour{\eqalign{
 \oint_{\Gamma} d\si_1\ \hat I(\{\si_i\})&=2i\; \sin(\pi \ap  s_{12})\ A^{(0)}(2,1,\ldots,n)+\ldots+\cr
 &+2i\;\sin\lf(\pi\ap  \sum\limits_{j=2}^{n-1}s_{1j}\ri)\ A^{(0)}(2,3,\ldots,n-1,1,n)=0\ .}}
The expression \sinuscontour\ is  just the sum of the two relations \treeMono\ and \treeMonoi, which adds up to zero. Hence, actually in this case there is no contribution from the residuum $\Rc$.

\subsec{Tree--level KLT relation and a double of contours}

The tree--level KLT relation expresses a multi--dimensional complex integral over the sphere
in terms of pairs of real integrals subject to some monodromy phases \KawaiXQ. The simplest case
describes four closed string scattering and is given in terms of the following single complex integral
\eqn\Ritzaualm{
\Ic=\int_{\IC}d^2z\  z^{\ap s+n_{12}}\ (1-z)^{\ap u+n_{23}}\ \bar z^{\ap s+\bar n_{12}}\ (1-\bar z)^{\ap u+\bar n_{23}}\ ,}
with some integers $n_{12},n_{23},\bar n_{12},\bar n_{23}\in\IZ$ and real numbers $s,u\in\IR$ (defined by 
$s=2p_1p_2,u=2p_1p_4$).
For the method proposed in \KawaiXQ\ one first writes $z=x+iy$. Then, the integrand in \Ritzaualm\ represents an analytic function in $y\in\IC$ with four branch points $\pm ix,\pm i(1-x)$ located along the imaginary axis $\Re(y)=0$. For the case $0<x<1$ the branch cut structure in the complex $y$--plane is depicted in Fig.~9. 
\iifig\SigmaWorldsheet{Deforming the real $y$ integration  contours $C,C'$}{to contours $C_+,C_-$ along the imaginary axis, respectively.}{\epsfxsize=0.5\hsize\epsfbox{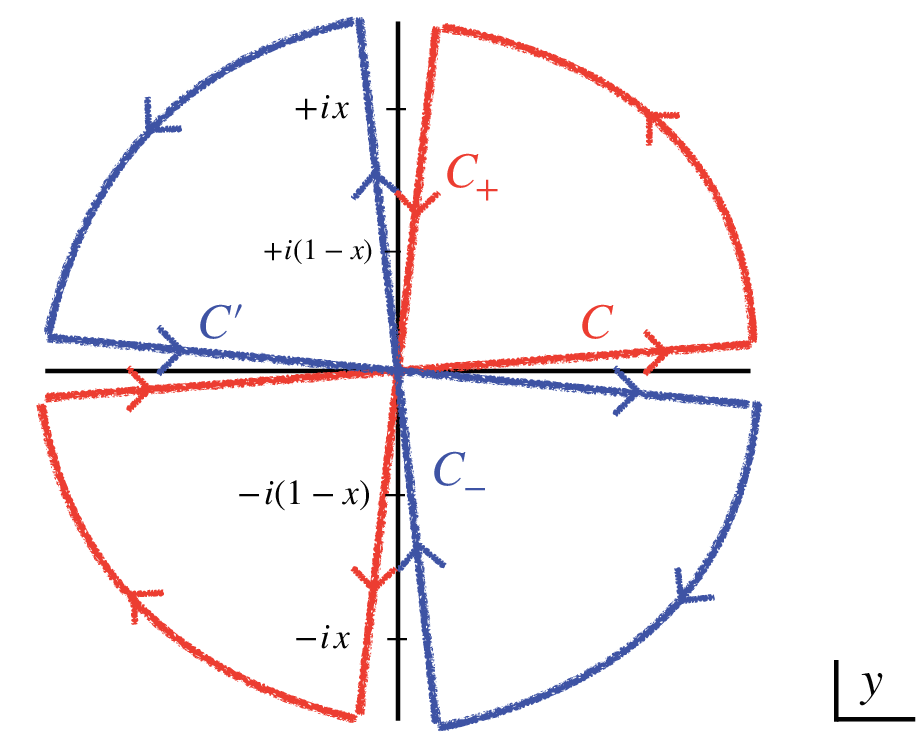}}
\noindent In the next step the original $y$ integration  along the real axis $C$ is deformed to  
the contour $C_+$ along the pure imaginary axis (depicted in red in Fig.~9) parameterized by $y=i\tilde y$ with $\tilde y\in (-\infty,\infty)$. After introducing the new coordinates
\eqn\Hinterkaiserhof{\eqalign{
\xi&=x+iy=x-\tilde y\ ,\cr
\eta&=x-iy=x+\tilde y}}
the integral  \Ritzaualm\ becomes \KawaiXQ
\eqnn\Vorderkaiserfeldenhutte{
$$\eqalignno{
\Ic_+&=\int_{-\infty}^\infty dx\int_{C_+} \hat I&\Vorderkaiserfeldenhutte\cr
&=\fc{i}{2}\ \int_{-\infty}^\infty d\xi \int_{-\infty}^\infty d\eta \ |\xi|^{\ap s+n_{12}}\ |1-\xi|^{\ap u+n_{23}}\ |\eta|^{\ap s+\bar n_{12}}\ |1-\eta|^{\ap u+\bar n_{23}}\ \Pi(\xi,\eta;s,u)\ ,}$$}
with the single--valued integrand $\hat I$ and the phase factor $\Pi$
\eqn\KLTphase{
\Pi(\xi,\eta;s,u)=e^{\pi i \ap s\{1-\theta(\xi\eta)\}}\ e^{\pi i \ap  u\{1-\theta[(1-\xi)(1-\eta)]\}}\ ,}
 rendering the correct branch of the integrand in the first and third quadrant.
 This phase structure can be accommodated by  appropriately passing the branch points along $C_+$. More, precisely  the latter are passed to the right along $C_+$. 
To illustrate the situation we briefly discuss the case $0<\xi<1$. Then, the phase \KLTphase\ becomes $\Pi=e^{\pi i\ap s}$ for $\eta<0$, $\Pi=1$ for $0<\eta<1$ and
$\Pi=e^{\pi i\ap u}$ for   $\eta>0$, respectively.
This phase structure is furnished  by the three contours $C_+$   displayed  (in red) in Fig.~10.
\iifig\BeispielVerstanden{Double of contours $C_+$ and $C_-$ for the integration $\eta$}
{for the region $0<\xi<1$ with relevant phases.}{\epsfxsize=0.75\hsize\epsfbox{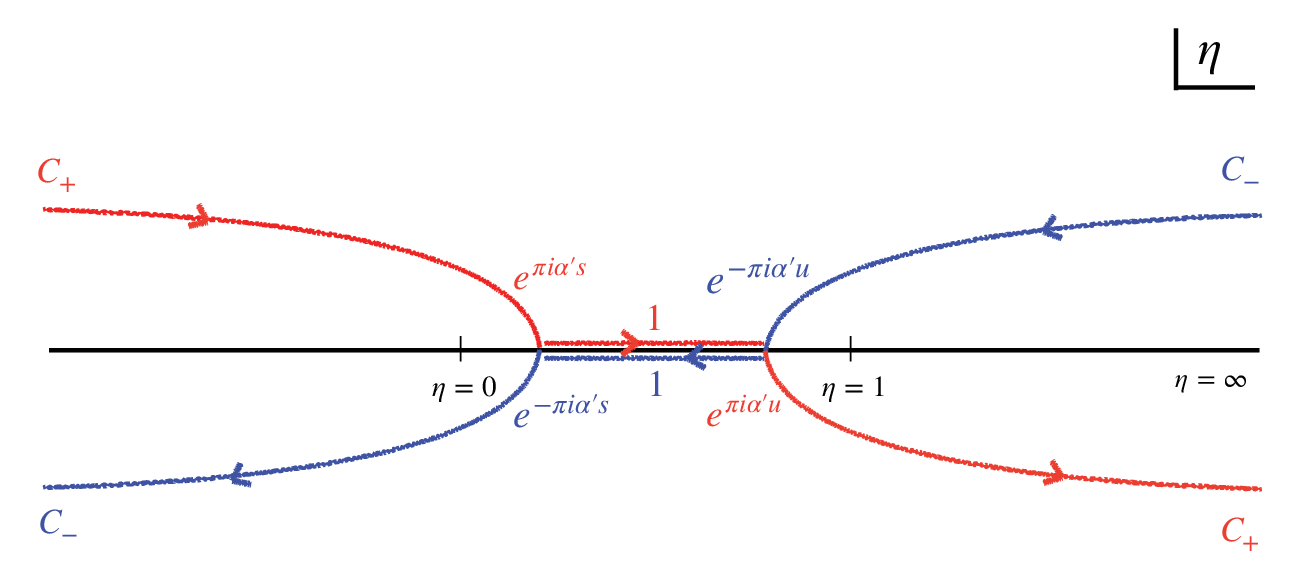}}
\noindent
In \Vorderkaiserfeldenhutte\ for $0<\xi<1$ the aforementioned regions contribute  the following three terms:
\eqn\termi{
\Ic_+\simeq\fc{i}{2}\ \int_0^1d\xi\lf\{e^{\pi i\ap s}\ \int_{-\infty}^0d\eta\ \ldots+\int_0^1d\eta\ \ldots+e^{\pi i\ap u}\ \int^{\infty}_0d\eta\ \ldots\ \ri\}\ .}
Explicitly computing $\Ic_+$ (by means of counter deformations) yields the KLT result $\Ic$, i.e. $\Ic_+=\Ic$. 

Alternatively, instead of the pair of contours $C$ and $C_+$ (depicted in red in Fig. 9) we may also consider the  pair of contours $C'$ and $C_-$ (depicted in blue in Fig. 9) leading to the integral:
\eqnn\Hinterkaiserfeldenalm{
$$\eqalignno{
\Ic_-&=\int_{-\infty}^\infty dx\int_{C_-} \hat I&\Hinterkaiserfeldenalm\cr
&=-\fc{i}{2}\ \int_{-\infty}^\infty\!\!\! d\xi \int_{-\infty}^\infty\!\!\! d\eta\  |\xi|^{\ap s+n_{12}}\ |1-\xi|^{\ap u+n_{23}}\ |\eta|^{\ap s+\bar n_{12}}\ |1-\eta|^{\ap u+\bar n_{23}}\ \Pi(\xi,\eta;s,u)\ .}$$}
Note, that in \Hinterkaiserfeldenalm\ the overall minus sign\foot{We emphasize, that in the complex $y$--plane the contribution of the contour $C_-$ does not cancel that of $C_+$, because  effectively the contour $C_-$ may be deformed and aligned to  the contour $C_+$ through infinity without crossing any branch points.} stems from the opposite direction of $C_-$ compared to $C_+$. Now the phase structure \KLTphase\ in the second and fourth quadrant is furnished 
by passing the branch points to the left and moving opposite to $C_+$.
Again, for the case $0\!<\xi\!<1$ the corresponding contours $C_-$ are drawn   (in blue) in Fig.~10.
The latter contribute to  \Hinterkaiserfeldenalm\ the following three terms:
\eqn\termii{
\Ic_-\simeq-\fc{i}{2}\ \int_0^1d\xi\lf\{e^{-\pi i\ap s}\ \int_{-\infty}^0d\eta\ \ldots+\int_0^1d\eta\  \ldots+e^{-\pi i\ap u}\ \int^{\infty}_0d\eta\ \ldots\ \ri\}\ .}
Eventually, after choosing the integration paths along $C_+$ and $C_-$ such that the phases \KLTphase\ are accommodated the two expressions $\Ic_+$ and $\Ic_-$ can be related as:
\eqn\Pfandlhof{
\Ic_-=\Ic_+^\ast\ .}
In \Pfandlhof\  the complex conjugation originates from the opposite direction of integration rendering the correct phases \KLTphase\ for each interval (between a pair of branch points) along $C_-$ compared to $C_+$. 
To summarize,  there are two different  contours $C_+$ and $C_-$ for deforming the original real integrations $C$ and $C'$ along $\Im(y)=0$, respectively i.e. two ways for analytic continuing  the complex  closed string coordinate $z\in\IC$. At any rate,  if we integrate along both $C_+$ and $C_-$ with $\Rc=C_+\cup C_-$ we obtain\foot{Note, that the sum of \termi\ and \termii\ gives the full (doubled) KLT result:
$\Ic_++\Ic_-=-\tilde A(1,2,3,4)$
$\times\lf\{\sin(\pi \ap s)\ A(2,1,3,4)+\sin(\pi\ap u)\ A(1,3,2,4)\ri\}=2\Ic\ .$} the real and single--valued expression, i.e.:
\eqn\Einserkogel{ 
\oint_\Rc \hat I=\Ic_++\Ic_-=\Ic_++\Ic_+^\ast=2\; \Ic\ .}

A double of contours exists also in  the multi--dimensional case for each complex coordinate
$z_t,\; t\!=\!1,\ldots,n$, i.e. now we are having $n$  pairs of contours $C_\pm$. After writing $z_t=x_t+iy_t$ we consider  contours in the complex $y_t$--planes each with  $n\!-\!1$  pairs of branch points  $z_t=z_l$ and $\bar z_t=\bar z_l$ located along the imaginary axes  $\Re(y_t)=0$.  Similarly  as in the complex one--dimensional case discussed above and depicted in Fig.~9 each coordinate $y_t$ may be integrated along either one of the cycles  $C_+$ and $C_-$ with opposite rotation directions for $C_+$ and $C_-$ due to their different orientations. Along the latter the real coordinates $\xi_t=x_t+iy_t,\eta_t=x_t-iy_t$ are introduced. The branch cut structure is specified by the hyperplanes $\xi_t-\xi_l=0$ and $\eta_t-\eta_l=0$ and can be described by the  phase factor \Phases
\eqn\KLTPhase{
\Pi(\xi_l,\xi_t,\eta_l,\eta_t;p_tp_l)=e^{\pi i \ap p_tp_l\{1-\theta[(\xi_l-\xi_t)(\eta_l-\eta_t)]\}}\ ,}
when moving along the contours $C_\pm$.
For a given ordering of the real points $\xi_s,\eta_s$ these phases are constants. As a consequence
the integrations along $C_\pm$ can be divided into pieces supplemented by the phases \KLTPhase.
The latter can be furnished by appropriately integrating along the contours $C_\pm$ and passing the branch points at $z_t=z_l, \ov z_t=\ov z_l$, i.e. at $\xi_t=\xi_l, \eta_t=\eta_l$.
If we integrate all $n$ coordinates along $C_+$ we obtain the original KLT result $\Ic_+= \Ic$. On the other hand, integrating all  $n$ coordinates along $C_-$ gives $\Ic_-=\Ic_+^\ast$.
For each coordinate $y_i$ both contours $C_\pm$ are closed at infinity. Hence, they can be combined  into the double of contours $\Rc=C_+\cup C_-$ and we may consider the $n$--fold combination 
\eqn\Elferkogel{ 
\oint_{\Rc}\ldots \oint_{\Rc}\hat I=2^{n-4}\ \lf(\Ic_++\Ic_-\ri)=2^{n-4}\ \lf(\Ic_++\Ic_+^\ast\ri)
=2^{n-3}\ \Ic\ ,}
with the single--valued integrand
\eqn\Veroni{
\hat I=V_{CKG}^{-1}(\Sc)\ \prod_{r<s}|\xi_s-\xi_r|^{\ap p_rp_s}\;|\eta_s-\eta_r|^{\ap p_rp_s}\ \Pi(\xi_r,\xi_s,\eta_r,\eta_r;p_rp_s)\ ,}
leading to a single--valued and real expression $\Ic$. In \Elferkogel\ the $n$--fold integration paths along $C_+$ amount to $\Ic_+$, while those along $C_-$ give rise to $\Ic_-$.
Finally, there is the   volume $V_{CKG}(\Sc) $ of the conformal Killing group on the sphere, which 
can be cancelled  by choosing   $\xi_1,\eta_1=0,\xi_{n-1},\eta_{n-1}=1$ and $\xi_n,\eta_n=\infty$.

In the following let us describe how to disentangle the $n$--fold contour integral \Elferkogel.
For given orderings $\sigma,\rho$ of the $2n$ real coordinates $\xi_r,\eta_r\in (-\infty,\infty)$ along the real line  the product of phase factors \KLTPhase\ entering \Veroni\ reduces  to a constant $e^{i\pi \ap\Phi_{\sigma,\rho}}$ depending only on the external momenta $p_s$.  Concretely, for a given ordering $\si$ of the $n$ coordinates $\xi_s$ the total phase $\Phi_{\sigma,\rho}$ is furnished by appropriately choosing for each of the $n\!-\!3$ (unfixed) coordinates $\eta_r$ arcs around the $n$ points $\eta_s$ with proper phases $e^{i\pi\ap\phi_{\sigma,\rho}(r)}$  such that $\Phi_{\sigma,\rho}=\sum_{r=1}^{n-3}\phi_{\sigma,\rho}(r)$ and connect the $n\!-\!3$ coordinates $\eta_r$ by performing iterated integrations subject to the orderings $\sigma,\rho$ and orientations of $C_\pm$.
In this setup the difference between integrating along $C_+$ or $C_-$ results in opposite angles  
$\phi_{\sigma,\rho}(r)$ and integration directions between a pair of points $\eta_r$.
 As a result for \Elferkogel\ we can write
\eqn\Verstanden{\eqalign{
\oint_{\Rc}\ldots \oint_{\Rc}\hat I&=\lf(\fc{i}{2}\ri)^{n-3}\sum_{\si,\rho\in S_{n-1}}^\prime  \prod_{t=1}^{n-3}\lf(e^{i\pi\ap\phi_{\sigma,\rho}(t)}-e^{-i\pi\ap\phi_{\sigma,\rho}(t)}\ri)\cr
&\times
  \tilde A^{(0)}(\sigma(1,\ldots,n-1),n)\  A^{(0)}(\rho(1,\ldots,n-1),n)=2^{n-3}\ \Ic\ ,}}
with the open string (tree--level) subamplitudes $\tilde A^{(0)},A^{(0)}$ defined in \openTreeAmp\ and 
accounting for the iterated integrations between the points $\xi_s,\eta_s$, respectively. Due to our choice of $\xi_1,\eta_1=0$ and 
$\xi_{n-1},\eta_{n-1}=1$ in \Verstanden\ the prime at the sum restricts to all 
$\h(n-1)!$ permutations $\si,\rho$ with $n-1$ to the right of~$1$. To conclude, with \Verstanden\ 
we have an expression of the tree--level KLT result $\Ic$ as multi--dimensional integral over 
doubles of contours $C_\pm$. 

E.g. in the case of $n\!=\!6$ for $\sigma\!=\!id$, i.e. $\xi_1<\ldots<\xi_6$   the 
ordering $\rho$ of points $\eta_i$ with $\eta_2<\eta_1<\eta_5<\eta_4<\eta_3<\eta_6$  in \Veroni\ gives rise to the following total phase factor $e^{i\pi \ap\Phi_{\si,\rho}}=e^{\pi i\ap s_{21}}
e^{\pi i\ap(s_{34}+s_{35})}e^{\pi i\ap s_{45}}$. Along $C_+$ the latter is realized if the coordinates $\eta_2,\eta_3$ and $\eta_4$ are integrated along arcs around the two points $\eta_1,\eta_5$ such that the three angles $\phi_{\si,\rho}(2)=s_{21}, \phi_{\si,\rho}(3)=s_{34}+s_{35}$ and $\phi_{\si,\rho}(4)=s_{45}$ are generated, respectively. Their paths $C_+$ are  depicted in red in Fig.~11.
\iifig\BeispielVerstanden{Double of contours $\Rc$ for the integrations $\eta_2,\eta_3,\eta_4$}
{for the region $0<\xi_2<\xi_3<\xi_4<1$ with relevant phases.}{\epsfxsize=0.75\hsize\epsfbox{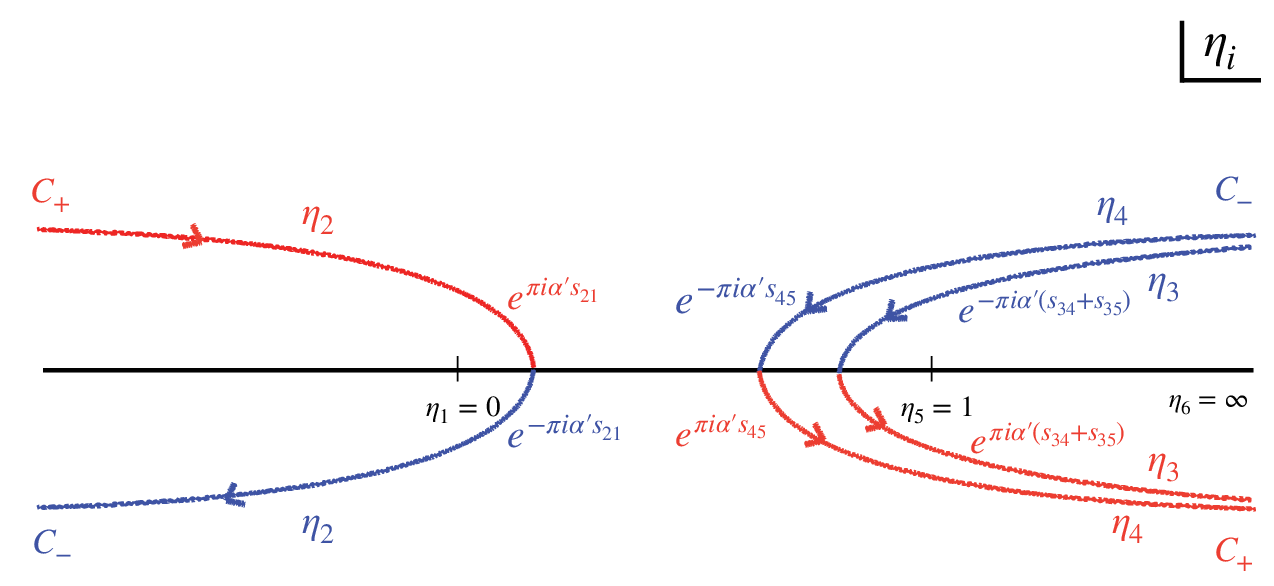}}
\noindent 
 Together with the other contours $C_-$ in \Verstanden\ they give rise to the following phases
$$\eqalign{\prod_{t=2}^{4}&\lf(e^{i\pi\ap\phi_{\sigma,\rho}(t)}-e^{-i\pi\ap\phi_{\sigma,\rho}(t)}\ri)\cr
&=\lf(e^{\pi i\ap s_{21}}-e^{-\pi i \ap s_{21}}\ri)\; \lf(e^{\pi i\ap(s_{34}+s_{35})}-e^{-\pi i \ap(s_{34}+s_{35})}\ri)\; \lf(e^{\pi i\ap s_{45}}-e^{-\pi i \ap s_{45}}\ri)\ ,}$$
which contribute the following term 
$$2^3\ \sin(\pi\ap s_{12})\;\sin(\pi \ap(s_{34}+s_{35}))\;\sin(\pi \ap s_{45})\ 
\tilde A^{(0)}(1,2,3,4,5,6)\; A^{(0)}(2,1,5,4,3,6)$$
to the full KLT result $2^3\Ic$.

\subsec{Open string one--loop monodromy relations on the torus}

We have concluded in \sinuscontour\ that
applying \Othercycle\ to the tree--level world--sheet disk does not give rise to a residuum contribution. On the other hand, as we shall demonstrate here this is not the case at one--loop.
The planar one--loop monodromy relations on the cylinder world--sheet $\Si_+=\Cc$ may be written in terms of the loop momentum $\ell$ \TourkineBAK
\eqnn\OneMono{
$$\eqalignno{
 A^{(1)}(1,2,\ldots,n)&+ e^{\pi i \ap s_{12}}\ A^{(1)}(2,1,\ldots,n)+\ldots+&\OneMono\cr
 &+e^{\pi i \ap \sum\limits_{j=2}^{n-1}s_{1j}}\ A^{(1)}(2,3,\ldots,n-1,1,n)
 =A^{(1)}(1|2,\ldots,n)[\vartheta_{p_1}]\ ,}$$}
with the planar one--loop open string subamplitudes
\eqn\subSH{\eqalign{
A^{(1)}(\rho(1,\ldots,n))&=2^{-d/2}\;\delta(p_1+\ldots p_n) \int_0^\infty\fc{dt}{t}\; V_{\rm CKG}^{-1}(\Cc)\cr
&\times\int_{-\infty}^\infty d^d\ell\int_{\Jc_\rho} \lf(\prod_{r=1}^n d\sigma^2_r\ri)\ I^{(1)}_P(\{\si^2_i\};\ell)\ ,}}
with $\si^2_i\in (0,t)$, the  domain of (iterated) integration
\eqn\Domain{
\Jc_\rho=\{(\si^2_1,\ldots,\si^2_{n})\in \IR^{n}\ |\ 0\leq \si^2_{\rho(1)}<\ldots<\si^2_{\rho(n)}< t\}\ ,}
and the planar one--loop integrand:
\eqn\tv{
I^{(1)}_P(\{\si^2_i\};\ell)= 
e^{-\h\pi \ap t \ell^2+2\pi  \ap\ell\sum\limits_{r=1}^n p_r \si^2_r}\ \prod_{r,s=1\atop r<s}^n\lf|\fc{\theta_1(i\si^2_{sr},\tau)}{\theta_1'(0,\tau)}\ri|^{2\ap p_rp_s}\ .} 
In addition, we have the non--planar subamplitude
\eqn\subSHT{\eqalign{
A^{(1)}(1|2,\ldots,n)[\vartheta_{p_1}]&=2^{-d/2}\;\delta(p_1+\ldots p_n) \int_0^\infty\fc{dt}{t}\; V_{\rm CKG}^{-1}(\Cc)\cr 
&\times\int_{-\infty}^\infty d^d\ell\ \vartheta_{p_1}\int_{\Jc} \lf(\prod_{r=1}^n d\sigma^2_r\ri) I^{(1)}_{NP}(\{\si^2_i\};\ell)\ ,}}
with the phase
\eqn\TVphase{
\vartheta_{p_1}:=e^{\pi i\ap \ell p_1}\ ,}
and the integration $\Jc$ region parameterized by $\si^2_1\in(0,t)$ and $\si^2_2<\ldots<\si^2_n$.
Furthermore, we have the non--planar one--loop integrand:
\eqnn\TV{
$$\eqalignno{
I^{(1)}_{NP}(\{\si^2_i\};\ell)
&=e^{-\h\pi \ap t \ell^2+2\pi  \ap\ell\sum\limits_{r=1}^n p_r \si^2_r}\cr
&\times\prod_{r,s=2\atop r<s}^n\lf(\fc{\theta_1(i\si^2_{sr},\tau)}{\theta_1'(0,\tau)}\ri)^{2\ap p_rp_s} \prod_{r=2}^n\lf(\fc{\theta_2(i\si^2_{r1},\tau)}{\theta_1'(0,\tau)}\ri)^{2\ap p_rp_1}.\quad\qquad&\TV}$$} 
The relation \OneMono\ is derived by considering  the closed contour $\gamma=C\cup C_+$ in the complex $\si^2_1$--plane, depicted in red in Fig.~12  and evaluate the contour integral \othercycle.
\iifig\SigmaWorldsheet{Monodromy cycles and branch cut structure}{on the doubled cylinder $\Si=\Si_-\cup\Si_+$.}{\epsfxsize=0.6\hsize\epsfbox{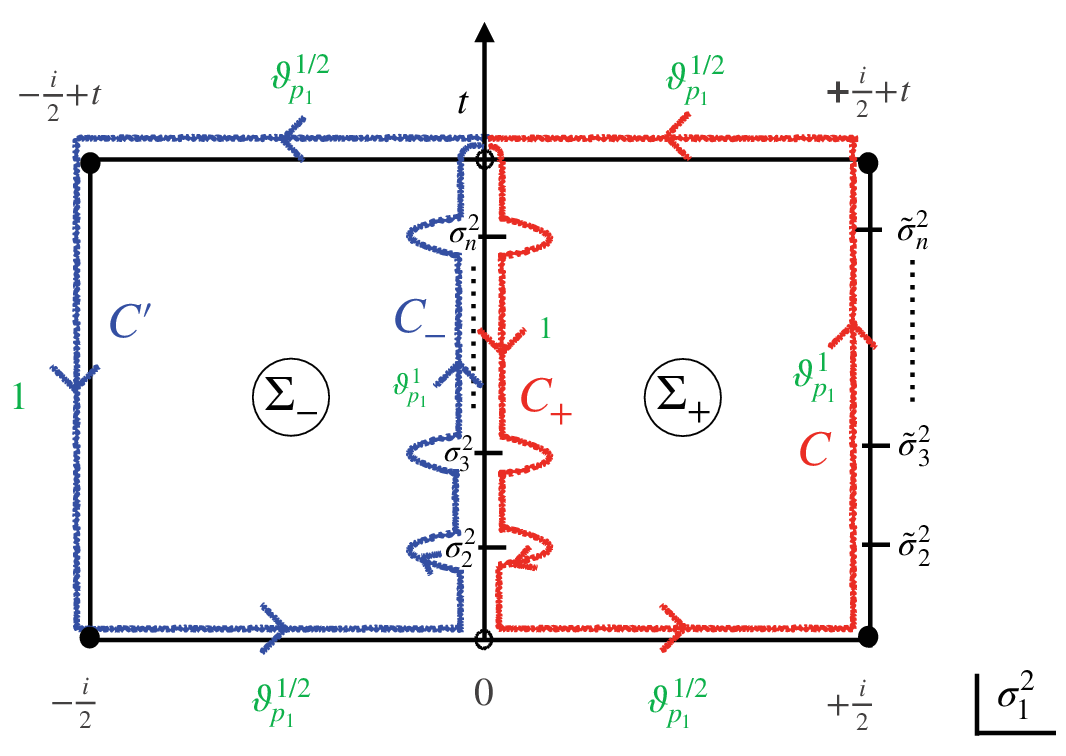}}
\noindent
There are contributions stemming from integrating along $\si_1^2\in(0,t)$ and 
$\si_1^2\in+\fc{i}{2}+(0,t)$. On the other hand, since
\eqn\Naunspitze{
I^{(1)}_{P}(\{\si^2_1+t,\si^2_i\};\ell)= I^{(1)}_{P}(\{\si^2_1,\si^2_i\};\ell-2p_1)\ ,}
the two contributions from the boundaries $\si^2_1=0$ and $\si^2_1=t$ cancel after loop momentum integration. 
In the expressions \tv\ and \TV\ by using (B.12) the loop--momentum dependent phase factor  can be rewritten as follows:
\eqn\expTV{
e^{2\pi \ap \ell\sum\limits_{r=1}^n p_r \si^2_r}=\prod_{r=1}^n\lf(\fc{\theta_1\lf(i\si^2_r+\h\ri)}
{\theta_1\lf(i\si^2_r-it+\h\ri)}\ri)^{\ap\ell p_r}=\prod_{r=1}^n\lf(\fc{\theta_1\lf(i\si^2_r-\h\ri)}
{\theta_1\lf(i\si^2_r-it-\h\ri)}\ri)^{\ap\ell p_r}\ .} 
In this form we evidence a pair of branch points at $\si^2_1=\fc{i}{2}$ and $\si^2_1=\fc{i}{2}+t$. Passing the latter along $C$ by a quarter arc produces the monodromy phases $\vartheta_{p_1}^{1/2}$ and $\vartheta_{p_1}^{-1/2}$, respectively. Similarly, at  $\si^2_1=-\fc{i}{2}$ and $\si^2_1=-\fc{i}{2}+t$.
Furthermore, within $\Si_+$  the branch cut structure at $\si^2_1=0,t$ can also be examined by noting:
\eqn\Regalpspitze{
e^{2\pi \ap \ell\sum\limits_{r=1}^n p_r \si^2_r}=\prod_{r=1}^n \lf(\fc{\theta_1\lf(i\si^2_r\ri)}
{\theta_1\lf(i\si^2_r-it\ri)}\ri)^{\ap\ell p_r}\ .}
Then, we associate a pair of branch points at $\si^2_1=0$  with monodromy phase 
$\vartheta_{p_1}$ and at $\si^2_1=t$ with 
phase $\vartheta_{p_1}^{-1}$, respectively. The corresponding phase structure is depicted in Fig.~12.

On the other hand, we may also consider \othercycle\ with the closed contour $\gamma'=C'\cup C_-$ on a second cylinder $\Si_-$ surface, depicted in blue in Fig.~12 (supplemented by  a global phase factor of $\vartheta_{p_1}$).
Again, there is no contribution from the boundaries $\si^2_1=0$ and $\si^2_1=t$. Thus, in addition to \OneMono\ we obtain the following open string one--loop relation:
\eqnn\OneMonoi{
$$\eqalignno{
 -A^{(1)}(1,2,\ldots,n)&- e^{-\pi i \ap s_{12}}\ A^{(1)}(2,1,\ldots,n)+\ldots+&\OneMonoi\cr
 &-e^{-\pi i \ap \sum\limits_{j=2}^{n-1}s_{1j}}\ A^{(1)}(2,3,\ldots,n-1,1,n)
 =-A^{(1)}(1|2,\ldots,n)[\vartheta^{-1}_{p_1}]\ .}$$}
Of course, \OneMonoi\ simply follows from \OneMono\ by complex conjugation and multiplication  by $(-1)$ taking into account the reversal of the contours. 
Thus, at the level of the integrands \tv\ and \TV\ we can write
\eqn\ObereRegalm{\eqalign{
\oint_{\gamma} d\si^2_1\ \hat I(\{\si^2_i\};\ell)&=-\int^t_0 d\si^2_1\ \hat I_P(\{\si^2_i\};\ell)+
\vartheta_{p_1}\ \int^t_0 d\si^2_1\ I^{(1)}_{NP}(\{\si^2_i\};\ell)=0\ ,\cr
\oint_{\gamma'} d\si^2_1\ \hat I(\{\si^2_i\};\ell)&=\int^t_0 d\si^2_1\ \hat I_P^\ast(\{\si^2_i\};\ell)-
\vartheta_{p_1}^{-1}\ \int^t_0 d\si^2_1\ I^{(1)}_{NP}(\{\si^2_i\};\ell)=0\ ,}}
which after integrating the remaining variables give rise to \OneMono\ and \OneMonoi, respectively.

Let us now consider a closed cycle $\Gamma$ on the doubled surface $\Sigma=
\Si_+\cup \Si_-$, which gives rise to a torus manifold $\Tc$. On the latter we choose the  closed cycle $\Gamma=C\cup C'$  to fulfill   \Othercycle.  After properly adjusting  phases  to render the correct branch when crossing the real axis $\Im(\si_1^2)=0$ (likewise imposing  gluing conditions for $C$ and $C'$) we  have
\eqn\giveindeed{\eqalign{
\oint_{\Gamma} d\si^2_1\ \hat I(\{\si^2_i\};\ell)&=\int_{C'} d\si^2_1\ 
\hat I(\{\si^2_i\};\ell)+\int_{C} d\si^2_1\ \hat I(\{\si^2_i\};\ell)\cr
&=-(1-\vartheta_{p_1})\ \int^t_0 d\si^2_1\ I^{(1)}_{NP}(\{\si^2_i\};\ell)}}
from the contributions of the edges $C$ and $C'$. 
Again, $C$ and $C'$ can be deformed towards the real axis $\Im(\si_1^2)=0$ and we expect some  contributions from the $n\!-\!1$ points $\si_i^2,\;i=2,\ldots,n$ located along the real axis. They can be collected by considering the path along $\Rc=C_+\cup C_-$  with closed cycles encircling the  $n\!-\!1$ points $\si^2_i$, which gives:
\eqn\Vorderkaiserfeldenhuette{\eqalign{
 \oint_{\Rc} d\si^2_1\ \hat I(\{\si^2_i\})&=\vartheta_{p_1}^{1/2}\,\lf\{\vartheta_{p_1}^{1/2}\,\int_{C_-} d\si^2_1\ \hat I(\{\si^2_i\};\ell)+\vartheta_{p_1}^{-1/2}\,\int_{C_+} d\si^2_1\ \hat I(\{\si^2_i\};\ell)\ri\}\cr
 &=\vartheta_{p_1}\int^t_0 d\si^2_1\ \hat I_P^\ast(\{\si^2_i\};\ell)-\int^t_0 d\si^2_1\ \hat I_P(\{\si^2_i\};\ell)\cr
&=\int^t_0 d\si^2_1\ I^{(1)}_{NP}(\{\si^2_i\};\ell)-\vartheta_{p_1}\ \int^t_0 d\si^2_1\ I^{(1)}_{NP}(\{\si^2_i\};\ell)\cr
&=-\oint_{\Gamma} d\si^2_1\ \hat I(\{\si^2_i\};\ell)\ .}}
While the second last  equality  derives from applying \ObereRegalm, the last equality follows from   \giveindeed\ and thus confirms  \Othercycle.
As a consequence, we may write:
\eqn\Schleierwasserfall{\eqalign{
\vartheta_{p_1}\int^t_0 d\si^2_1\ \hat I_P^\ast(\{\si^2_i\};\ell)&-\int^t_0 d\si^2_1\ \hat I_P(\{\si^2_i\};\ell)\cr
&=\lf(1-\vartheta_{p_1}\ri)\; 
\int^t_0 d\si^2_1\ I^{(1)}_{NP}(\{\si^2_i\};\ell)\ .}}
After integrating the latter over the remaining variables we obtain a non--vanishing linear combination of the two relations \OneMono\ and \OneMonoi. Hence, in contrast to the tree--level case \sinuscontour\ at $\si_1^2=0$ there is a  non--vanishing residuum \Vorderkaiserfeldenhuette\ contributing to  \Othercycle\ and arising from closing  the contours $C_+$ and~$C_-$. Alternatively, from \ObereRegalm\ we can also derive the two relations
\eqnn\Going{
$$\eqalignno{
\vartheta_{p_1}^{-1}\int^t_0 d\si^2_1\ \hat I_P(\{\si^2_i\};\ell)&-\vartheta_{p_1}^{1}\int^t_0 d\si^2_1\ \hat I_P^\ast(\{\si^2_i\};\ell)=0\ ,&\Going\cr
\int^t_0 d\si^2_1\ \hat I_P(\{\si^2_i\};\ell)&-\int^t_0 d\si^2_1\ \hat I_P^\ast(\{\si^2_i\};\ell)=
2i\sin\lf(\pi\ap\ell p_1\ri)\; \int^t_0 d\si^2_1\ I^{(1)}_{NP}(\{\si^2_i\};\ell)\ ,}$$}
which after integrating  over the remaining variables give rise to one--loop monodromy relations 
on the doubled surface $\Si=\Si_-\cup\Si_+$ combining \OneMono\ and \OneMonoi.

\comment{Note, the same conclusions are reached without using the loop momentum formalism
by following the approach performed in \HoheneggerKQY.}

\subsec{One--loop KLT relation and a double of contours}

In Subsection 3.1  we have considered in the complex $\si^2_t$--plane a closed contour  composed out of the polygon \edges\ and the path \edgeR\ and applied the Cauchy integral formula \Cauchy, cf. also Fig.~2. We have found:
\eqn\Melan{
(1-\varphi_t^2)\int_{C_1}I+\varphi_t^2\int_{C_+}\hat I+\int_{C_-}\hat I+\varphi_t\int_{C_2}\hat I+\varphi_t\int_{C_2'}\hat I=0\ .}
For this choice one conclusion has been the cancellations of the (non--planar) contributions from the edges $C_2$ and $C_2'$ such that \Melan\ becomes:
\eqn\Melanie{
(1-\varphi_t^2)\int_{C_1}I+\varphi_t^2\int_{C_+}\hat I+\int_{C_-}\hat I=0\ .}
Then, in the complex $\si^2_t$--plane the contours $C_\pm$ can always be aligned along the imaginary axis $\Re(\si^2_t)=0$ such that $C_-=-C_+$ while avoiding  the other  $n-1$ branch points. Thus, \Melanie\ 
leads to the  result \Contour:
\eqn\Regalpturm{
\int_{C_1}I-\int_{C_+}\hat I=0\ .}
In this subsection we shall discuss an alternative set of polygons leading to the same results \Contour\ and \Regalpturm, but  bypassing the edges $C_2$ and $C_2'$.
Concretely, we may consider the edges $\Gamma=C_1\cup C_1'$ and the path $\Rc=C_+\cup C_-$, depicted in Fig.~13.  
\ifig\Region{Alternative contour in the complex $\si^2_t$--plane.}{\epsfxsize=0.5\hsize\epsfbox{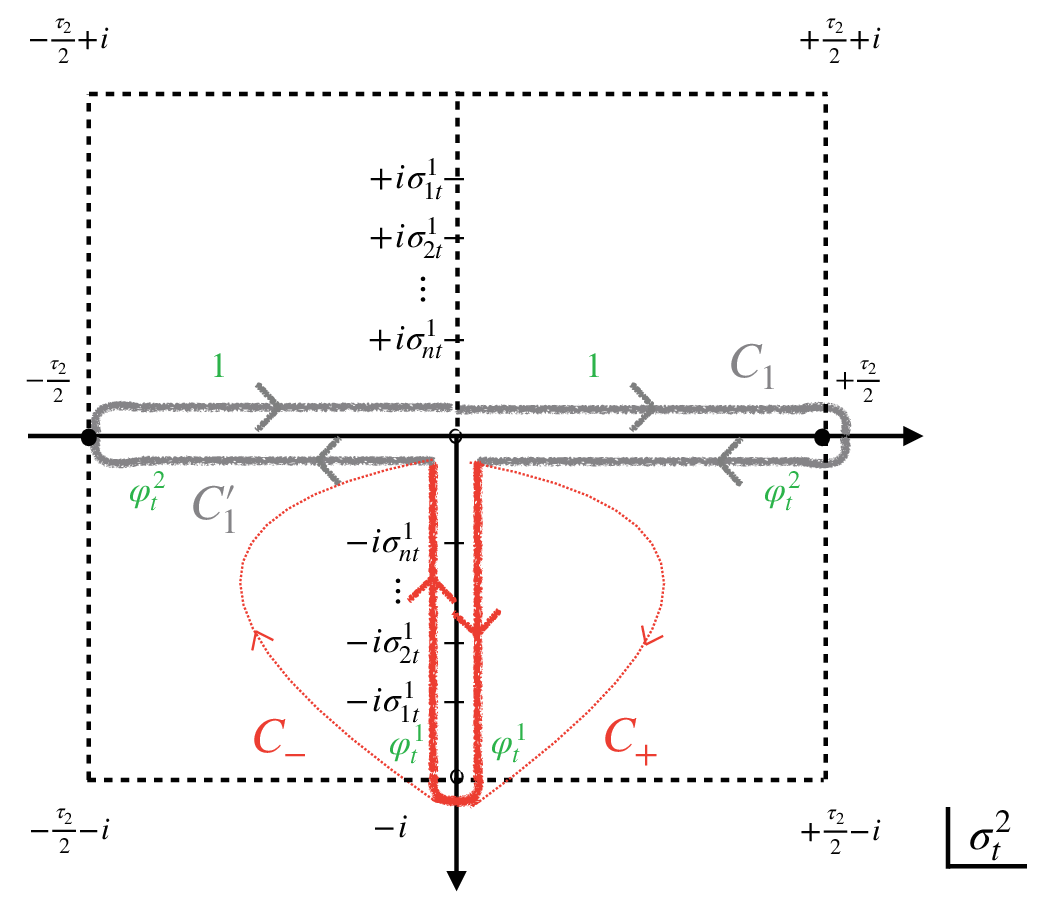}}
\noindent
To investigate the branch cut structure along the closed cycle $\Gamma\cup\Rc$ we look at the theta--function representations \Regalpwand\ and \Antoni. As a consequence, passing the branch point 
$\si_t^2=\fc{\tau_2}{2}$ in clockwise orientation by a semi arc we encounter the phase $\varphi_t^2$.
Similarly, the phase $\varphi_t^{-2}$ is picked up at the point $\si_t^2=-\fc{\tau_2}{2}$.
The cycle $\Rc$ is chosen such that it avoids all the  $n\!-\!1$ branch points located within  the interval  $-1\leq\im(\si^2)\leq0$   by clockwise encircling them, cf. Fig.~13. Again, the cut at $\Re(\si^2_t)=0$ prevents the contributions of $C_+$ and $C_-$ from cancelling out and $\Rc$ amounts to \Peterskoepfl, while the contribution of $\Gamma=C_1\cup C_1'$ gives the first line of 
\Pyramidenspitze. As a consequence, with \Cauchy\ we verify \Contour\ and \Regalpturm.

As a final remark we point out that the  closed contour, depicted in Fig.~13  looks very similar to what we have discussed in Fig.~9  for the tree--level case. However, in the latter case the contours $C_+$ 
and  $C_-$ are joined only at infinity.

\goodbreak
\appendix\appB{Jacobi theta functions and  identities}

A meromorphic function $f(z,\tau)$ of $z$ is defined to be elliptic if it is doubly periodic on a torus, i.e. $f(z+1,\tau)=f(z,\tau)$ and $f(z+\tau,\tau)=f(z,\tau)$.
Elliptic functions having a prescribed set of zeroes and poles in a period parallelogram can be constructed as quotients of Jacobi theta functions.
The latter  are holomorphic and multi--valued functions on $\IC/(\IZ+\tau\IZ)$. In this Appendix we collect some useful identities for Jacobi theta function.

In the computation of one--loop string amplitudes the latter are encountered\foot{We refer the reader to  \Kaidi\ for a detailed and recent presentation with applications.} with characteristics $a,b$ and are defined on $z\in\IC$ as:
\eqn\DefTheta{
\theta\big[^a_b\big](z,\tau)=\sum_{n=-\infty}^\infty e^{\pi i\left(n-\fc{a}{2}\right)^2\tau}\,e^{2\pi i\left(z-\fc{b}{2}\right)\left(n-\fc{a}{2}\right)}\ ,\ \ \ {\rm for}\ \ a,b=0,1\ .}
Throughout the text we will use the notation
\eqn\Thetaone{\eqalign{
\theta_1(z,\tau)&=\theta\big[^1_1\big](z,\tau)\ ,\ \ \
\theta_2(z,\tau)=\theta\big[^1_0\big](z,\tau)\ ,\cr
\theta_3(z,\tau)&=\theta\big[^0_0\big](z,\tau)\ ,\ \ \ 
\theta_4(z,\tau)=\theta\big[^0_1\big](z,\tau)\ .}}
Sometimes, the following representation is useful:
\eqnn\ThetaOne{
$$\eqalignno{
\theta_1(z,\tau)&=2\ e^{\fc{\pi i\tau}{4}}\sum_{n=0}^\infty(-1)^n\ e^{\pi i\tau n(n+1)}\ \sin[\pi (2n+1) z]\cr
&=2q^{1/8}\sin(\pi z)\prod_{n=1}^\infty (1-q^n)(1-q^ne^{2\pi i z})(1-q^n e^{-2\pi i z})\ ,\cr
\theta_2(z,\tau)&=2\ e^{\fc{\pi i\tau}{4}}\sum_{n=0}^\infty e^{\pi i\tau n(n+1)}\ \cos[\pi (2n+1) z]\cr
&=2q^{1/8}\cos(\pi z)\prod_{n=1}^\infty (1-q^n)(1+q^ne^{2\pi i z})(1+q^n e^{-2\pi i z})\ ,&\ThetaOne\cr
\theta_3(z,\tau)&=1+2\sum_{n=1}^\infty e^{\pi i\tau n^2}\ \cos(2\pi n z)\cr
&=\prod_{n=1}^\infty (1-q^n)(1+q^{n-\h}e^{2\pi i z})(1+q^{n-\h} e^{-2\pi i z})\ ,\cr
\theta_4(z,\tau)&=1+2\sum_{n=1}^\infty (-1)^n\ e^{\pi i\tau n^2}\ \cos(2\pi n z)\cr
&=\prod_{n=1}^\infty (1-q^n)(1-q^{n-\h}e^{2\pi i z})(1-q^{n-\h} e^{-2\pi i z})\ .}$$}
Under the modular transformation $(z,\tau)\rightarrow \left(\fc{z}{\tau},-\fc{1}{\tau}\right)$, the theta-function $\theta\big[^a_b\big](z,\tau)$ transforms in the following way
\eqn\TrafoTheta{
\theta\big[^a_b\big]\left(\fc{z}{\tau},-\fc{1}{\tau}\right)=\sqrt{-i\tau}\,e^{\fc{i\pi}{2}\,ab+i\pi\,\fc{z^2}{\tau}}\,\theta\big[^{\ b}_{-a}\big](z,\tau)\ .}
The derivative of the theta function $\theta[^1_1\big](z,\tau)=\theta_1(z,\tau)$ with respect to the first argument can be related to the Dedekind eta function as
\eqn\Dedekind{
\theta'_1(0,\tau)=2\pi\,\eta^3(\tau)\ ,}
where
\eqn\ProductDede{
\eta(\tau)=q^{\fc{1}{24}}\ \prod_{n=1}^\infty(1-q^n)\ ,\ \ \ q=e^{2\pi i\tau}\ ,}
which transforms as:
\eqn\TrafoEta{
\eta(-1/\tau)=\sqrt{-i\tau}\,\eta(\tau)\ .}

\def\frac{\fc}
Furthermore, under shifts of the first argument, the Jacobi theta functions transform as \MOS
\eqn\IdentityShiftTheta{
\theta\big[^a_b\big]\left(z+\frac{\epsilon_1}{2}\,\tau+\frac{\epsilon_2}{2},\tau\right)=e^{-\frac{i\pi\tau}{4}\,\epsilon_1^2-\frac{i\pi\epsilon_1}{2}(2z-b)-\frac{i\pi}{2}\,\epsilon_1\epsilon_2}\,\theta\big[^{a-\epsilon_1}_{b-\epsilon_2}\big](z,\tau)\ .}
With $\theta\big[^a_b\big](z\pm\h,\tau)=\theta\big[^a_{b\mp 1}\big](z,\tau)=e^{\mp i\pi a} \theta\big[^a_{b\pm 1}\big](z,\tau)$ this leads to:
\eqn\halfshifttheta{\matrix{
&\theta_1(z+\h,\tau)=\theta_2(z,\tau), 
&\theta_1(z-\h,\tau)=e^{+ i\pi}\ \theta_2(z,\tau)\ ,\cr
&\theta_2(z+\h,\tau)=e^{- i\pi}\ \theta_1(z,\tau),
&\theta_2(z-\h,\tau)=\theta_1(z,\tau)\ .}}
In particular, we have 
$\theta\big[^a_b\big](z\pm1,\tau)=e^{\mp i\pi a} \theta\big[^a_b\big](z,\tau)$, i.e.:
\eqn\thetashift{\matrix{
&\theta_1(z\pm1,\tau)=e^{\mp i\pi}\ \theta_1(z,\tau),
&\theta_2(z\pm 1,\tau)=e^{\mp i\pi}\ \theta_2(z,\tau)\ ,\cr
&\theta_3(z\pm 1,\tau)=\theta_3(z,\tau),
&\theta_4(z\pm 1,\tau)=\theta_4(z,\tau)\ ,}}
Furthermore, with $\theta\big[^a_b\big](z\pm\fc{\tau}{2},\tau)=e^{-\fc{\pi i\tau}{4}}e^{\mp i\pi (z-\fc{b}{2})} \theta\big[{a\mp 1\atop b}\big](z,\tau)$, we have:
\eqn\shifttheta{\eqalign{
\theta_1(z\pm \fc{\tau}{2},\tau)&=e^{-\fc{i\pi\tau}{4}}\ e^{\mp i\pi z}\ e^{\pm\fc{i\pi}{2}}\theta_4(z,\tau)\ ,\cr
\theta_4(z\pm \fc{\tau}{2},\tau)&=e^{-\fc{i\pi\tau}{4}}\ e^{\mp i\pi z}\ e^{\pm\fc{i\pi}{2}}\theta_1(z,\tau)\ ,}}
while with $\theta\big[^a_b\big](z\pm\tau,\tau)=e^{-\pi i\tau}e^{\mp i\pi (2z-b)} \theta\big[{a\mp 2\atop b}\big](z,\tau)$, we obtain:
\eqn\shiftthetaa{\eqalign{
\theta_1(z\pm \tau,\tau)&=e^{-i\pi\tau}\ e^{\mp 2i\pi z}\ e^{\pm i\pi}\theta_1(z,\tau)\ ,\cr
\theta_4(z\pm \tau,\tau)&=e^{-i\pi\tau}\ e^{\mp 2i\pi z}\ e^{\pm i\pi}\theta_4(z,\tau)\ ,\ \cr
\theta_2(z\pm \tau,\tau)&=e^{-i\pi\tau}\ e^{\mp 2i\pi z}\ \theta_2(z,\tau)\ ,\cr
\theta_3(z\pm \tau,\tau)&=e^{-i\pi\tau}\ e^{\mp 2i\pi z}\ \theta_3(z,\tau)\ .}}
Finally, for $\tau=i\tau_2$ we can write $\overline{\theta\big[^a_b\big]( z,\tau)}=\theta\big[^{a}_{b}\big](\bar z,\tau)$, i.e.:
\eqn\complextheta{\eqalign{
\overline{\theta_1( z,\tau)}&=\theta_1(\bar z,\tau)\ ,\ 
\overline{\theta_2( z,\tau)}=\theta_2(\bar z,\tau)\ ,\cr
\overline{\theta_3( z,\tau)}&=\theta_3(\bar z,\tau)\ ,\ 
\overline{\theta_4( z,\tau)}=\theta_4(\bar z,\tau)\ .}}

\listrefs
\end